\documentclass[a4paper,11pt]{article}
\pdfoutput=1 

\usepackage{jcappub} 

\usepackage[T1]{fontenc} 
\usepackage{amsmath}
\usepackage{amssymb}

\definecolor{darkgreen}{rgb}{0,0.5,0}
\definecolor{darkred}{rgb}{0.7,0,0}
\definecolor{buddhistmonkrobe}{rgb}{0.966,0.42,0.23}
\definecolor{mermaid}{rgb}{0,0.72,0.42}

\newcommand{\diff}[0]{\mathrm{d}}

\usepackage{booktabs}
\usepackage{mathrsfs}
\usepackage{enumitem}
\usepackage{xifthen}

\DeclareMathAlphabet{\mathpzc}{OT1}{pzc}{m}{it}

\newcommand{\mcol}[3]{\multicolumn{#1}{#2}{#3}}
\newcommand{\erfc}{\ensuremath{\mathop{\rm erfc}}}

\newcommand{\aprop}{\mathrel{\raise.423ex\hbox{$\propto$}\kern-.76em\lower.8ex\hbox{$\approx$}}}
\newcommand{\simprop}{\mathrel{\raise.42ex\hbox{$\propto$}\kern-.76em\lower.72ex\hbox{$\sim$}}}

\newcommand{\vka}{\ensuremath{\varkappa}}

\newcommand{\vro}{\ensuremath{\varrho}}
\newcommand{\vfi}{\ensuremath{\varphi}}

\newcommand{\Eta}{\ensuremath{\mathrm{H}}}

\newcommand{\Kappa}{\ensuremath{\mathrm{K}}}

\newcommand{\ionfrac}[2][env]{\ensuremath{\phi_{#2}^{\ifthenelse{\equal{#1}{}}{}{\scriptscriptstyle\langle\mathrm{#1}\rangle}}}}
\newcommand{\ionfracUp}[2][env]{\ensuremath{\Phi_{#2}^{\ifthenelse{\equal{#1}{}}{}{\scriptscriptstyle\langle\mathrm{#1}\rangle}}}}
\newcommand{\fraction}[2][env]{\ensuremath{f_{\ifthenelse{\equal{#2}{j}\OR\equal{#2}{k}}{#2}{\rm #2}}^{\ifthenelse{\equal{#1}{}}{}{\scriptscriptstyle\langle\mathrm{#1}\rangle}}}}
\newcommand{\numspec}[2][env]{\ensuremath{N_{\ifthenelse{\equal{#2}{j}\OR\equal{#2}{k}}{#2}{\rm #2}}^{\ifthenelse{\equal{#1}{}}{}{\scriptscriptstyle\langle\mathrm{#1}\rangle}}}}
\newcommand{\temperature}[2][env]{\ensuremath{T_{\ifthenelse{\equal{#2}{j}\OR\equal{#2}{k}}{#2}{\rm #2}}^{\ifthenelse{\equal{#1}{}}{}{\scriptscriptstyle\langle\mathrm{#1}\rangle}}}}
\newcommand{\density}[2][env]{\ensuremath{n_{\ifthenelse{\equal{#2}{j}\OR\equal{#2}{k}}{#2}{\rm #2}}^{\ifthenelse{\equal{#1}{}}{}{\scriptscriptstyle\langle\mathrm{#1}\rangle}}}}

\newcommand{\abundance}[2][]{\ensuremath{\vfi_{\ifthenelse{\equal{#2}{j}\OR\equal{#2}{k}}{#2}{\scriptscriptstyle\rm #2}}^{\ifthenelse{\equal{#1}{}}{}{\scriptscriptstyle\langle\mathrm{#1}\rangle}}}}
\newcommand{\Enhance}[2][]{\ensuremath{\Eta_{\ifthenelse{\equal{#2}{j}\OR\equal{#2}{k}}{#2}{\scriptscriptstyle\rm #2}}^{\ifthenelse{\equal{#1}{}}{}{\scriptscriptstyle\langle\mathrm{#1}\rangle}}}}
\newcommand{\pdf}{\ensuremath{\mathpzc{p}}}
\newcommand{\cv}{\ensuremath{\sf cv}}
\newcommand{\icv}[1][1]{\ensuremath{{\sf cv\small\phantom{\big|}}^{\scriptscriptstyle\!\!\!\!{-}\!#1}}}
\newcommand{\maxval}[1]{\ensuremath{\widehat{#1}}}

\newcommand{\av}[2][env]{\ensuremath{\left\langle#2\right\rangle}^{\ifthenelse{\equal{#1}{}}{}{\!\scriptscriptstyle\langle\mathrm{#1}\rangle}}}
\newcommand{\Prob}[3][]{\ensuremath{\mathscr{P}^{#3}_{\!\mathrm{\scriptscriptstyle #1}#2\,}}}
\newcommand{\edens}[3][]{\ensuremath{\mathscr{U}^{#3}_{\mathrm{#1}#2\,}}}
\newcommand{\cdiff}[3][]{\ensuremath{\mathscr{D}^{#3}_{\mathrm{#1}#2\,}}}
\newcommand{\length}[3][]{\ensuremath{\mathscr{L}^{#3}_{\!\mathrm{#1}#2\;}}}

\newcommand{\tscale}[3][]{\ensuremath{\tau^{#3}_{\mathrm{#1}#2}}}
\newcommand{\lscale}[3][]{\ensuremath{\lambda^{#3}_{\mathrm{#1}#2}}}

\newcommand{\rigidity}[3][]{\ensuremath{R^{#3}_{\mathrm{#1}#2}}}
\newcommand{\energy}[3][]{\ensuremath{E^{#3}_{\mathrm{#1}#2}}}
\newcommand{\area}[3][]{\ensuremath{\mathscr{A}^{#3}_{\mathrm{#1}#2}}}
\newcommand{\Ekin}[2]{\energy[kin]{#1}{#2}}

\newcommand{\envind}{{\scriptscriptstyle\rm \langle env\rangle}}
\newcommand{\gsaind}{{\scriptscriptstyle\rm \langle gsa\rangle}}
\newcommand{\kB}{\ensuremath{k_{\scriptscriptstyle\mathrm{B}}}}
\newcommand{\pinj}{\ensuremath{p_{\rm inj}}}
\newcommand{\xinj}{\ensuremath{\zeta_{\rm inj}}}
\newcommand{\adindex}[2][\langle env\rangle]{\ensuremath{\gamma_{\rm ad#2}^{\scriptscriptstyle\rm#1}}}
\newcommand{\rcomp}{\ensuremath{\vro_{\rm sh}}}
\newcommand{\bsh}{\ensuremath{\beta_{\rm sh}}}
\newcommand{\Mach}{\ensuremath{\mathscr{M}_{\rm sh}}}

\newcommand{\down}{\ensuremath{{\scriptscriptstyle -}}}

\newcommand{\pspecind}{\ensuremath{\alpha}}
\newcommand{\turbind}{\ensuremath{\vka}}
\newcommand{\Bturb}{\ensuremath{\delta\!B}}

\newcommand{\turbspec}[2][\Bturb]{\ensuremath{\mathscr{I}_{\!{#1}}}}
\newcommand{\turbscale}{\ell_0}
\newcommand{\cutoff}{\mathscr{C}}
\newcommand{\enhgas}{\eta_{ji}}
\newcommand{\enhdust}{\eta_{\rm\scriptscriptstyle dust}}
\newcommand{\enhmax}{\maxval{\eta}_{\rm\scriptscriptstyle gas}}
\newcommand{\rLarmor}{r_{\rm\scriptscriptstyle L}}

\newcommand{\cm}{\ensuremath{\mathrm{cm}}}

\newcommand{\pc}{\ensuremath{\mathrm{pc}}}

\newcommand{\erg}{\ensuremath{\mathrm{erg}}}
\newcommand{\eV}{\ensuremath{\mathrm{eV}}}

\title{Nonthermal element abundances at astrophysical shocks}

\author[a,1]{Bj\"{o}rn Eichmann,\note{Corresponding author.}}
\author[b]{J\"{o}rg P. Rachen}

\affiliation[a]{Ruhr Astroparticle and Plasma Physics Center (RAPP Center), Ruhr-Universit\"at Bochum, Institut f\"ur Theoretische Physik IV, 44780 Bochum, Germany}
\affiliation[b]{Astrophysical Institute, Vrije Universiteit Brussel (VUB), Pleinlaan 2, 1050 Brussels, Belgium}

\emailAdd{eiche@tp4.rub.de}
\emailAdd{Jorg.Paul.Rachen@vub.be}

\abstract{The nonthermal source abundances of elements play a crucial role in the understanding of cosmic ray phenomena from a few GeV up to several tens of EeV. We present a first systematic approach to describe the change of the abundances from the thermal to the nonthermal state via diffusive shock acceleration by a temporally evolving shock. We consider hereby not only ionization states of elements contained in the ambient gas, which we allow to be time dependent due to shock heating, but also elements condensed on solid, charged dust grains which can be injected into the acceleration process as well. Our generic parametrized model is then applied to the case of particle acceleration by supernova remnants in various ISM phases, for which we use state-of-the-art computation packages to calculate the ionization states of all elements. The resulting predictions for low energy cosmic ray (LECR) source abundances are compared with the data obtained by various experiments. 

We obtain excellent agreement for shocks in warm ionized ISM environments, which include HII regions, if dust grains are injected into the diffusive shock acceleration process with a much higher efficiency than ions. Less dependence of the fit quality is found on the mass-to-charge ratio of ions. For neutral environments, assuming that there are shocks in the weakly ionized component, and for the hot ionized medium we obtain generally inferior fits, but except for the cold neutral medium we do not exclude them as subdominant sites of Galactic cosmic ray production. The key challenge is found to be putting the LECR abundance of pure gas phase elements like neon and the (semi-)volatile elements phosphorus, sulfur and chlorine into the right balance with silicon, calcium and elements of the iron group. We present a brief outlook to the potential consequences of our results for the understanding of the composition around the second knee or the cosmic ray spectrum, or for the viability of explaining ultra-high energy cosmic rays with a dominant contribution by radio galaxies. 
}

\begin{document}
\maketitle
\flushbottom

\section{Introduction}
\label{sec:intro}
The spectrum of cosmic rays extends over more than ten orders of magnitude in total energy and is largely described as a power law with multiple breaks \cite{Matthiae2019}. Longest known are the steepening at around 1 PeV \cite{KulikovKhristiansen1959} commonly called ``the knee''. Later, a long-suspected flattening at around 4 EeV has been experimentally confirmed \cite{Bird+1994}, and became known as ``the ankle'' to complete the anatomic analogy, which was then stretched by the discovery of another steepening at around 100 PeV \cite{KASKADE-Grande2006}, now called ``the second knee'' \cite{Hoerandel2007}. All these features are thought to hide information about the origin of cosmic rays, but unfortunately their meaning is disputed: The knee could be explained by a source cutoff \cite[e.g.,][]{LagageCesarsky1983,Caprioli+2010_contrSNR} or a transport effect in the Galaxy \cite{Ptuskin1993,Candia+2002}, in which case the second knee would mark the Galactic source cutoff. However, the latter could also be caused by an emerging second Galactic \cite{Thoudam:2016syr} or an extragalactic component. So, the ankle, traditionally understood as a transition from a steep Galactic to a flatter extragalactic cosmic ray spectrum, could also be a feature of the extragalactic cosmic ray spectrum itself, caused by Bethe-Heitler losses of protons in the cosmic microwave background \cite{Berezinsky+2005}, spallation of heavy nuclei in compact sources \cite{UFA2015}, Bethe-Heitler modified acceleration of primordial gas at large scale shocks around clusters of galaxies \cite{Kang+1997, Murase+2009, Thoudam:2016syr}, different classes radio galaxies as cosmic ray accelerators \cite{Eichmann_2019}, or by interactions with the gas around the cosmic ray sources \cite{Kachelriess+2017, FangMurase2018} --- to mention just some of the options. 

A key role in resolving these ambiguities has always been played the measurement of the chemical composition of high energy cosmic rays. While for long time this has been known in detail only at low ($\sim$ GeV) energies, improvements in our understanding of the cosmic ray shower development in our atmosphere have opened composition studies up to the highest energies \cite{Gaisser+1993}. In fact, the first attempt to model the transition from Galactic to extragalactic cosmic rays has been based on separating a rather light and and rather heavy air showers from early experiments under use of a predicted extragalactic spectrum, yielding a self-consistent result \cite{Rachen:1993gf}.   
Meanwhile model independent results of many modern experiments on the base of measurements of different shower components on the ground (e.g., \cite{Kascade2003,KascadeGrande2004}), and/or of the shower maximum $X_{\rm max}$ with air fluorescence \cite{AugerComposition2014,TAComposition2018} and radio techniques \cite{Buitink+2014,Huege2016}, have shown that the ``heaviness'' of the cosmic ray spectrum, usually measured by the average logarithm of nuclear mass, $\av[]{\ln A}$, follows a similarly complex pattern as the spectrum itself. While this information is now commonly used to characterize models for the knee and the second knee to ankle regimes on experimental grounds \cite{Hoerandel2004,GaisserStanevTilav2013}, most astrophysical models have the problem that their predictions regarding composition are often based on pure heuristics. This brings us back to one of the main problems of cosmic ray theory: understanding the injection of non-relativistic matter into the acceleration process. 

It is widely believed that the origin of the low-energy cosmic rays (LECR) is first seeded by stellar activity ejecting a solar like composition at about a few MeV, that is subsequently accelerated to GeV till TeV energies by the passing shock wave of a supernova remnant (SNR) via the mechanism of diffusive shock acceleration  (DSA, \cite{Drury:1983zz} and references therein). In this simple picture, the first ionization potential (FIP) is the key parameter that sets the resulting abundances after the acceleration process, as the shock is only able to process charged particles in the usual tenuous, collisionless astrophysical environments. Observations can confirm this dependency to a certain extend, however, it has also been realized since several decades that the volatility of elements, i.e.\ its ability to condense into solid compounds, correlates in a similar manner. Thus, refractory elements that are locked in dust grains are preferentially accelerated compared to those in the gas phase yielding a second explanation of the LECR abundances. A first, detailed investigation on this ambiguity has been performed by Ellison, Meyer and Drury \cite{MeyerDruryEllison1997, EllisonDruryMeyer1997}, showing that both volatility and the mass-to-charge ($A/Q$) ratio among those volatile elements need to be taken into account. They are able to explain the general trend of the LECR abundances, but they neither discuss the influence of the ambient SNR environment nor details on the different states of ionization of the elements or the development of the shock environment.

An important aspect of the injection mechanism is its dependence on the mass and charge of the species: A common approach is to suppose that a certain factor of the particle's Larmor radius yields the critical quantity that defines the minimal momentum $\pinj$ to inject particles into the acceleration, leading to $\pinj\propto Q$ for a given momentum. Based on this presumption Malkov \cite{Malkov1998} showed in a simplified 1D model for small wave amplitudes of the turbulence, that the injection efficiency increases dependent on the wave amplitude as a function of $(A/Q)$. However, recent 2D particle-in-cell (PIC) simulations \cite{Caprioli+2017, Hanusch+2019} indicate that $\pinj\simprop A$. Further, these work also determine the mass-to-charge dependence of the injection efficiency. Though, their results are quite different as discussed in more detail in Sect.~\ref{sec:particleInjection}, both models provide some agreement to the observed LECR data. Still, these models can only provide a gross representation of LECR abundances, as solely light, single ionized gas elements have been included in these simulations. 

In this paper, we provide a joined description of the acceleration elements considering their injection \textit{both} through ionized gas and dust grains. We consider the individual ionization states of the elements in the ambient gas around the SNR for different ISM phases, and account for their temporal changes downstream of the shock as well as the influence of the SNR evolution. We also take into account state of the art knowledge on gas and dust fractions for all elements in different ISM phases. While our goal is to provide an explanation of the detailed structure of cosmic ray element abundances at low energies, the motivation of this paper goes farther: By developing a universal, parametric model of the DSA injection process --- gauged by explaining LECR measured in our Galaxy from ISM properties, but generally applicable to DSA in any cosmic environment --- we pave the way to a theoretical source modeling of cosmic ray composition up to the highest energies, for both Galactic and extragalactic sources. 

The paper is organised as follows: In Sect.~\ref{sec:basics} we introduce the basic physics of particle acceleration, that are needed to determine the source abundances of different elements. In that, we discuss the injection timescale for different magnetic field models and provide the ionization states upstream and downstream of the shock for different SNR environments. Subsequently, we apply our model to the LECR data in Sect.~\ref{sec:results}, and conclude with an outlook to further applications of the model in Section~\ref{sec:conclusions}.

\section{Elemental abundances in diffusive shock acceleration}
\label{sec:basics}

\subsection{The scale-invariant DSA spectrum}

Diffusive shock acceleration is a process which populates phase space with a nonthermal spectrum ${\mathpzc f}(p) \aprop p^{-4}$ by repeatedly scattering charged particles back and forth across a shock front in a self similar process \cite{Drury:1983zz}. Under ideal conditions, i.e., as long space and time limitations do not play a role and we can consider the process stationary, this power law strictly holds above an injection momentum \pinj\ up to arbitrarily high momenta $p$, and the power law index depends only on properties of the background plasma and \textit{not} on any properties of the particle, like its charge or mass. In this paper, we will not use ${\mathpzc f}(p)$ but the isotropic \textit{differential number density} in momentum, $dN/dp = 4\pi p^2 {\mathpzc f}(p) \propto p^{-\pspecind}$, with
\begin{equation}
\pspecind \;=\; \frac{\rcomp+2}{\rcomp-1} \;=\; \frac{3\adindex{}-1+4\Mach^{-2}}{2-2\Mach^{-2}}\quad,
\end{equation}
for a shock with a velocity $\bsh c$, Mach number $\Mach = \bsh \sqrt{\av{mc^2}\!\!/\adindex{}\kB\temperature{}}$ and compression ratio $\rcomp = (\adindex{}+1)/(\adindex{}-1+2\Mach^{-2})$, where \adindex\ is the adiabatic index of the background plasma, $\av{mc^2}$ the average rest mass energy of its particles and $\kB\temperature{}$ their thermal (i.e., kinetic) energy. For a monoatomic non-relativistic gas ($\adindex{}\!\!=\frac53$) and strong shocks ($\Mach\to\infty$) we then obtain the well-known result $\pspecind=2$, but note that low Mach numbers can make the spectrum significantly steeper while a lower adiabatic index can make it significantly flatter.

It is important to note that the DSA differential number spectrum is a strict power law only if given in momentum; the number spectrum in kinetic energy is $dN/d\Ekin{}{} \propto \Ekin{}{-3/2}$ for $\alpha=2$ in the non-relativistic regime, only at ultra-relativistic energies $dN/d\Ekin{}{} \simeq (1/c)\,dN/dp$. The same relation at high energies holds for the differential number spectrum in total energy, $dN/dE$, while the entire low energy part of the spectrum is squeezed into a narrow peak around $E=\av[]{mc^2}$. The differential number spectrum in particle rigidity $\rigidity{}{}=pc/Q e$ is $dN/d\rigidity{}{} \propto dN/dp$ only if the charge $Q e$ of the particle does not change during acceleration -- an assumption we \textit{cannot} make in the context of this paper where we consider acceleration from a low ionized state at injection to a potentially fully ionized state at very high energies.

\subsection{Turbulent magnetic fields and particle diffusion}

A turbulent magnetic field $\Bturb$ generated at a critical wavenumber $k_0$ can be described for $k\ge k_0$ by its inertial spectrum
\begin{equation}\label{turbspec}
\turbspec{}(k) \equiv \turbspec[0]{} \left(\frac{k}{k_0}\right)^{\!\!-\turbind} =\; 
	\displaystyle\frac{\edens{\Bturb}{}\Kappa}{k_0} \left(\frac{k}{k_0}\right)^{\!\!-\turbind} \qquad \text{for}\quad \turbind\ge 1\;.
\end{equation}
with the normalisation factor
\begin{equation}
    \Kappa = \max\!\big(\,\turbind{}-1\;,\;1/\ln(\,\maxval{k}/k_0)\,\big)\;.
\end{equation}
Writing it in this way ensures that even for a continuous transition $\turbind\to 1$ the normalisation remains continuous and reasonable for all values, although in practice only few discrete values for $\turbind$ will be important, like $\turbind=1$ for the often used assumption of Bohm-diffusion, or  $\turbind=\frac53$ for Kolmogoroff-turbulence, which has been explored in the context of DSA by Biermann \& Strittmatter \cite{BiermannStrittmatter1987}, hereafter referred to as BS87. 

For $k\le k_0$, simulations of dynamo processes that are shown to work even in tenuous astrophysical conditions \cite{Rincon+2016} suggest that the saturated spectrum in $\turbspec{}(k)$ is roughly flat (\turbind=0) between $k_0$ and the inverse outer scale $1/\length{}{}$ \cite{BrandenburgSubramanian2005,TobiasCattaneo2008}, although the theoretical expectation is rather that of a Kazantsev spectrum ($\turbind=-\frac32$). This allows us to use a simple approach to estimate the energy density of the regular field that is assumed in linear DSA theory to determine the particle gyration, as
\begin{equation}\label{regularB}
\turbspec[0]{} = \frac{\edens{B}{}\length{}{}}{\turbscale-1} \;. 
\end{equation}
\length{}{} is the characteristic size of the system and we introduce the dimensionless critical scale $\ell_0 = \length{}{}k_0$ for which we can assume a canonical value $\turbscale\sim 10$. In our approach we consider diffusion only for Larmor radii $\rLarmor \le 1/k_0$, and as postulated above the spectrum for $k<k_0$ acts as a regular field 
\begin{equation}\label{B-dB-connection}
    B=\sqrt{8\pi\edens{B}{}}\;=\;\sqrt{8\pi\edens{\Bturb}{}(\turbscale-1)/\turbscale} 
\end{equation}
where the latter relation to the turbulent field is implied by continuity of the turbulence spectrum. The Larmor radius is then given by $\rLarmor = R/B \le \length{}{}/\turbscale$, if $R=pc/eQ$ is the rigidity for an ion of charge number $Q$. 

The resonant scattering condition then implies that in each Larmor-period the pitch angle $\vfi_{\rm p}$ of the particle with the mean field is changed by a small amount $\Delta\vfi_{\rm p} = \sqrt{8 \pi k\turbspec{}(k)}/B$, so that in a random walk a ``scattering event'' occurs after ${\sim}\Delta\vfi_{\rm p}^{-2}$ periods, hence the scattering time scale is $\tscale[s]{}{} \sim \rigidity{}{}/B c \Delta\vfi_{\rm p}^2 $ \cite{Drury:1983zz}, and we obtain a diffusion coefficient
\begin{equation}\label{Cdiff_gen}
\cdiff{}{}(\rigidity{}{}) \;=\; \frac13 \tscale[s]{}{}(\rigidity{}{})\,\beta\, c^2 \;=\; \frac{(\turbscale-1)}{3\Kappa\turbscale^2}\,\length{}{}\!\,\beta\,c\left(\frac{\rigidity{}{}}{\rigidity{0}{}}\right)^{\!\!2-\turbind}
\end{equation}
for a particle velocity $\beta\,c$ and a limiting rigidity $\rigidity{0}{} = B/k_0$ above which DSA breaks down. Note that the strength of the magnetic field enters only through $\rigidity{0}{}$, which results from connecting $B$ and $\Bturb$ via Equation~\ref{B-dB-connection}. We replaced $k_0$ by the observable $\length{}{}$ and the canonical number $\turbscale$ for this case. 
If this connection is dropped and $B_{\rm env}$ and $\Bturb_{\rm env}$ are given by the ambient medium, e.g.\ the Galactic magnetic field, the expression reads
\begin{equation}\label{Cdiff_env}
\cdiff{}{}(\rigidity{}{}) \;=\; \frac{\beta\,c\, \lscale[coh]{,\Bturb}{\envind} }{3 \Kappa}\,\frac{\edens{B}{\envind}}{\edens{\Bturb}{\envind}}\left(\frac{\rigidity{}{}}{\rigidity{0}{}{}}\right)^{\!\!2-\turbind}\quad,
\end{equation}
where $\lscale[coh]{,\Bturb}{\envind}$ is the empirical coherence length of the turbulent field. For the spiral arms of the Galaxy, there is \cite{IMAGINE2018}
\[
\edens{B}{\gsaind} \sim  3\times 10^{-13}\,\erg/\cm^3\;,\quad\edens{\Bturb}{\gsaind} \sim 10^{-12}\,\erg/\cm^3\quad{\rm and}\quad\lscale[coh]{,\Bturb}{\gsaind} \gtrsim 1\pc\,. 
\]
Note that $\lscale[coh]{,\Bturb}{\envind} \sim \maxval\rLarmor$, thus for $\turbind>1$ Eq.~\ref{Cdiff_env} is identical to the expression found in BS87 up to some little understood factors of order $1$.  Note also that for the generated regular field, Eq.~\ref{regularB}, the assumption $\turbscale\gg 1$ leads to $b\simeq 1$ in BS87 notation.

In the most general case, we have to expect that ambient and generated turbulent fields arising from different mechanisms co-exist with different indices $\turbind$. Downstream of the shock, $(\rcomp-1)/\rcomp$ of the kinetic energy $\edens[kin]{}{}$ is converted into internal energy, and plasma turbulence is likely to be the transmitter from large scale motion to thermalization at small scales. Such plasma turbulence can invoke dynamo action \cite{BrandenburgSubramanian2005,Rincon+2016} and thus produce turbulent magnetic fields with some efficiency $\xi_B$, i.e.
\begin{equation}\label{BturbTD}
    \edens{\Bturb}{\rm TD} = \xi_B\,\edens[kin]{}{}\,(\rcomp-1)/\rcomp = {\textstyle\frac{1}{2}}\, \density{ion}{}\av{mc^2}\,\bsh^2\,\xi_B\,(\rcomp-1)/\rcomp\;. 
\end{equation}
Note that only the charged part of the ambient particle density \density{ion}{} is actually able to interact with the shock. Further, charged hydrogen and helium determine the dynamics of the system, so that the dynamical energy density $\density{ion}{}\av{mc^2}$ is derived from these two gas elements. 
Such turbulent dynamo spectra would typically show a Kolmogorov spectrum \cite{BrandenburgNordlund2011}, $\turbind=\frac53$, and simulations suggest that about a fifth of the energy in plasma turbulence is converted into magnetic fields \cite{Seta+2020}, so if we assume that about one third of the internal energy is in the turbulent cascade, $\xi_B \sim 0.05\ldots 0.1$ may be considered a reasonable guess (see also the recent work of Chamandy and Shukurov \cite{ChamandyShukurov2020}, note that they compare with the total energy in supernovae and not just the kinetic energy of the shock). Upstream of the shock, the only agent which could produce turbulence beyond the ambient small scale field are the cosmic rays themselves. Bell \cite{Bell2004} suggested a mechanism of turbulence generation by cosmic rays for the case of strong turbulence, $B\sim \Bturb$, now commonly called the Bell-instability, which saturates at   
\begin{equation}\label{BturbBI}
    \edens{\Bturb}{\rm BI} = \textstyle{\frac12}\edens[cr]{}{}\bsh = 
    {\textstyle\frac{1}{4}}\, \density{ion}{}\av{mc^2}\,\bsh^3\,\xi_{\rm cr}\,(\rcomp-1)/\rcomp\;. 
\end{equation}
with $\edens[cr]{}{}=\xi_{\rm cr}\,\edens[kin]{}{}\,(\rcomp-1)/\rcomp$, where $\xi_{\rm cr}\sim 0.1\ldots0.3$ denotes the efficiency of conversion of internal energy into cosmic rays --- a value range long known to be required for supernovae to produce the observed cosmic ray flux that is now also confirmed by simulations \cite{Caprioli2012}, and as well expected from basic equipartition arguments for nonthermal processes \cite{RachenManifesto2019}. This wave turbulence would typically inherit the flat energy spectrum of the generating cosmic rays, i.e., $\turbind = 1$. Upstream this generated turbulent field will typically dominate the magnetic energy density as long $\bsh$ does not become too small, but downstream it is in competition with the turbulent dynamo field. Note that although $\xi_{\rm cr}\gtrsim\xi_B$, with the additional factor $\bsh$ in Eq.~\ref{BturbBI} we always have $\edens{\Bturb}{\rm BI} \ll \edens{\Bturb}{\rm TD}$; nevertheless, because of the flatter spectrum we can expect that Bohm diffusion caused by the Bell instability dominates at the smallest scales downstream, governing cosmic ray injection. 

To calculate the total diffusion coefficient as a function of particle rigidity, we can recall that in the ultrarelativistic case it is simply a scattering time scale times $\beta c$, and as time scales are inversely additive, diffusion coefficients are as well so that we write the total diffusion coefficient as
\begin{equation}
\cdiff{}{}(\rigidity{}{}) \simeq \left[ \sum_m \cdiff{m}{-1}(\rigidity{}{}) \right]^{-1}\;.
\end{equation}
We can then determine the $\cdiff{m}{}$ from Eq.~\ref{Cdiff_gen} (for generated fields) or \ref{Cdiff_env} (for ambient fields where we have to consider that these are compressed downstream of the shock) for each involved process separately, and of course we have to do this for the upstream and downstream regime separately as different processes may contribute on both sides of the shock.

\subsection{Particle injection and elemental enhancement}
\label{sec:particleInjection}

The fully normalised spectrum of cosmic rays of species $j$ (we use here a single counting index over all relevant stable isotopes) with nuclear mass $A_j$ and nuclear charge $Z_j$ injected into DSA during a time interval $\Delta t$ can be written as 
\begin{equation}
\left[\frac{dN_j}{dp}\right]_{\Delta t} = \frac{(\pspecind-1)\,\chi_j\,\Enhance{j}}{\pinj}\,  \left(\frac{p}{\pinj}\right)^{-\pspecind} \cutoff\big(R_j(p)\big)\;.
\label{eq:dNdp}
\end{equation}
Here we have put all the complexity of the physics in the breakdown regime around the critical rigidity into a cutoff function $\cutoff(\rigidity{j}{}(p))$ that depends only of particle rigidity $\rigidity{j}{}(p)$, and note that for typical values $\alpha \simeq 2$ the shape of this cutoff function has virtually no impact on the normalisation to the total particle number. Up to modifications due to particle charge and mass, this is basically given by
\begin{equation} 
\chi_j = \Prob{0}{} \density{}{}\, \abundance{j} \,\area[sh]{}{}(t)\, \bsh(t)\,c\,\Delta t\;,
\end{equation}
where $\density{}{}$ is the number density of (charged \emph{and} neutral) atoms in the environment under consideration, $\abundance{j}$ the normalized cosmic abundance of element species $j$ ($\sum_j\abundance{j} =1$), $\area[sh]{}{}$ the shock area and $\bsh c$ its velocity, which both may be time dependent (e.g., in a supernova remnant). $\Prob{0}{}$ is the probability of a proton crossing the shock to be injected into DSA, so if all hydrogen is ionized, $\chi_{j{=}1}$ is indeed the number of protons injected. To consider the mass and charge dependence of injection 
\begin{align}
\Enhance{j} \: = \:& \:\fraction{j}\,\ionfracUp{j0}\,\sum_{i=1}^{Z_j}  \frac{\enhgas}{\tscale[inj]{,ji}{}}\int_0^{\Delta t} \ionfrac{ji}(t)\, \exp\!\left(
-\frac{t}{\tscale[inj]{,ji}{}}
\right)dt \nonumber\\
&\:+ \; \fraction{j}\,(1-\ionfracUp{j0})\,\sum_{i=1}^{Z_j}  \frac{\enhgas}{\rcomp\tscale[inj]{,ji}{}}\int_0^{\Delta t} \ionfrac{ji}(t)\, \exp\!\left(
-\frac{t}{\rcomp\tscale[inj]{,ji}{}}
\right)dt \nonumber\\
&\:+\; \xinj^{1-\pspecind}\,\left(1-\fraction{j}\right) \enhdust\;.%
\label{enhancement}
\end{align} 
Here $\fraction{j}$ is the fraction of the elemental species that is present in gas in a given environment\footnote{
We use this notation in equations and note that the index $j$ runs over all relevant isotopes. Generally the gas fraction is given per element and valid for all isotopes of this element, and we denote it simply with the element name (e.g., $\fraction{Ar}$). The environment may be explicitly specified (e.g., $\fraction[WNM]{Ar}$). In general theoretical discussions without reference to a specific element, we also use the notation $\fraction{gas}$.}
with a fraction \ionfracUp{j0} of neutral gas in the upstream medium. The first term of the sum of Eq.~\ref{enhancement} represents the contribution of the neutral fraction of gas in the upstream that passes the shock and gets ionized and inserted into the downstream flow at a distance $x$ from the shock, whereas the second term provides the contribution of the already ionized gas particles. 
The last term in the sum describes the injection of dust grains for which we assume a constant injection probability $\Prob[\!dust]{,j}{} = \enhdust\Prob{0}{}$, i.e., independent of the mass, charge or structure of the dust grain. 
From empirical considerations, essentially by the demand to fit the cosmic ray enhancement ratio of calcium, an element which is depleted from the gas phase by ${>}99\%$ in all ISM phases except maybe the hottest ones, over hydrogen, which is only found in gas and has no mass-to-charge dependent injection efficiency, we expect $\enhdust \gg 1$ \cite[see also][]{Simpson1983, Voyager2016}.

While charged dust grains can be assumed to be injected into the DSA process with the momentum they carry from the upstream, $m_{\rm g} \bsh c$, ions thermalize to a downstream temperature $T_{-}$ and only supra-thermal particles can be injected into DSA. We write this as 
\begin{equation}
 \pinj/A_j \equiv p_0 = \xinj m_p \bsh c
\end{equation}
with  $\xinj\sim 3$ being found in simulations \cite{Caprioli2012}. Hence, we treat all elements \textit{to be injected with the same momentum per nucleon}, no matter whether they are injected via the gas or dust phase, and correct then the normalisation of the dust part by $\xinj^{1-\alpha}$ to account for the shift in injection momentum. This simple assumption also frees us from all details regarding dust acceleration, as no matter how and on which time scales a grain is sputtered during the acceleration process, whatever breaks out remains with the same momentum per nucleon. We therefore consider dust grains just as ensembles of atoms, which are injected together but otherwise accelerated like single ions. 

The time integral in Eq.~\ref{enhancement} requires some explanation. We treat here the ionization fraction $\ionfrac{ji}(t)$ in the downstream medium as a time dependent function, to allow the shock to ``breed'' its own ions for DSA injection even when most of the matter is neutral. Obviously, collisionless MHD shocks do not see neutral material and vice versa -- a neutral atom will not react on the shock front at all. 
So, the speed of uncharged particles is not effected by the shock whereas the speed of charged particles is reduced by a factor $1/\rcomp$. 
According to the basic principles of the acceleration process \cite{Drury:1983zz}, the probability to be injected into DSA is
$\propto \exp\big(-\bsh c x/ \rcomp \cdiff{\down}{}\!(\pinj)\big)$. This leads to the integral in Eq.\,\ref{enhancement} if we define the injection timescale of uncharged particles as
\begin{equation}
\tscale[inj]{,ji}{} = \frac{\rcomp\,\cdiff{\down}{}\!(\pinj)}{\xinj\,\bsh^2 c^2}\,,
\label{eq:injectionTime}
\end{equation}
where we want to choose $\Delta t \gg \tscale[inj]{,ji}{}$ for consistency. 
In general, $\tscale[inj]{,ji}{} \propto (A/Q)^{2-\turbind}$ so that --- dependent on the diffusion coefficient --- $\tscale[inj]{,ji}{}$ can be of the order of years in an external field of Kolmogoroff type, but it can also be down to minutes if there is efficient generation of magnetic turbulence by cosmic rays via the Bell-instability. 
For the calculation of the time dependent ionization fractions $\ionfrac{ji}$ and a final discussion see Sect.~\ref{sec:shockHeating} and \ref{sec:shockEvol}.
 
Finally we have to turn to the last yet unexplained quantity in  Eq.~\ref{enhancement}, the charge and mass dependent enhancement factors $\enhgas$ for gas-phase ions. 
These factors have been found in simulations, although there is some debate on its actual value: 

Caprioli et al.~\cite{Caprioli+2017}, hereafter C+17, compared the nonthermal spectra of different elements at the same kinetic energy per charge $\energy[k/Q]{}{} \equiv \energy[kin]{}{}/Q = p^2/(2 m_p A Q)$ and showed that $dN/d\energy[k/Q]{}{} \propto (A/Q)^2$ at $\Mach\geq 10$ without a saturation for high $(A/Q)$ values. Certainly, as enhancement is related to an injection probability which cannot be larger than $1$, in fact cannot be expected to be even close to $1$, so there must be a maximum value $\enhmax\ll \Prob{0}{-1}$. 
Simulations have shown that at $\Prob{0}{} \gtrsim 10^{-4}$ the part of the energy density in cosmic rays becomes almost constant between $(10\dots30)\%$ leading to an efficient CR acceleration \cite{Caprioli2012}. Thus, we should assume a maximum value $\enhmax \sim 10^3$, so that we obtain 
\begin{equation}
    \enhgas = \min\left[ \left(\frac{A_j}{Q_{ji}}\right)^{\!\!\frac{5-\alpha}{2}}, \enhmax \right]\quad\text{ for } \Mach\geq 10\;,
    \label{eq:C+17enhanc}
\end{equation}
where we used that
\begin{align}
 \frac{dN}{dp}\,\frac{dp}{d\energy[k/Q]{}{}}\: & = \: \frac{(\pspecind-1)\,\chi_j}{2\,(2m_p)^{\frac{\pspecind-1}{2}}}\,\Enhance[gas]{j}\,\left(\frac{A_{j}}{Q_{ji}}\right)^{\frac{\pspecind-1}{2}}\,\energy[k/Q]{}{{-\frac{1+\pspecind}{2}}}\,p_0^{\pspecind-1}\,\,\cutoff(R_j(\energy[k/Q]{}{})) \nonumber \\ & = \: \frac{dN}{d\energy[k/Q]{}{}} \propto (A_j/Q_{ji})^2
 \label{eq:deriveScaling}
 \end{align}
 and approximate that the $(A/Q)$ scaling of the gas dependent enhancement function $\Enhance[gas]{j}\equiv \Enhance{j}(\fraction{j}=1)$ is determined by $\Enhance[gas]{j}\propto \enhgas$. 
 Note that at $\Mach<10$ the simulations show a flattening of the $(A/Q)$ scaling, which we do not incorporate --- though in this work the Mach numbers are mostly significantly higher than 10 --- and with respect to the following prediction we refer to Eq.~\ref{eq:C+17enhanc} as the \emph{steep gas enhancement}.  
 
 Hanusch et al.~\cite{Hanusch+2019}, hereafter H+19, compared the resulting nonthermal tail of the spectra above a certain kinetic energy per nucleon $\energy[k/A]{}{} \equiv \energy[kin]{}{}/A = p^2/(2 m_p A^2)$ yielding $dN/d\energy[k/A]{}{} \propto (A/Q)$ with a saturation at a factor of $\sim 10$, and then a decline towards larger $A/Q$. That the latter is, what they claim, a continuous trend towards a vanishing enhancement for neutral particles, can hardly be believed due to the empirical need of $\enhdust\gg 1$ for dust, which is expected to be in a range $A/Q\sim 10^2\ldots10^{12}$ (see Sect.~\ref{sec:dust}). 
 Therefore, we suggest to use the shallow enhancement that is expected from H+19, but keep the maximal value $\enhmax$ where saturation starts as a free parameter within the physical reasonable boundaries that have been discussed before. Thus, our second approach for the case of a \emph{shallow gas enhancement} is given by 
\begin{equation}
    \enhgas = \min\left[ \frac{A_j}{Q_{ji}}, \enhmax \right]\,,
    \label{eq:H+19enhanc}
\end{equation}
based on the same consideration as before (\ref{eq:deriveScaling}). 

This procedure allows to compare the different predictions from C+17 and H+19, and at the same time it provides the possibility to test the impact of the saturation of the gas enhancement at arbitrary $(A/Q)$ ratios.

\section{Gas and dust phases in the ISM} 

\subsection{ISM phases in a nutshell}

The previously introduced injection model is applicable to an arbitrary astrophysical acceleration environment. In principle, the shock within our Galaxy develops into the diffuse gas of the interstellar medium (ISM), which is observed to exist in four dominant phases: a cold neutral medium (CNM), a warm neutral (WNM) and ionized (WIM) medium as well as a hot ionized medium (HIM). In addition, the cold ISM is often subdivided into the dense, molecular phase of the ISM that is observed to be distributed in the shape of discrete molecular clouds (MC) and the cold atomic gas surrounding these clouds that is usually labeled as the CNM. In the vicinity of young, massive stars --- typically O or B stars --- there is another ISM phase known as \textit{HII-regions}, where the dense clouds in which star formation is taking place are ionized by high-energy UV-radiation emitted by the central star. These regions show a particle density $\density[HII]{}{}=(600\pm400)\,\text{cm}^{-3}$ that is similar to the MC, although its ionized part scales with the inverted distance to the star \cite{HuntHirashita2009}. 
The temperature of the HII-region is similar to the one of the ionizing star, so about $10^4\,\text{K}$, however it also decreases with distance. The hot, ionized gas of the HII-region is typically expanding supersonically outwards into a cavity of subsonic, neutral gas --- most likely a MC ---leading to the formation of shock waves. 
HII regions have a low volume filling factor in the Galaxy, but as they are typically observed where star formation is taking place, and also supernovae explosions are expected to happen near star forming regions due to the low lifetime of their progenitor stars, a significant fraction of supernova remnants are expected and also observed in this environment \cite{Crowther2013}. Moreover, recent observations \cite{GeyerWalker2018} indicate a further sub-phase of the ISM: a hot, dense companion of the WIM, coined the \textit{dense warm ionized medium (DWIM)} with densities quite similar to the CNM, but much higher temperatures, which is supposed to exist mainly at very low scale height at the central Galactic disk. Although not yet established, we decide to include it in our study as a presents an interesting intermediate case between the regular WIM and HII regions. An accurate characterisation of the ISM, e.g.\ \cite{HeilesTroland_2003}, is surely more complicated, but in the context of this work we can stick to the simplified picture presented so far. For more details on the physics of the ISM the interested reader is referred to the review article by Ferri\`{e}re \cite{Ferriere2001_ISMreview}.

In this work, we are predominantly interested in the resulting ionization fractions $\ionfracUp{ji}$ of the different gas elements dependent on the ISM phases. For this we use the photoionization code \texttt{Cloudy}\footnote{\url{https://www.nublado.org/}} and perform a simulation that includes the characteristic temperatures and densities of the phases as given in column 1--4 of Table \ref{AParameters}. We treat the gas in all phases in collisional ionization equilibrium and include the impact by the cosmic ray background as well as the local interstellar radiation field by Black \cite{Black1987} --- more details on the simulation setup are given in Appendix \ref{app:cloudySim}. Hereby, we ensure that the resulting ionization fraction $\ionfracUp{11}$ of hydrogen, as given in the fifth column of Table \ref{AParameters}, matches the observations. In addition, the table provides the average ionization fraction $\av{\ionfracUp[]{}}=\sum_{j}\abundance{j}(1-\ionfracUp{j0})$ using normalized solar abundances, and the charged particle density $\density{ion}{}$. 
It is worth to note that there is a tremendous difference between the ionization fractions of different elements, especially in the cold phases of the ISM. In the warm ionized phases of the ISM $\ionfrac{11}$ becomes similar to the ionization fraction of the more heavy elements, except for helium that provides $(1-\ionfracUp[WIM]{20})=0.968$ for the WIM, and $(1-\ionfracUp[DWIM]{20})=0.227$ for the DWIM. 
In the case of an HII-region, $\ionfracUp{ji}$ depends on the distance $d_*$, effective temperature $T_*$ and photon luminosity $Q_*(H)$ of the star whose radiation field is the predominant source of ionization. We use the typical values of an O6 star \cite{OsterbrockFerland2006_book} with $T_*=43\,600\,\text{K}$ and $Q_*(H)=10^{49.34}\,\text{s}^{-1}$. There is no generic distance value as the HII-region is an expanding structure that shows $0.1\,\text{pc}\lesssim d_*\lesssim 0.1\,\text{kpc}$, with huge consequences on the resulting ionization fraction. However, if the O6 star is on length scales of $d_*$ close to the center of the SNR --- which is trivially the case in the scenario that the hot star that created the HII region is also the progenitor of the SNR --- we can suppose in the following that $d_*\simeq r_{\rm sh}$, where $r_{\rm sh}$ is the shock radius. Hence, the ionization fraction $\ionfracUp[HII]{ji}$ of an HII-region changes with time dependent on the temporal evolution of the shock. We return to the discussion of possible consequences of this simplifying assumption in Section~\ref{sec:conclusions}. 
\begin{table*}[t]
\centering
\footnotesize
\caption{Properties of the ISM phases.}
  \begin{tabular}{l c c c c c c c}
  \toprule
           &   $T_+^{\langle \mathrm{env} \rangle}\,[\text{K}]$  & $\Delta T_+^{\langle \mathrm{env} \rangle}\,[\text{K}]$  &
$\density{}{} \,[\text{cm}^{-3}]$  & $\Delta \density{}{} \,[\text{cm}^{-3}]$ & $\ionfracUp{11}$ & $\av{\ionfracUp[]{}}$ & \density{ion}{} \\ 
  \midrule 
   CNM & $125$ & $\phantom{0}75$ & $\phantom{0}42$ & $38$ & $3.0\times 10^{-6}$ & $1.5\times 10^{-4}$ & $0.010$\\
   WNM & $5\,750$ & $2\,750$ & $0.35$ & $0.25$ & $0.011$ & $0.011$ & $0.0096$ \\
   WIM & $8\,000$ & $2\,000$ & $0.20$ & $0.01$ & $0.997$ & $0.994$ & $0.20$\\
   DWIM & $20\,000$ & $2\,000$ & $42$ & $38$ & $0.934$ & $0.879$ & $37$\\
   HIM & $550\,000$ & $450\,000$ & $0.0055$ & $0.0045$ & $1.0$ & $1.0$ & $0.0055$\\
   \bottomrule
\end{tabular}
  \label{AParameters}
\end{table*}

\subsection{Shock heating and downstream ionization time scales}
\label{sec:shockHeating}

The ionization fractions $\ionfracUp{ji}$ do not account for the heating of the ISM by the the passing shock. Downstream of the shock, the ionization fraction $\ionfrac{ji}$ in the shocked medium needs to account for the temperature shift from the upstream to the downstream medium. In principle, the Rankine-Hugoniot conditions provide that
\begin{equation}
\temperature{-} \approx {\textstyle\frac{3}{16}} \av{mc^2} \bsh^2/\kB\,,
\end{equation}
for $\adindex{}\!\!=\frac53$ and $\Mach\gg 1$, so that the downstream temperature only depends on the shock velocity and not on the upstream temperature of the different ISM environments.\footnote{Note that this relation only holds in the case of an adiabatic shock, wheres for isothermal conditions the shift towards $T_{-}^{\langle \mathrm{env} \rangle}$ is only valid on length scales of the cooling length.} We suppose in the following that collisional ionization determines the ionization fractions of the downstream gas, since we will focus on nonradiative shocks with $\bsh>0.001$, so that $\temperature{-} > 10^5\,\text{K}$.

\begin{figure}[htbp]
\centering
\includegraphics[height=.2342\textheight]{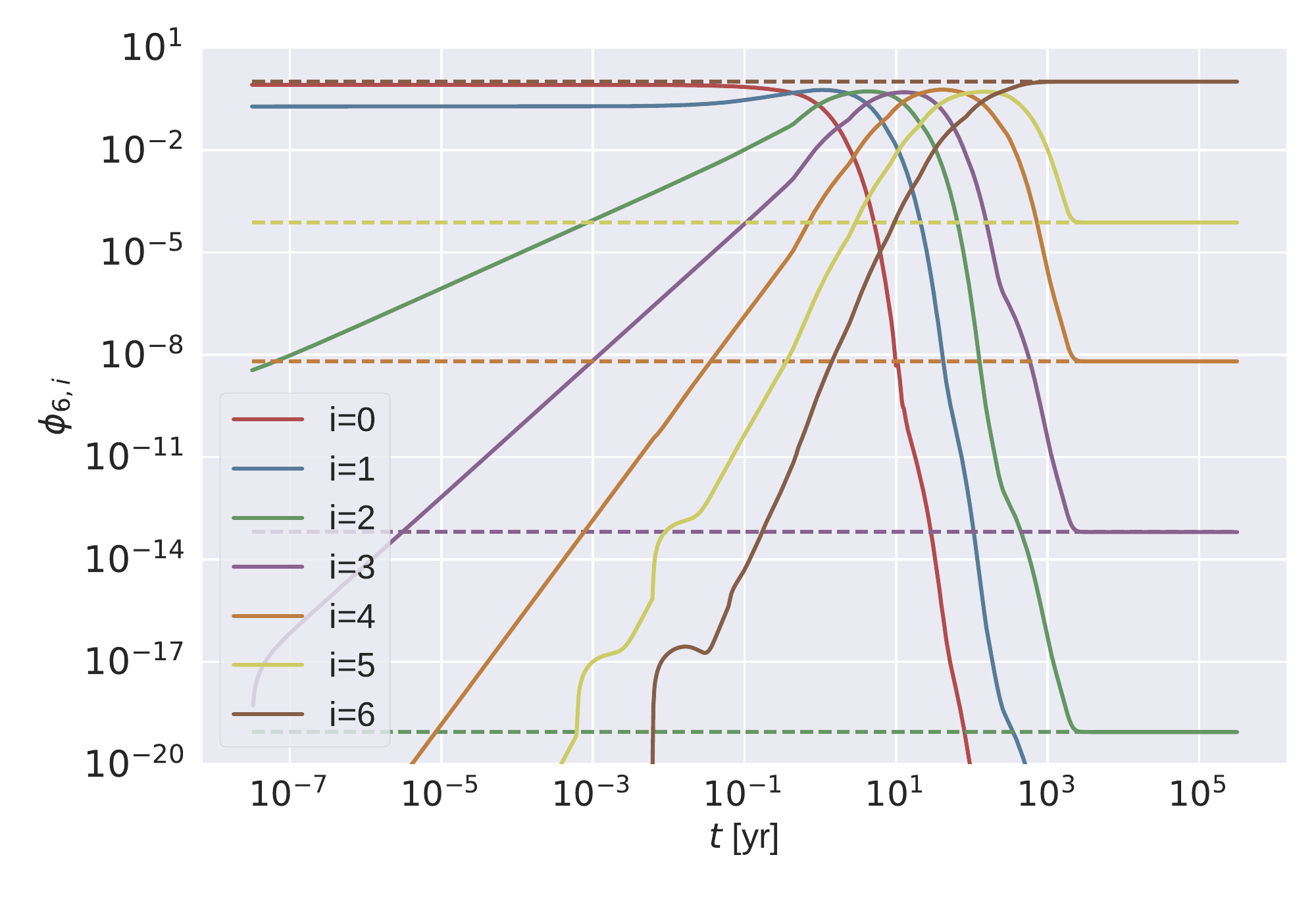}
\includegraphics[height=.23\textheight]{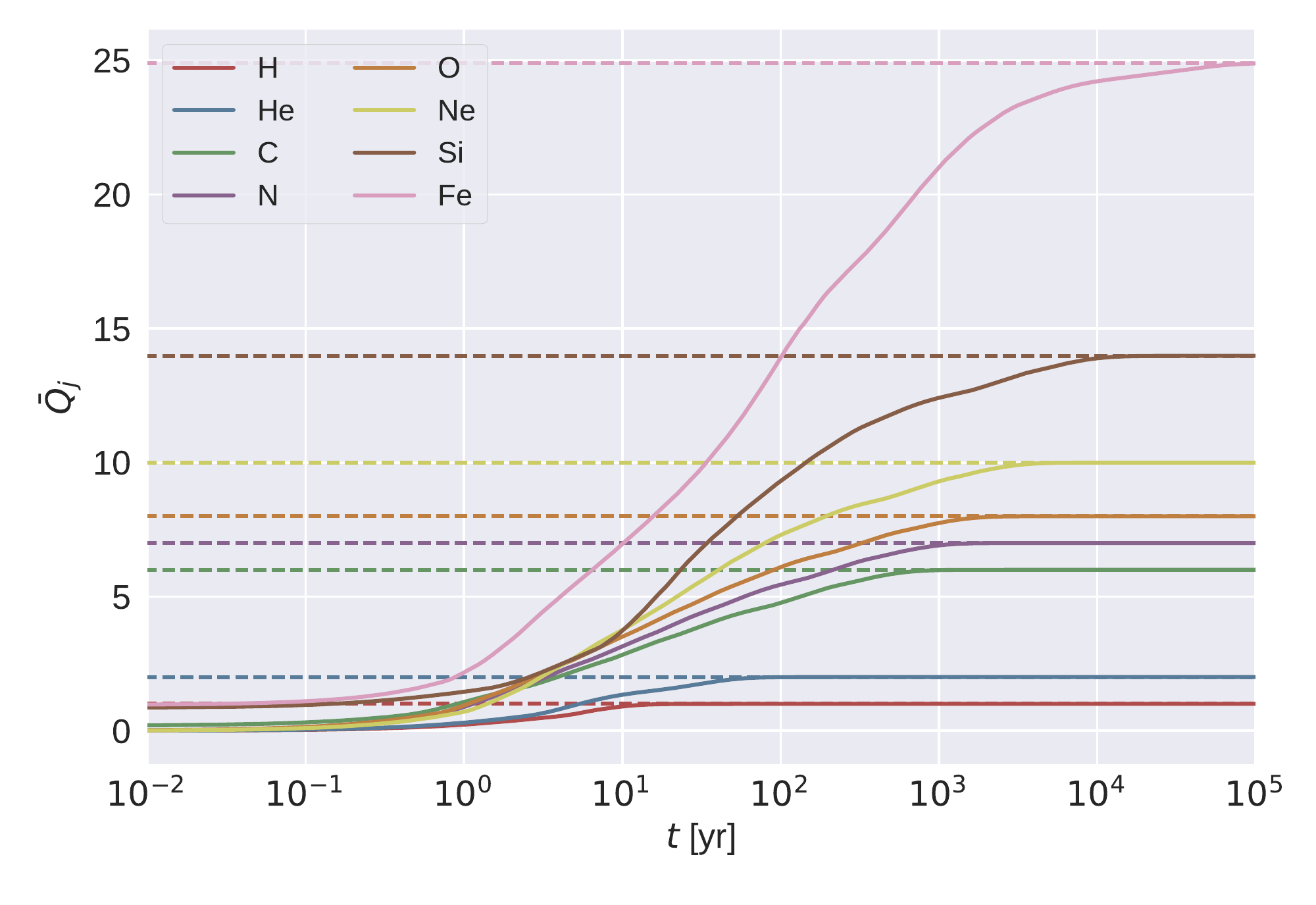}
\caption{Temporal evolution of the ion fraction \ionfrac{6i}(t) of carbon (\emph{left}) and the mean charge number (\emph{right}) downstream of the shock. Therefore, the typical values of the CNM --- as given in Table \ref{AParameters} --- are used upstream, and a typical downstream temperature of $T_{-}=10^8\,\text{K}$. The solid lines are derived from Eq.~\ref{eq:rateEq} and the dashed lines indicate the resulting values in the case of a collisional ionization equilibrium in the downstream ---- using the available algorithm from \texttt{Chianti}.}
\label{fig:temEvol}
\end{figure}

The injection into the acceleration process according to Eq.~\ref{eq:injectionTime} happens on timescales that are usually much smaller needed to accomplish an equilibrium state. Therefore, we have to solve the rate equation
\begin{equation}
    \frac{\diff \ionfrac{ji}}{\diff t} \simeq
    \begin{cases}
    \ionfrac{j,i-1}\,\omega_{j,i-1}^{\rm ion} - \ionfrac{ji}\,\omega_{ji}^{\rm rec}\, \quad&\text{for }i>\hat{Q}(t)\,,\phantom{\Big|} \\
    \ionfrac{j,i+1}\,\omega_{j,i+1}^{\rm rec} - \ionfrac{ji}\,\omega_{ji}^{\rm ion}\, \quad&\text{for }i<\hat{Q}(t)\phantom{\Big|} 
    \end{cases}
    \label{eq:rateEq}
\end{equation}
with the boundary condition that $\sum_{i=0}^{Z_j} \ionfrac{ji}=1$, to obtain the ionization fraction $\ionfrac{ji}(t)$ at a given time $t$ in the downstream medium. Here, we do not include all individual ionization and recombination stages of the ion, but use a simple two-level --- so-called \emph{coronal} --- approximation, where the equilibrium state evolves from the balance of collisional ionization and (radiative, dielectronic) recombination --- independent of the given electron density. However, this approach is only suitable at densities $\lesssim 10^7\,\text{cm}^{-3}$, whereas at much higher densities ($\gtrsim 10^{16}\,\text{cm}^{-3}$) the collisional ionization is followed by a three-body recombination leading to the familiar Saha equation for the ionization balance. So, the rate equation (\ref{eq:rateEq}) considers two adjacent ionization stages based on the maximally occupied ionization state $\hat{Q}(t)$. The latter is not known from first principles if the equilibrium state is not accomplished, but for a sufficiently small $\diff t$ the ionization fractions hardly change, so that $\hat{Q}(t)\approx \hat{Q}(t-\diff t)$, and latter can be derived from the previous calculation step. The total ionization and recombination rate $\omega_{ji}^{\rm ion}$ and $\omega_{ji}^{\rm rec}$, respectively, are taken from the \texttt{Chianti} atomic database\footnote{\url{https://www.chiantidatabase.org/}} which is also part of \texttt{Cloudy}. This procedure adopts some of the basic ideas of the ChiantiPy v0.9.5\footnote{\url{https://chiantipy.readthedocs.io/en/latest/}} realization of the collisional ionization equilibrium \cite{2009A&A...498..915D}, and ensures that the equilibrium values are obtained for $t\rightarrow\infty$, as displayed in Figure~\ref{fig:temEvol}. 

It is shown that significant changes of the ion fraction due to the heating by the shock happen on timescales of the order of years. A thermal equilibrium is accomplished at $t\sim 10\,\text{yr}$ for hydrogen and at $t\sim 10^5\,\text{yr}$ for iron. Note that there is some slight dependence on the given ionization fraction in the upstream medium and downstream temperature. Still, for injection timescales $\ll 0.1\,\text{yr}$ the ionization fraction that enters DSA is in principle already given by the ionization fraction \ionfracUp{ji} of the ambient upstream medium. 

\subsection{Shock evolution}
\label{sec:shockEvol}
The acceleration of Galactic CRs is typically associated with supernova remnants (SNRs), i.e., at shocks that result from the interaction of stellar material ejected by a supernova with the ambient gas. In the early stages of the SNR, the \emph{ejecta-dominated (ED) phase} and the \emph{Sedov-Taylor (ST) phase}, radiative losses from the SNR are insignificant. First, the ejecta from the stellar progenitor are highly supersonic preceding a so-called \emph{blast-wave shock} which compresses and heats the ISM environment. The counteraction of the shocked ISM gas leads to the formation of a \emph{reverse shock}, that initiates the ST phase where the expansion is driven by the thermal pressure of the shocked gas yielding an adiabatic expansion. At an age of about $10^5\,\text{yr}$ radiative losses become significant so that the adiabatic expansion breaks down and the SNR enters the so-called \textit{snowplow phase}. 

\begin{figure}[htbp]
\centering
\includegraphics[width=.495\linewidth]{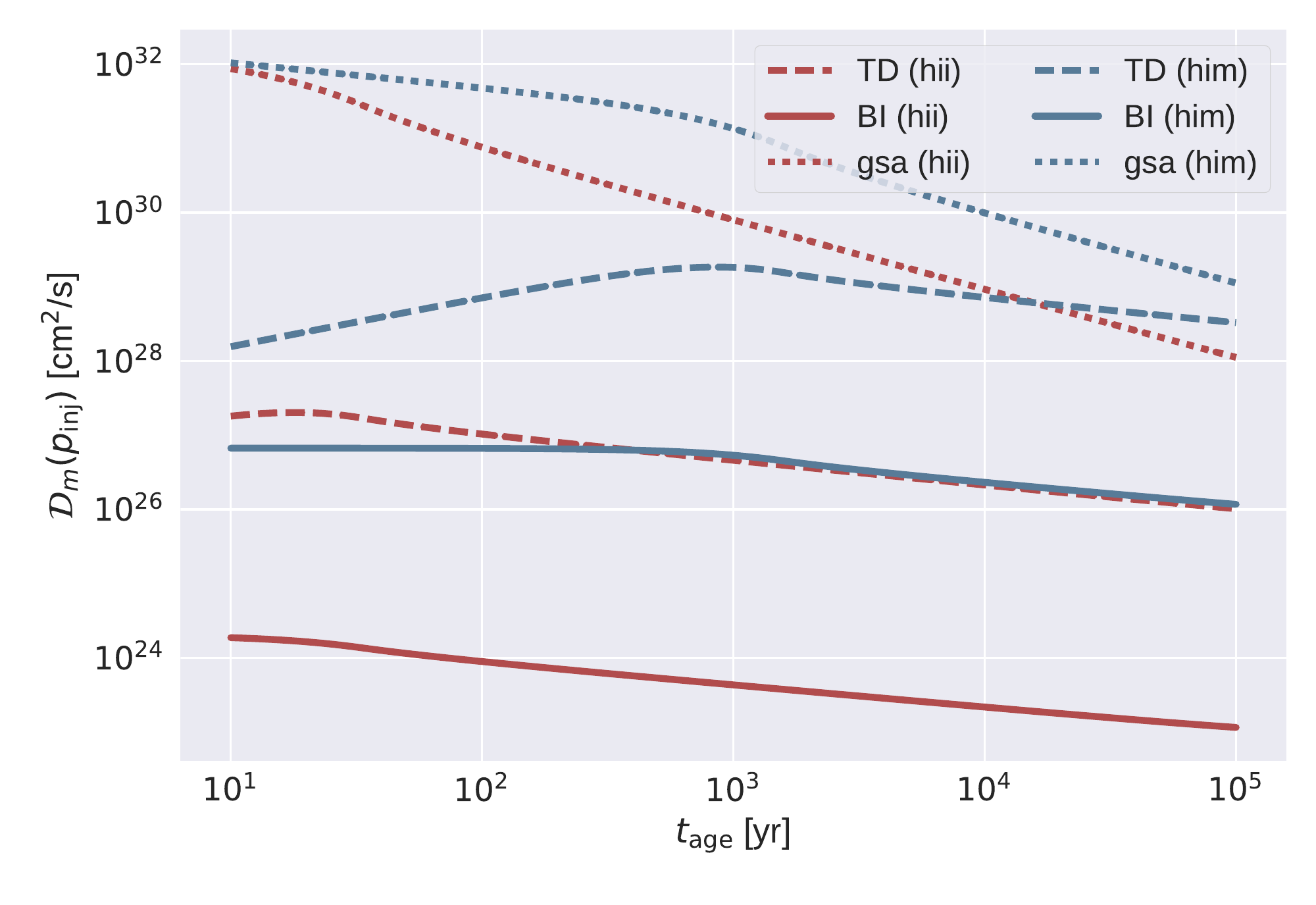} 
\includegraphics[width=.495\linewidth]{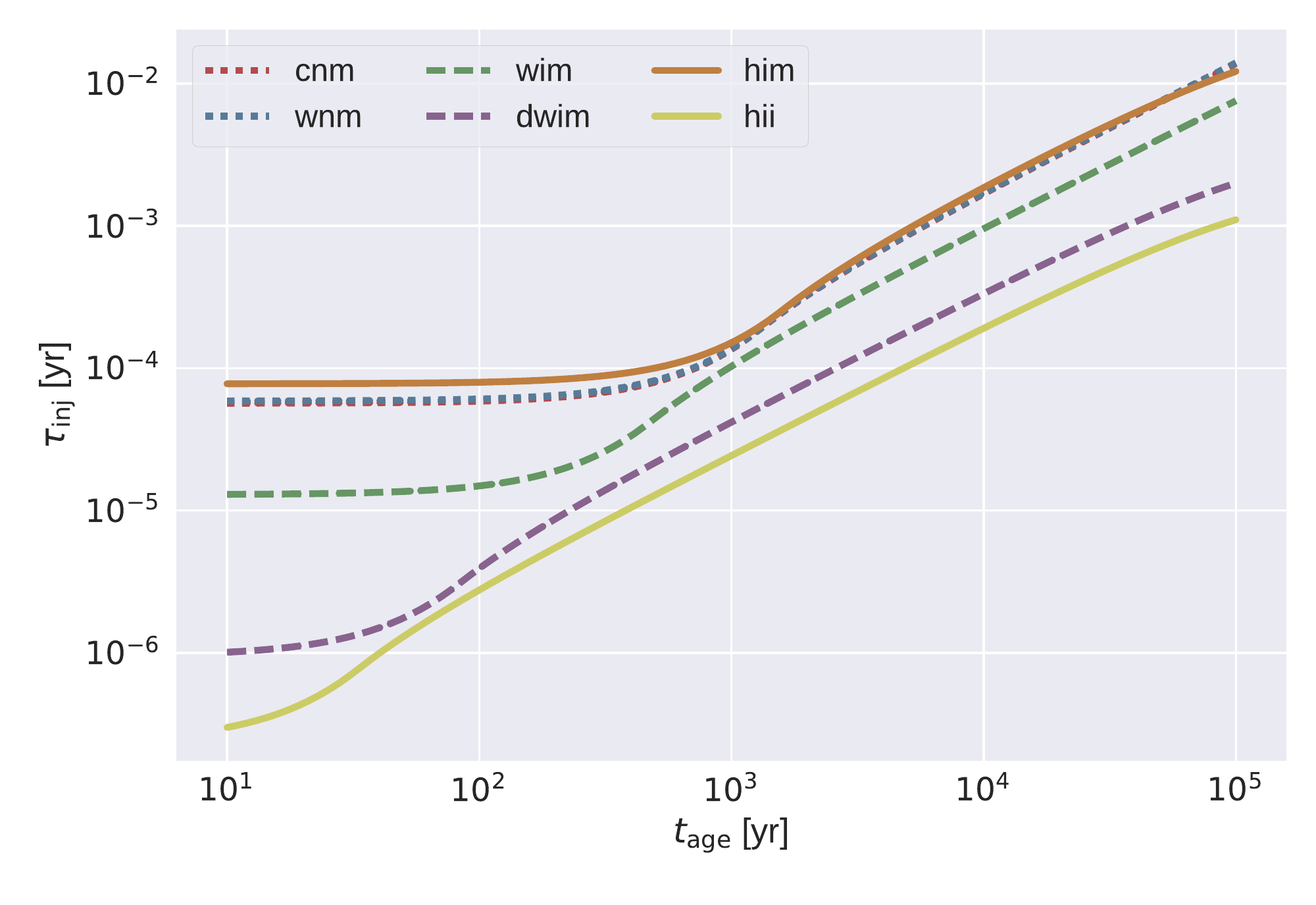}
\caption{Temporal evolution of different physical quantities downstream of the shock dependent on the age of the SNR using $E_{\rm ej}=10^{51}\,\text{erg}$, $M_{\rm ej}=1.4\,M_\odot$. 
\emph{Left}: The diffusion coefficient $\cdiff{m}{}(p_{\rm inj})$ of protons that results from the Bell-instability (BI), a turbulent dynamo (TD), and the magnetic field of the spiral arms of the Galaxy (gsa), respectively. The chosen environments span the range of possible values for all of the considered ISM phases.  
\emph{Right}: The injection time of protons dependent on the ambient ISM phase.}
\label{fig:shockEvol}
\end{figure}

In this work we will focus on the case of a uniform ambient medium and uniform ejecta, so that the motion of the blast-wave shock can be approximated by \cite{TrueloveMcKee1999}
\begin{align}
    r_{\rm sh}(t<t_{\rm ST})\:&\simeq\: 2.01\,(t/t_{\rm ch})\,[1+1.72\,(t/t_{\rm ch})^{3/2}]^{-2/3}\,r_{\rm ch}\,,\\
    \bsh(t<t_{\rm ST})\:&\simeq\: 2.01\,[1+1.72\,(t/t_{\rm ch})^{3/2}]^{-5/3}\,\beta_{\rm ch}
    \label{eq:EDphaseEQ}
\end{align}
during the ED phase and 
\begin{align}
    r_{\rm sh}(t\geq t_{\rm ST})\:&\simeq\: (1.42\,(t/t_{\rm ch})-0.254)^{2/5}\,r_{\rm ch}\,,\\
    \bsh(t\geq t_{\rm ST})\:&\simeq\: 0.569\,(1.42\,(t/t_{\rm ch})-0.254)^{-3/5}\,\beta_{\rm ch}
    \label{eq:STphaseEQ}
\end{align}
during the ST phase. The transition happens at $t_{\rm ST}\simeq 0.495\,t_{\rm ch}$ and the characteristic quantities are given by 
\begin{align}
    t_{\rm ch}\:&=\: E_{\rm ej}^{-1/2}  \, M_{\rm ej}^{5/6} \, \left(\av{m}\density{ion}{}\right)^{-1/3}\, ,\\
    r_{\rm ch}\:&=\: M_{\rm ej}^{1/3} \, \left(\av{m}\density{ion}{}\right)^{-1/3}\, ,\\
    \beta_{\rm ch} \:&=\: r_{\rm ch} / (t_{\rm ch}\,c)\,.
\end{align}
Due to the differences in the charged particle density \density{ion}{} the shock characteristics differ by more than an order of magnitude across different environments, but the conditions in the shocked downstream medium --- e.g., the diffusion coefficient, the downstream temperature and the injection timescales --- differ much more. In the ED phase --- that lasts a few years for HII-regions, and almost 1000\,years in the case of the HIM --- all environments are heated to a temperature of a few $10^9\,\text{K}$ downstream of the shock and in the ST phase the temperature decreases according to $\temperature{-}\propto t^{-5/6}$. The temporal development of the diffusion coefficient at $p_{\rm inj}$ strongly depends on the considered magnetic field. As shown in the left Fig.~\ref{fig:shockEvol}, the total diffusion coefficient $\cdiff{}{}$ is at all ages of the non-radiative phases of the SNR governed by the Bohm diffusion caused by the Bell instability, although $\edens{\Bturb}{\rm BI} \ll \edens{\Bturb}{\rm TD}$ by about an order of magnitude. 

Therefore, particles with a low $A/Q$ ratio yield for all of the ISM phases injection times $\ll 0.1\,\text{yr}$, as displayed in the right panel of Figure~\ref{fig:shockEvol}. Only in the case of old SNRs within the ST phase, particles with $A/Q\gg 1$ can spend enough time in the downstream medium to suffer from minor changes of the ionization fraction before entering the DSA. During the ED phase, $\tau_{\rm inj}$ remains constant but differs across the ISM environemnts by more than two orders of magnitude. When entering the ST phase, $\tau_{\rm inj}$ starts to increase in an approximately linear fashion, and the differences between the environments are diminished to one order of magnitude because the ST phase is entered the earlier the higher the density is. In principle, only SNRs with a small shock velocity --- hence, a small Mach number ($\Mach\sim 10$) --- show significant changes of the ionization fraction due to the heating of the ambient medium by the passing shock.

Note that we do not account for the ionization imprint by the progenitor star of the SNR, as most of this radiation is absorbed by the interior of the SNR. Though, at about the end of the ST phase an enhancement of the ambient UV radiation field is expected at the outer regions of the SNR, due to so-called reverse-fluorescence processes \cite{Sauer+2008, Lucy1999}. This UV radiation with typical photon energies $\sim 3\,\eV$ at the shock surface could ionize a surrounding neutral medium, causing the ionization fractions to become similar to the warm ionized medium. We will return to this issue when discussing the resulting abundances of the neutral ISM phases.

\subsection{Injection of dust grains}
\label{sec:dust}

About $1\%$ of the total mass of interstellar matter resides in dust grains, i.e., compounds of a few hundreds up to ${\sim}\,10^{12}$ atoms \cite{SavageMathis1979,Draine2003}. The small size end of this range is hereby represented by macromolecules, in particular Polycyclic Aromatic Hydrocarbon (PAH) molecules, which bind about $10\ldots20\%$ of the carbon in ISM and seem to be tremendously stable against photoionic or collisional excitation \cite{PugetLeger1989,Tielens2008}. Recently it has been suggested that also a significant part of the iron could reside in macro-molecular compounds \cite{Tarakeshwar+2019}. On the large side, there are micron-sized graphite or silicate grains, on which chemically refractory elements like Mg, Ca, and elements of the iron group tend to deposit. Dust exists in virtually all ISM phases, but with increasing temperatures elements bound in it tend to evaporate, and in the HIM dust grains get reduced to resilient FeNi cores \cite{Tielens1998}, while most other elements, including silicon and even the most refractory element, calcium, can be expected to return largely into the gas phase.  

Except maybe in very cold environments, in particular large dust grains can be generally assumed to be charged, with charge numbers $Q\sim 1\ldots10$ \cite{WeingartnerDraine2001}. Given the large range of possible masses, this implies a mass-to-charge ratio $A/Q\sim 10^2\ldots10^{12}$. 
This means that, extrapolating the $A/Q$ dependence of the enhancement factor to the large range for dust, our assumption that dust injection can be described by a single factor $\enhdust$ implies that the enhancement factor saturates to a constant for values $A/Q$ beyond the range possible for single ions -- we return in the discussion of our results to the question of how far this assumption is justified. 

Once a charged dust grain is injected into the DSA process, it will increase its momentum with repeated shock crossings just like any other particle injected into DSA. As discussed in Ref.~\cite{EllisonDruryMeyer1997}, the grain can survive several shock crossings before getting sputtered in interactions with ambient gas, electrons and photons. 
A condition for its successful injection into the process is hereby, that the gyro-period of the grain needs to be smaller than the acceleration timescale, which needs to be smaller than the age of the SNR. This leads --- with some dependency on the grain type --- to the constraint  $\beta_{\rm sh}\gtrsim 0.001$, a condition which is generally satisfied for SNR shocks in all environments at all ages.
If we assume that atoms eventually removed from the grain remain ionized, they will continue the DSA process with the same momentum per nucleon as the remaining dust grain. Effectively, the result is a compound of DSA spectra of elements $j$ contained in the dust grain, starting from the injection momentum $A_j p_0$; the only role of the dust grain is to shift the atoms bound in it across the threshold for supra-thermal injection before they start to get accelerated individually. For this reason, there is no factor $\xinj$ in Eq.~\ref{eq:dNdp} for dust injection, neither is there any influence of the individual elements on the enhancement --- elements injected through the ``dust channel'' retain in the nonthermal spectra the abundances in that they appear in interstellar dust for a given environment.       

This leads us to the question to which fraction $\fraction{dust}$ individual elements are bound in dust. This is usually answered by the counter question, i.e., what is the \textit{gas fraction} $\fraction{gas}$ of an element in a given environment and using the assumption $\fraction{gas}+\fraction{dust}=1$, as implicitly assumed already in Eq.~\ref{enhancement}. The gas fraction is determined empirically from observations of absorption lines of various elements compared to a reference element, usually hydrogen. This method requires a well defined, dense patch of \textit{mostly neutral} ISM in the line of sight of a bright star, and the results for different lines of sight differ significantly even for similar ISM temperatures. A standard source of error in this determination is that either hydrogen is partly ionized and thus its total column density underestimated (which leads to overestimates of $\fraction{gas}$ for all measured elements), or that the elements themselves hide in different ionization states than investigated, leading to underestimation of $\fraction{gas}$, hence overestimation of $\fraction{dust}$.

The largest collection of results of such observations has been presented by Jenkins \cite{Jenkins2009}, who managed to combine the results of hundreds of observations with apparently deviating results into a systematical scheme. Jenkins defined an abstract general depletion factor $F_*$ and found that the (logarithmic) depletion of individual elements can be represented as a linear dependence on $F_*$, with a slope dependent on the individual element. He defined hereby $F_*=1$ as typical for the CNM, while $F_*=0$ is used for the lowest collective depletion found in his sample \cite[see][for details]{Jenkins2009}. As gas fractions are usually determined in mostly neutral patches of matter, we interpret the element depletion values found at the $F_*=0$ borderline as typical for WIM. For the average WNM, we use the best fit values given for $F_*=0.5$, while for the HIM we extrapolate Jenkins' fit down to $F_*=-0.5$. We are fully aware that doing this identification between Jenkins' $F_*$ and the ISM phases is rather an educated guess and prone to systematics, nevertheless it delivers the probably best estimates for gas fractions in ionized environments currently possible.  

Jenkins' analysis is presented for the elements C, N, O, Mg, Si, P, Cl, Ti, Cr, Mn, Fe, Ni, Cu, and Zn, where we strictly use his linear fits to derive our gas fractions within reasonable accuracy, except for nitrogen (see App.~\ref{app:GasFrac_individual} for an explanation). For other elements, we refer to the previous review by Savage and Sembach \cite{SavageSembach1996} or specialized papers \cite{Barker+1984_Al,Crinklaw+1994_Ca,SofiaJenkins1998,Kemp+2002_NaK,Kimura+2003_LIC}. Additionally, data from the microscopic analysis of stratospheric micrometeorites \cite{Schramm+1989}, dust particles mostly formed in the solar system and not in the ISM, allow at least qualitative conclusions on grain structure and thus the tendency of elements to be released from grains with increasing temperatures. For example, they show that sulfur, an element with highly uncertain and contradictory depletion properties, is bound even in smooth a potentially resilient grain structures, while the highly depleted element calcium is mostly present in coarse structure which can be expected to evaporate easily from grains. Finally, if no data are present at all, we can use chemical similarities between elements to provide estimates. Details on our approach and a complete table with the best-guess element gas fractions and their assigned confidence for all ISM phases are given in Appendix \ref{app:GasFrac_approach}. 

\subsection{Total abundances of elements and isotopes}
\label{sec:abundances}

The final question to determine cosmic ray abundances from our model and the above-mentioned information about the ISM is the question about the total cosmic abundances of element-isotopes. First of all, it is indeed the isotope abundances we need as the acceleration enhancement factors depend on nuclear mass $A$, while the chemical properties of elements determining ionization potential or gas-to-dust relation depend almost entirely on nuclear charge $Z$. In our notation we write this as:
\begin{equation}
   \abundance{j} = \abundance[iso]{j}\times 10^{[X_j]}\left/\;\sum_k\,\abundance[iso]{k}\times 10^{[X_k]}\right.
\end{equation}
where the for ISM studies common notation $[X]$ denotes the decadic logarithm of the cosmic abundance of element $X$ (in arbitrary normalization), and as usual the index $j$ runs over all relevant isotopes with $X_j$ being the corresponding element, and \abundance[iso]{j} is the relative abundance of isotope $j$ within element $X_j$ (i.e., $\sum_{X_j=X} \abundance[iso]{j} = 1$ for all $X$). We consider elements up to Zn ($Z=30$), the abundance of which is low but still larger than that of all heavier elements summed up, and for each element all stable isotopes which contribute more than $0.5\%$, rounded to full percent values such that the normalisation of \abundance[iso]{j} is maintained, using the values given in the \textit{WWW table of radioactive isotopes} \cite{isotopesTableOnline}.  

What remains is the question of total cosmic abundances of the elements. Here the common approach is to use abundances found in the solar system, as they are best known. There are two standards: abundances observed in the solar photosphere, and abundances determined from certain types of meteorites, so-called \textit{CI chondrites}. The latter are generally thought to give insight into the pre-solar abundances of the cloud from which the solar system formed, but the disadvantage is that there are large systematic underestimations of abundances of volatile elements. We use here generally the photospheric solar abundances reported in the \textit{Treatise of Geochemistry, 2nd edition} \cite{PalmeLoddersJones2014} as a reference, except for Lithium which is known to be destroyed in the sun and where we take the CI chondrite value reported in the same work. Several authors have pointed out that the sun may not be representative for cosmic abundances and that instead B-star abundances should be used. Looking at the ``cosmic standard'' abundances obtained by Nieva and Przybilla \cite{NievaPrzybilla2012} from B-star observations, however, we find that for the investigated elements N, O, Ne, Mg, Si and Fe their values agree with the solar ones within the error bars, and only for carbon and helium deviations are found on a $\gtrsim 3\sigma$ level: C seems to be over-respresented in the sun by about 30\%, while for He the cosmic abundance is found to be 15\% higher than solar. We keep these deviations in mind, but use for consistency solar abundances for all elements. 

The element abundances we use in this paper and all isotope masses we consider are listed in Table \ref{galactic_abundances} in Appendix \ref{app:GasFrac}, with the values for element abundances normalized to $10^{12}$ hydrogen atoms.  

\section{Predicting low energy cosmic ray abundances}
\label{sec:results}

The base assumption for explaining the cosmic ray flux observed is not only that SNR can efficiently convert thermal to nonthermal matter, but also that DSA finds the conditions to produce flat source indices $\pspecind\approx 2$. This requires that strong shocks with compression ratio $\rcomp\approx 4$ can form in the \textit{ionized material} of the environment. It is not clear whether these conditions are fulfilled in all ISM phases. In the contrary, for molecular clouds (MC) detailed studies of the dynamic of shock formation in \emph{both} neutral and ionized matter have shown that there is no jump in the ionized material expected whatsoever. A general theory of interstellar shocks in mixed neutral and ionized media \cite{DraineMcKee1993} shows that those shocks that do form, form only in the neutral material and are largely continuous (so-called C-type shocks). They are of tremendous importance for the chemistry of molecular clouds, but that cannot support DSA as this requires a strong discontinuity (called J-type shocks in this context). We therefore will not discuss the MC case in this study. For other neutral material, the above-mentioned general theory could be used to check whether and to which extent J-type shocks in the ionized material contained in the CNM or WNM can form and contribute to the LECR flux. We decided, however, not to make this check. Rather, we use the opposite approach: We simply assume that such shocks exist with the required large compression ratio in both the CNM and WNM, and compare the results to cosmic ray data. The same we do for the other, generally ionized ISM phases, and for all of them we present their ``LECR fingerprint'', i.e., the LECR composition that would be observed if all cosmic rays originate from just one ISM phase. These ``fingerprints'' we can then use to identify the likely culprit responsible for the production of low energy cosmic rays (see Sect.~\ref{sec:conclusions}).

Low energy cosmic rays are usually compared in $dN/d\energy[k/A]{}{}$ at the same $\energy[k/A]{}{} = \energy[kin]{}{}/A = p^2/2 m_p A^2$. 
Therefore, we obtain from a transformation similar to Eq.~(\ref{eq:deriveScaling}), that the abundance of dust elements does not change with mass or charge number, whereas $dN/d\energy[k/A]{}{}$ of the gas elements is essentially $\propto \enhgas$ if we account for all dependencies on $A$ and $Q$. 

Here, we compare our model with the most recent estimates on the source abundances of elements with $Z\geq 6$ based on the observations of the
Cosmic Ray Isotope Spectrometer (CRIS) on board of the NASA Advanced Composition Explorer spacecraft \cite{ACE-CRIS2018}. Since these data do not cover all elements, we also use the best-fit source abundances (and upper limits) of phosphorus, potassium, chromium, manganese (and fluorine, chlorine, titanium, vanadium) from the Ulysses High-Energy Telescope (HET) observations \cite{ULYSSES1996}. In addition, the source abundances of copper and zinc are used from the HEAO3-C2 data \cite{1981ICRC....9..118P}, yielding a set of 21 data points $O_j$ with a variance $\sigma^2(O_j)$. These source abundances haven been derived after applying corrections to the observed composition for solar modulation as well as interstellar propagation, and are normalized to silicon. Note that these predictions depend on the assumption that all species have essentially the same energy spectra at their sources --- at least at low energies. 
We use our model (\ref{eq:dNdp}) and determine the total low energy abundance of a given element \emph{at the same kinetic energy per nucleon}, with $\cutoff(\rigidity{j}{}(p))\simeq 1$, at an escape time $t$, where the accelerated particles leave the acceleration region, yielding
\begin{align}
    \numspec{j}(t)\,&=\,\sum_{k=1}^n \left[\frac{\diff N_j}{\diff\energy[k/A]{}{}}\right]_{\Delta t_k}\,\Lambda_{\rm ad}(t,t'_k)\nonumber \\ 
    &\propto\, \sum_{\Delta t}\,\xinj^{\pspecind-1}\,\chi_j\,\Enhance{j}\,\left( \frac{t}{t_{\rm ch}} \right)^{-\alpha_{\rm ad}(t)}\,\left( \frac{t'_k}{t_{\rm ch}} \right)^{\alpha_{\rm ad}(t'_k)}\,,
\label{eq:totalLECRabund}
\end{align}
with
\begin{equation}
    t'_k=\sum_{l=1}^{k}\Delta t_l\quad\text{and}\quad \alpha_{\rm ad}(t)=
    \begin{cases}
     1\,\quad&\text{ for } t<t_{\rm ch}\,,\\
     0.4\,\quad&\text{ for } t\geq t_{\rm ch}\,.
     \end{cases} \nonumber
\end{equation}
Here, $\Lambda_{\rm ad}(t,t'_k)$ refers to the adiabatic losses of particles that get injected into DSA at a time $t'_k<t$ --- all details that lead to this factor as well as the time dependent index $\alpha_{\rm ad}(t)$ are given in  Appendix \ref{app:AdiabaticLosses}. Note, that efficient particle acceleration happens during the nonradiative phases of the SNR, so that we suppose that $t\equiv\sum_{k=1}^{n}\,\Delta t_k = 10^{5}\,\text{yr}$.

We quickly recap the influence of various parameters contain in our model. For the main parameter that determines the enhancement of a given gas element, $\enhgas$, we allow a shallow and a steep dependence on ion mass-to-charge ratio $A/Q$ up to a maximum $\enhmax$, as explained in Section \ref{sec:particleInjection}. Another 
strong influence on the resulting abundances arises from the supposed gas fractions $\fraction{j}$. As discussed in Sect.~\ref{sec:dust} and Appendix \ref{app:GasFrac_approach}, there is still significant uncertainty assigned to these values in spite of rich data material at least for a number of elements. Clearly suggested by the data, however, is that $\fraction{j}$ is mostly determined by the temperature of the ISM phase, so that we can assume that  $\fraction[HII]{j}\simeq\fraction[DWIM]{j}\simeq\fraction[WIM]{j}$ as these environments mostly differ in density, but only little in temperature. 

\begin{figure}[htbp]
\centering
\includegraphics[width=.99\textwidth]{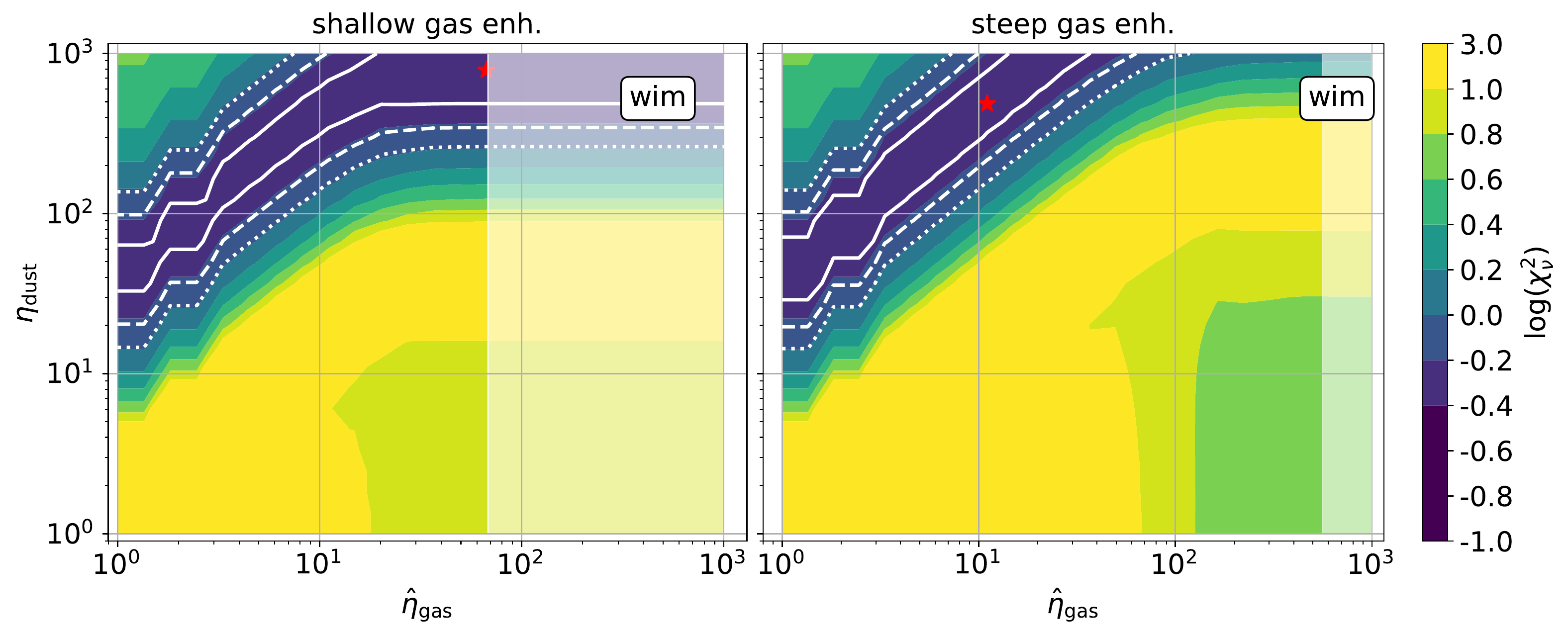}
\includegraphics[width=.99\textwidth]{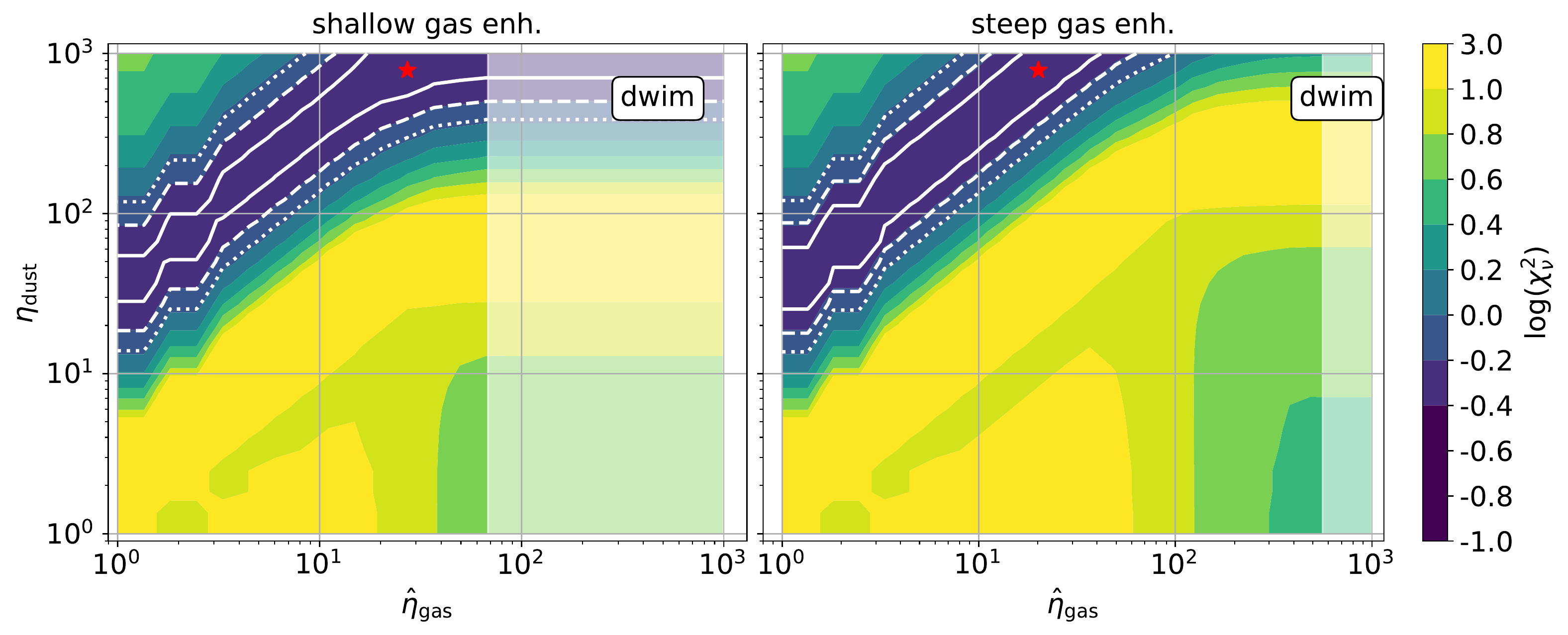}
\includegraphics[width=.99\textwidth]{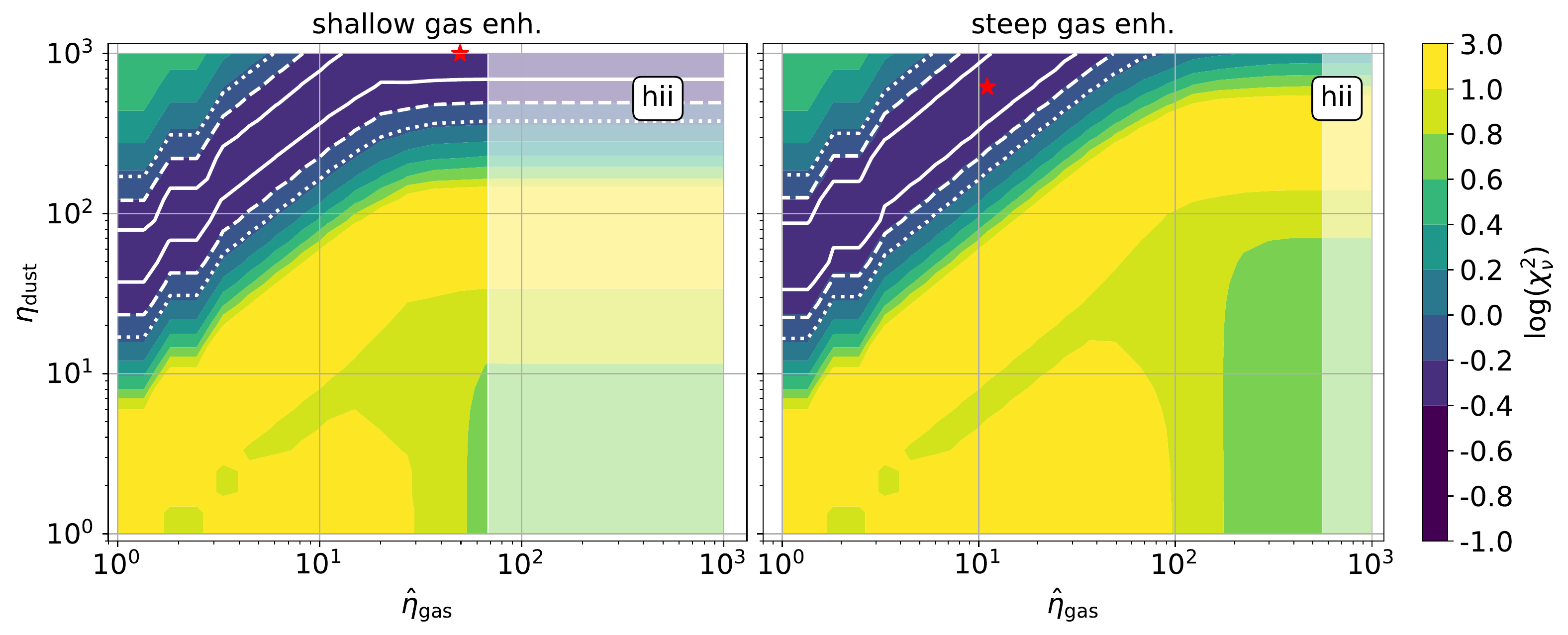}
\caption{Warm ISM phases and HII-regions: The reduced chi-squared values within the possible parameter space of $\enhmax$ and $\enhdust$ using a shallow gas enhancement (\emph{left panels}) and  a steep gas enhancement (\emph{right panels}), respectively. The red star marks the best-fit value and the white lines mark the $1\sigma$ (solid), $2\sigma$ (dashed), $3\sigma$ (dotted) confidence range with respect to the best fit values in the limiting case of only one degree of freedom, i.e.\ $\nu=1$.}
\label{fig:chi2_ionized}
\end{figure}

\begin{figure}[tbp]
\centering
\includegraphics[width=.99\linewidth]{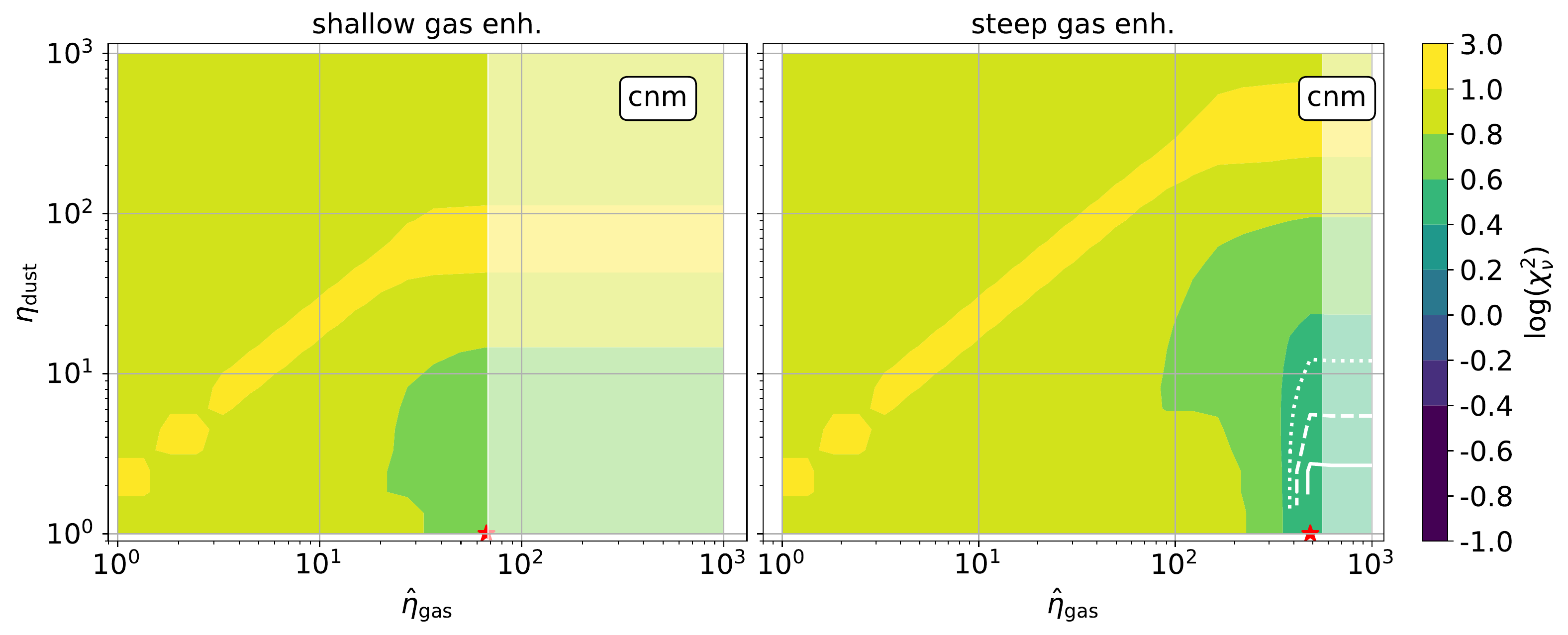}
\includegraphics[width=.99\linewidth]{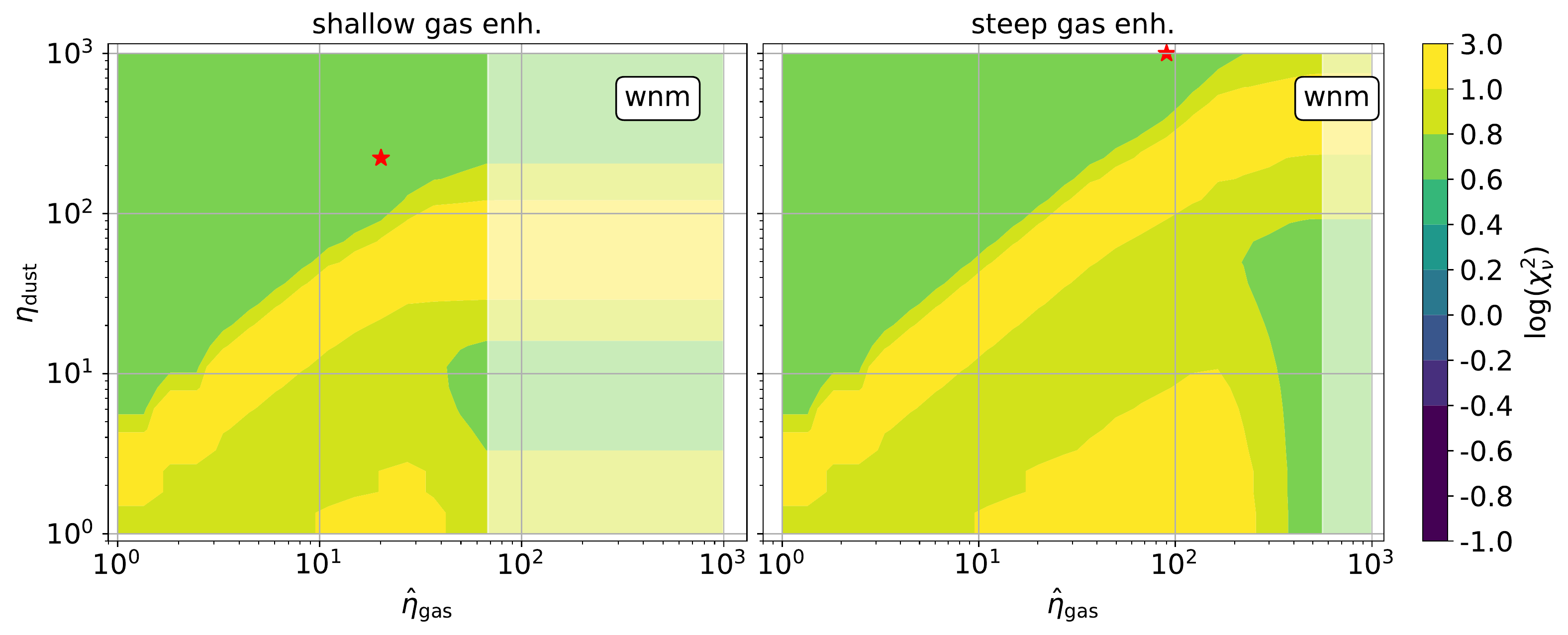}
\includegraphics[width=.99\linewidth]{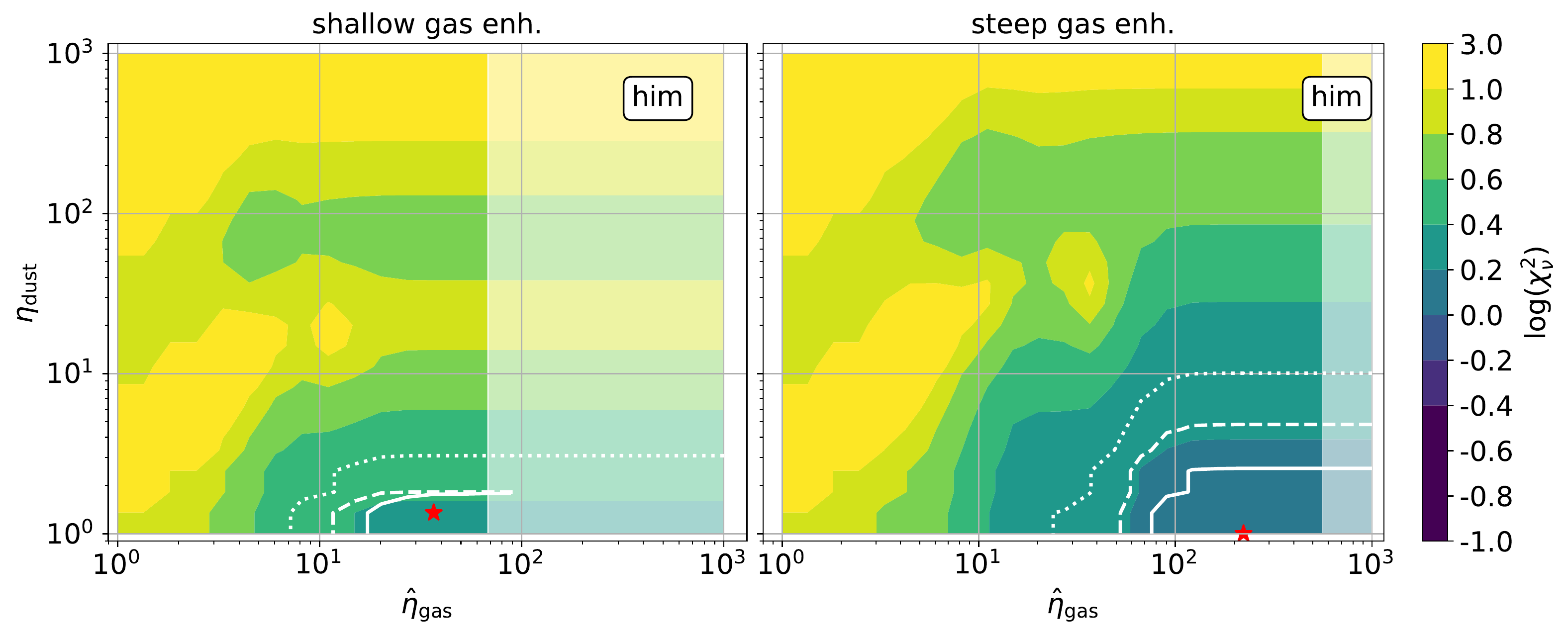}
\caption{The reduced chi-squared values of the HIM and the neutral ISM phases using the shallow (\emph{left panels}) and the steep (\emph{right panels}) gas enhancement, respectively. The red star marks the best-fit value and for the HIM the white lines mark the $1\sigma$ (solid), $2\sigma$ (dashed), $3\sigma$ (dotted) confidence range with respect to the best fit values in the limiting case of only one degree of freedom, i.e.\ $\nu=1$.}
\label{fig:chi2_neutral}
\end{figure}

So, we determine $\numspec{j}$ from the best guess values of $\fraction{j}$ (see Table \ref{galactic_abundances}), and the upstream conditions $\temperature{+}\!$, $\density{}{}$ as given in Table \ref{AParameters}. In addition, we use the given uncertainties of these parameters to determine their variance $\sigma^2(\numspec{j})$. Here, 50 uniform distributed values within the interval $\temperature{+}\pm\Delta \temperature{+}$ and $\density{}{}\pm \Delta\density{}{}$, respectively, are used, as well as 50 different values of $\fraction{j}$. These are drawn from a truncated log-normal distribution with a mean value $\fraction{j}$ and a standard deviation $(\sigma_j=\fraction{j}\times\cv)$ for $\icv>0$ and from a uniform distributed between 0 and 1 for $\icv=0$ ($\cv$ is called the coefficient of variation of the distribution, see  App.~\ref{app:GasFrac_approach} for details). 
Finally, we normalize our model with respect to the silicon abundance to compare it with the data. 
Some other parameters of our model can be fixed without loss of generality: $\turbscale=10$, $\lscale[coh]{,\Bturb}{\envind}=1\,\text{pc}$, $\xi_{\rm B}=0.1$, $\xi_{\rm cr}=0.1$ and $\xinj=3$. 
The main influence on the resulting abundances comes from the chosen values of $\enhmax$ and $\enhdust$. From first principle, we know that its values need to be within the interval $[1.,\,1000]$, although H+19 suggested that $\enhmax\sim 10$ --- in contradiction to the C+17 results. The simulations from C+17 and H+19 only consider $A/Q\ll 100$, so that their results can hardly be used to constrain the enhancement of dust grains, which has a typical $A/Q\sim 10^8$. So, we perform a chi-squared fitting on a 2D grid as shown a Fig.~\ref{fig:chi2_ionized} and \ref{fig:chi2_neutral}, and determine the best-fit values for a given environment. Here, the reduced chi-squared value is given by
\begin{equation}
    \chi_{\nu}^2 =\frac{1}{\nu}\,\sum_j\, \left(\Delta \numspec{j}\right)^2 = \frac{1}{\nu}\, \sum_j\,\frac{\big(O_j-\numspec{j}\big)^2}{\sigma^2\big(O_j\big)+\sigma^2\big(\numspec{j}\big)}\,,
\end{equation}
where the degrees of freedom $\nu=18=21-2-1$ is calculated based on the used number of data points minus the two fit parameters, as well as the normalization.\footnote{Note that the model prediction for the different elements is hardly linearly independent, hence, the supposed degrees of freedom is rather an upper limit and $1\leq \nu \leq 18$.}
Using the $\chi^2$ density function\footnote{\url{https://docs.scipy.org/doc/scipy/reference/generated/scipy.stats.chi2.html}} for the given degrees of freedom, we determine those parameter configurations that are within a given confidence limit of the best-fit values.

Figure~\ref{fig:chi2_ionized} shows that SNR in partly ionized environments, i.e., (D)WIM and HII regions, provide an excellent agreement to the data within a broad range of appropriate parameter values, indicating the tendency that $\enhdust$ increases proportionally with $\enhmax$ while being about one order of magnitude larger, until its maximally plausible value (here set to $10^3$) is reached. Also for $\enhmax$ there is a maximum that is determined from the mass-to-charge dependence of the gas enhancement (\ref{eq:H+19enhanc}) and (\ref{eq:C+17enhanc}) and the heaviest isotope that is taken into account. For the shallow enhancement, single ionized zinc is at most enhanced with a factor of $A/Q=68$, whereas for the steep enhancement it is at most enhanced with a factor of $(A/Q)^{1.5}=561$. For better comparability, we show in Fig.~\ref{fig:chi2_ionized} the results for $\enhmax\leq 10^3$, but note that any differences of $\chi^2_\nu$ at $\enhmax>68$ and $\enhmax>561$, respectively, can only result from numerical artifacts which is why this parameter range is dimmed in the Figure~\ref{fig:chi2_ionized}. 
Note, that the best fit-value of $\enhmax$ agrees quite well to what is expected from H+19, but the confidence intervals indicate that also a significantly higher value of $\enhmax$ yields almost the same agreement to the data. Very interesting to note is that the optimal $\enhdust$ tends to saturate at a constant value already before $\enhmax$ reaches its maximum, likely due to the fact that mostly refractory elements remain for $A/Q\gtrsim30$, rendering gas-phase enhancement less relevant. Moreover, the best-fit models of the different scaling of the gas enhancement are hardly distinguishable, hence, we present here only the results based on the shallow function --- the interested reader is referred to Appendix~\ref{app:AddResults} for the corresponding predictions in the case of a steep gas enhancement.

Contrarily, the HIM and especially the neutral phases of the ISM (see Fig.~\ref{fig:chi2_neutral}) are generally disfavoured, since $\text{min}(\chi^2_\nu)\gg 1$ for most parts of the parameter space especially if $\nu<18$. For the HIM, there is still a significant range of the parameter space allowing for a moderate agreement with the data in particular for the steep gas enhancement model, which results from the large uncertainty on its gas fractions, temperatures and densities. 
The data is described preferably for a low acceleration efficiency of dust grains, i.e.\ $\enhdust\lesssim 10$, but for this case shocks in the (D)WIM and HII-regions cannot describe the data. As $\enhmax$ and $\enhdust$ are thought to be universal parameters of DSA, and most supernovae occur at low galactic scale heights where the HIM is not present, this argues against a dominant contribution from SNR shocks in the HIM. Conversely, the latter environments give an excellent fit for rather large values of $\enhdust$, and here we still have some agreement also for the HIM and the WNM. Therefore, a subdominant contribution of the HIM and WNM to cosmic ray production otherwise dominated by (D)WIM and HII regions cannot be excluded --- see also the discussion in  Section~\ref{sec:conclusions}.

\begin{figure}[tbp]
\centering
\includegraphics[width=.99\linewidth]{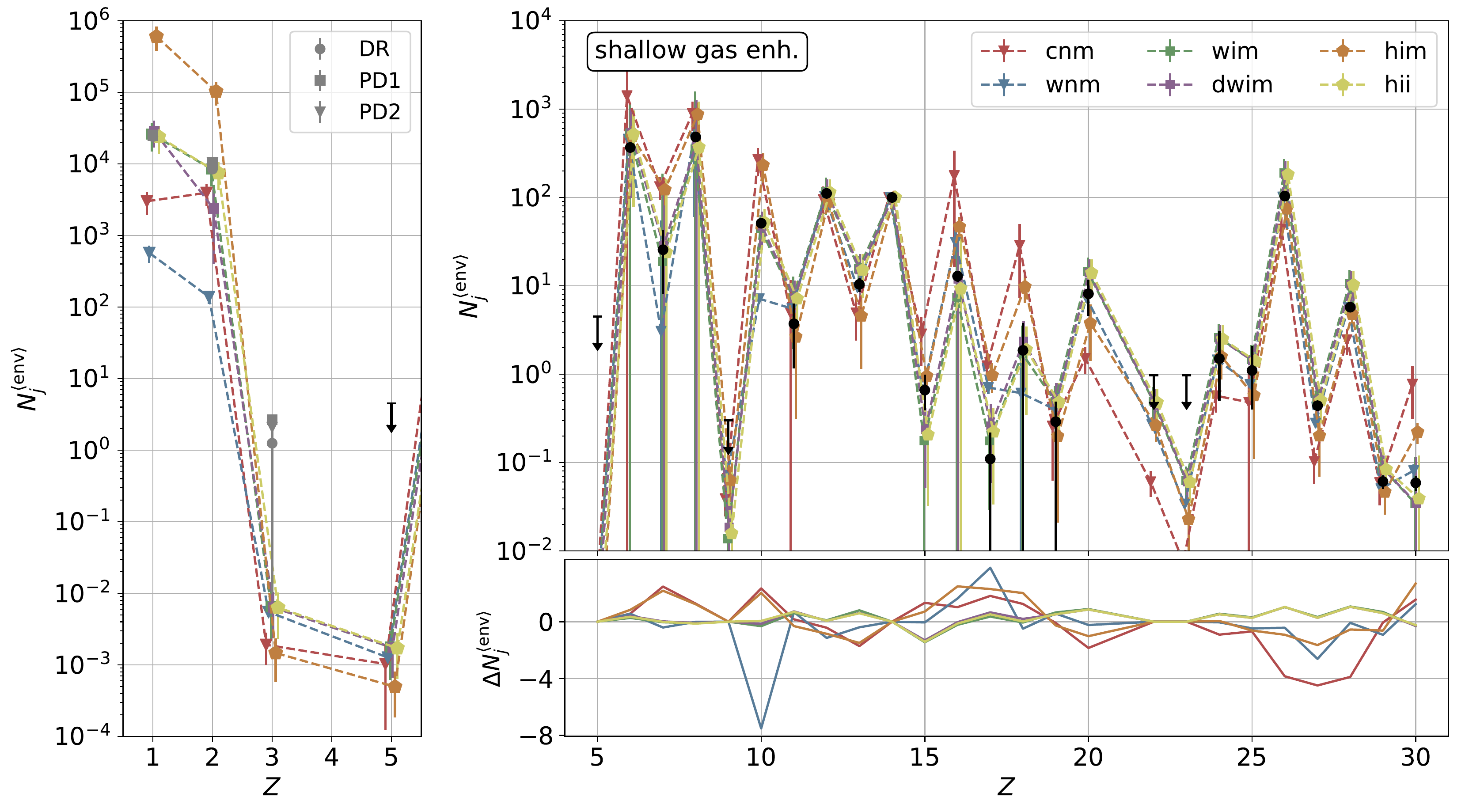}
\includegraphics[width=.99\linewidth]{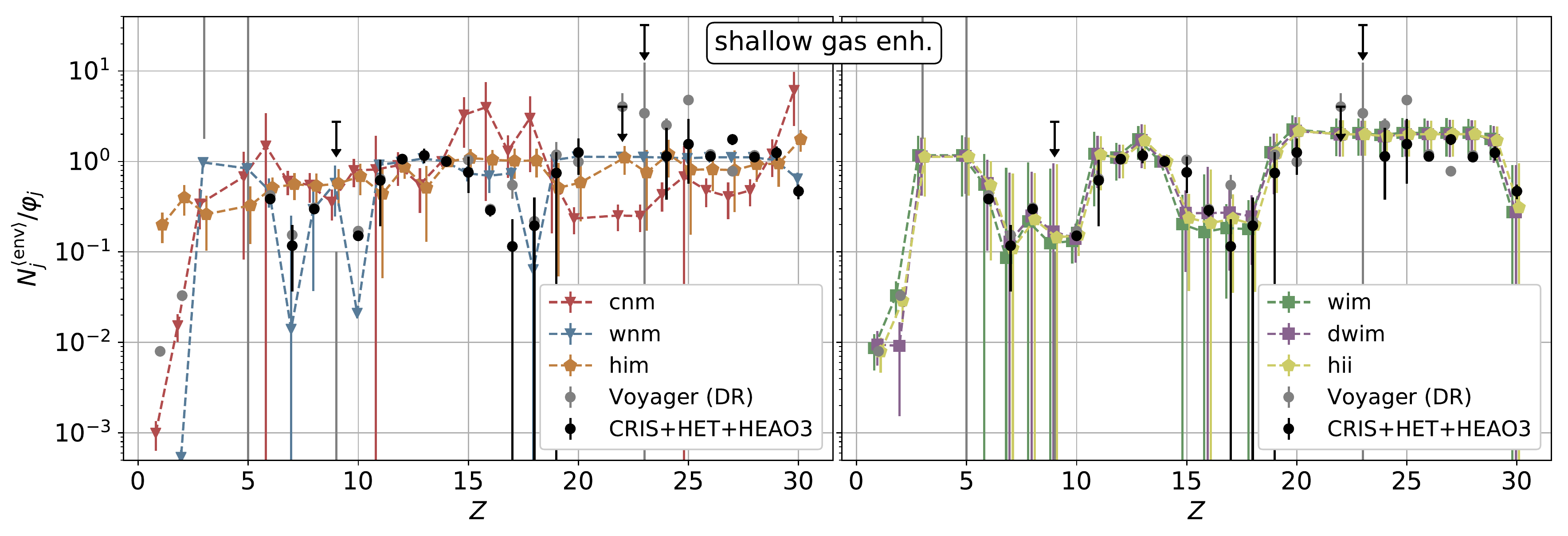}
\caption{The best-fit model using the shallow gas enhancement (\ref{eq:H+19enhanc}).
\emph{Upper left panel:} For light Elements with $1\leq Z\leq 5$. In addition, the expected source abundances from different GALPROP models (DR, PD1, PD2) based on the Voyager 1 observations \cite{Voyager2016} are shown for hydrogen, helium and lithium. 
\emph{Upper right panel:} For heavy Elements with $Z\geq 6$, that have also been included in the chi-squared fitting. The lower panel displays the residuum $\Delta N_j^{\langle \mathrm{env}\rangle}$ of the individual elements.
\emph{Lower panel:} The total enhancement of elements, i.e.\ the ratio of the total source abundance over the solar abundances. For comparison the total enhancement based on the Voyager 1 data using the GALPROP DR model (black), as well as the data used for the fitting (grey) is shown.}
\label{fig:bestfit_Hanusch}
\end{figure}

For a proper discussion of the CNM and WNM, it is important to look at the detailed ``fingerprint'' these environments leave for individual elements. Differentiating between heavy elements that require stars or supernovae to be produced, and the light ones that are either primordial ($Z=1,2$) or result mostly from cosmic ray spallation ($Z=3,4,5$), the upper panels of Fig.~\ref{fig:bestfit_Hanusch} show the resulting best-fit distribution of the LECR abundances for the shallow gas enhancement model (results for the steep gas enhancement model are hardly different and shown in Appendix \ref{app:AddResults}). The lower panels of this figure show the enhancement factors for all heavy elements, left for the less good fits obtained for the CNM, WNM and HIM, and right the ones for (D)WIM and HII regions. When reading these figures, we have to keep in mind two things: (a) that results are shown in each environment using $\enhmax$ and $\enhdust$ as marked by the red star in Fig.~\ref{fig:chi2_ionized} and \ref{fig:chi2_neutral}, which can be very different for different environments, and (b) the normalization to the silicon abundance need to be taken into account, which means that changes in the enhancement of some element when moving from one environment to the next are also influenced by changes in the behavior of silicon in this transition.

With this in mind, and restricting ourselves to the heavy elements for now, the first comparison shall be made between the extremes of the ISM phases, the CNM and the HIM, for which we find a surprising correspondence. For both, the best fit parameters fulfill $\enhmax \sim 100 \gg \enhdust \sim 1$, and for most of the elements they arrive at similar predictions for the enhancement factors. In particular, both overestimate the abundance of the pure gas phase element neon and the (semi-)volatile elements sulfur, chlorine, argon and zinc. Additionally, the CNM underestimates, in part severely, the enhancement of elements in the range Ca to Fe. Also for the HIM, where all elements are in the gas phase at least to a significant fraction, we remain with the problem of an apparent over-enhancement of neon, which can here be understood by an under-enhancement of the reference element silicon and all other refractory elements. The same effect is seen for hydrogen and helium, which are also purely in gas, and also for the above mentioned semi-volatile elements. That means that for both extremes of ISM phases we see that without efficient dust acceleration refractory elements cannot be brought in the right balance with volatile elements. 

The opposite case, i.e., $\enhdust \sim 300 \gg \enhmax \sim 10$ is explored by our results for the WNM and also the various warm ionised phases. For the WNM, we now see the opposite effect: neon is under-enhanced, and the reason for this might be that (a) silicon, still residing mostly in dust in this phase, is over-enhanced, and/or (b) that the ionisation fraction of neon with its FIP of $21.6\,\eV$ is too small. Another problem is found for sulfur and chlorine that are over-enhanced in our prediction, probably due to a too strong condensation of these elements in dust in this environment. For all other elements, the ``fingerprint'' matches quite well, and we discuss in Sect.~\ref{sec:conclusions} whether and how the consideration of UV photons emitted by the SNR can heal the mismatch seen for Ne, S and Cl. Such photons are not needed for the warm ionized environments. In fact, the WIM, the DWIM and the HII-regions all yield similar enhancement factors. All of them provide the observed LECR abundances of the heavy elements within the errors --- except for phosphorus, which may point to a systematic error in the determination of its depletion factors (see App.~\ref{app:GasFrac_individual}, but also the discussion in Section~\ref{sec:conclusions}). Note that the goodness of the fit would even increase if we take into account that the ISM abundance of carbon is likely smaller than the solar abundances by $\sim 30\%$ \cite{NievaPrzybilla2012}.
Further, we checked the dependence of the HII-region on the ionizing star type, obtaining that the resulting abundances change only at the order of a few percentage. The largest difference appears for the helium abundance, where a B0 star ($T_*=33\,300\,\text{K}$, $Q_*(H)=10^{48.16}\,\text{s}^{-1}$) yields about 50\% less helium from an HII-region. 

The source abundances of light elements with $Z\leq 3$ are unfortunately not well determined from observations, as the observed cosmic ray abundances of these mostly primordial elements are not necessarily connected to SNR.  In addition, cosmic ray spallation is another source of the observed cosmic ray abundances of elements with $Z=2\ldots5$, in particular $^3\text{He}$, $^9\text{Be}$ and $^{10,11}\text{B}$. By comparison, lithium, beryllium and boron are rare in the solar abundances and the attempts to derive their source abundances have mostly led to upper limits \cite{ULYSSES1996} or large uncertainties \cite{Voyager2016}. 
In the case of hydrogen and helium Cummings et al.~\cite{ULYSSES1996} used a source spectrum composed of a triple power law with different spectral indices and different spectral break points in order to derive the source abundances. Apart from the questionable physical motivation for such an approach, it makes their predictions hard to compare to our model. Still we display in the left Fig.~\ref{fig:bestfit_Hanusch} the results of hydrogen, helium and lithium from three different GALPROP models --- that differ for each element --- next to our model in order to give an idea of the expected source abundance of these elements. 
Though, it is quite surprising that without fitting to the data, the warm ionized environments (including HII-regions) yield a perfect agreement with the expected source abundances of hydrogen and helium, if $\enhdust\approx 800$. For lithium our results are about two orders of magnitude too small, however, this might be due to the fact that not SNR but classical novae are the source of lithium in Galactic cosmic rays \cite{Starrfield+2020}. 

\begin{figure}[tb]
\centering
\includegraphics[width=.99\linewidth]{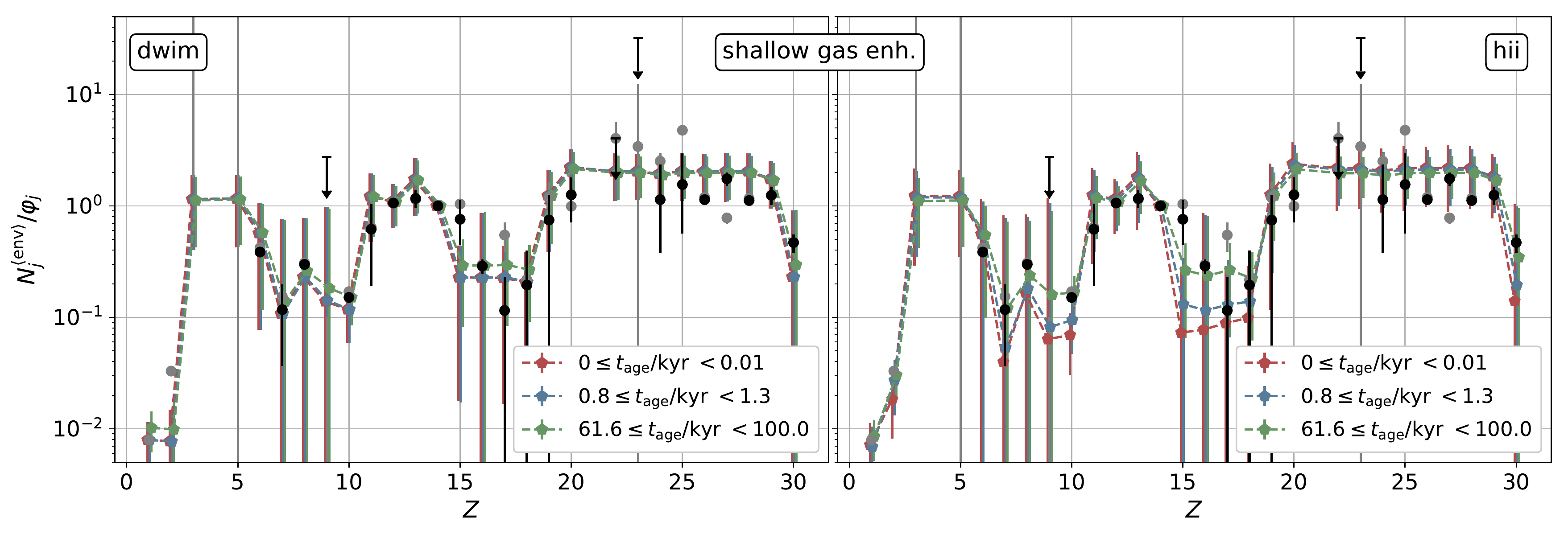}
\caption{The enhancement of elements for the DWIM (\emph{left}) and HII-regions (\emph{right}) at three different time intervals using the best-fit model for the shallow gas enhancement (\ref{eq:H+19enhanc}). For comparison the total enhancement based on the Voyager 1 data using the GALPROP DR model (black), as well as the data used for the fitting (grey) is shown.}
\label{fig:bestfit_Evol_Hanusch}
\end{figure}

While the choice of $\enhmax$ and $\enhdust$ along with the gas fractions $\fraction{j}$ have the dominant impact on the nonthermal abundances, the differences of the ionization fractions \ionfracUp{ji} and \ionfrac{ji} provide only a small correction factor, which can be exposed in parts by analyzing the temporal evolution of the total enhancement factor. Here, the volatile elements, especially oxygen, neon and argon, show an increasing enhancement with the SNR age that is the strongest for a dense, cold environment. At late times, the injection times of heavy (but hardly ionized) elements become sufficiently long, so that the ionization fractions are significantly modified downstream of the shock, in particular for the cold, neutral ISM phases. As these environments are clearly disfavoured from the previous investigations, Fig.~\ref{fig:bestfit_Evol_Hanusch} depicts the enhancement evolution for the DWIM and HII-regions. Among the warm and hot environments, HII-regions have the strongest evolution of the enhancement which is not surprising as its upstream ionization fractions $\ionfracUp{}$ decrease significantly with increasing $t_{\rm age}$. In general, the late stages of the ST phase provide the main contribution to the total abundances due to the large shock surface. 

In total, only a partly ionized environment with a significant efficiency of dust grain acceleration can describe the observed source abundance of LECR. Here, the (D)WIM and HII-regions yield almost identical abundances, and minor differences only occur for the volatile elements.

\section{Discussion, conclusions and outlook}
\label{sec:conclusions}

The source abundances of CRs are a hot research topic since several decades, though recent progress in the decoding of the observed composition of CRs at the highest energies increases the need for a reliable physical model, that explains the change of abundances from the thermal to the nonthermal energy regime. Driven by this, we developed a model that (a) considers the detailed ionization fractions of gas-elements in the shock environment, (b) allows for the injection of refractory elements in the acceleration process via charged dust grains, and (c) accounts for the temporal evolution of the shock as well as the resulting change of the ionization fractions of the ambient media due to the heating by the passing shock. 

Applied to the ISM, the model allows to predict a ``cosmic ray fingerprint'' each known ISM phase, which are characterized by their temperature, density and gas/dust ratios for all relevant elements. Comparing these predictions to data obtained for low energy cosmic rays (LECR), we can bring down our results to one, main message: The (D)WIM and HII fingerprint gives an excellent representation of the data for the case of \textit{a very efficient acceleration of dust grains}. This agrees perfectly to observations from other galaxies, like M31 and M33 \cite{Duric2000proc,Asvarov2014}, which show that the WIM is a favoured SNR environment. 
The HIM as an SNR environment produces rather poor fits, but a sub-dominant contribution of SNR in this environments cannot and should not be excluded --- it might even improve the fit results. Elaborating a realistic, mixed contribution scenario to really explain all the details known about LECR within our model certainly requires additional knowledge that to achieve is beyond the scope of this paper: A realistic picture of the supernova-history of our Galaxy as well as detailed Galactic cosmic ray propagation simulations for different spectra with different compositions injected at different scale heights and times. 

For the neutral ISM phases we see a significant mismatch in the ``LECR fingerprint'', in particular for CNM where this applies almost all elements. Additionally, with the low abundance of dust and the also low ionisation fraction of the gas elements, only a very small fraction of the thermal matter would be accelerated in this scenario, probably leading to significant energetics problems when it comes to explaining the total LECR flux. Together with the problem that it is not clear whether strong J-type shocks that would allow for DSA can actually form in the weak ion-component of the CNM, we therefore think that we can safely exclude this environment as a significant contributor to cosmic ray production.

For the WNM, the situation looks somewhat better as the mismatch is confined essentially to a few elements: the pure gas phase elements H, He, N, and Ne, and the semi-volatile elements S and Cl. It is interesting to ask whether UV photons due to the reverse fluorescence process illuminating the shock site in the late ST phases, as introduced in Sect.~\ref{sec:shockEvol}
could solve this problem. First of all, an ionisation of hydrogen that is generally assumed to be the case, would make it easier for strong J-type shock to form. 
\begin{figure}[b]
\centering
\includegraphics[width=.75\linewidth]{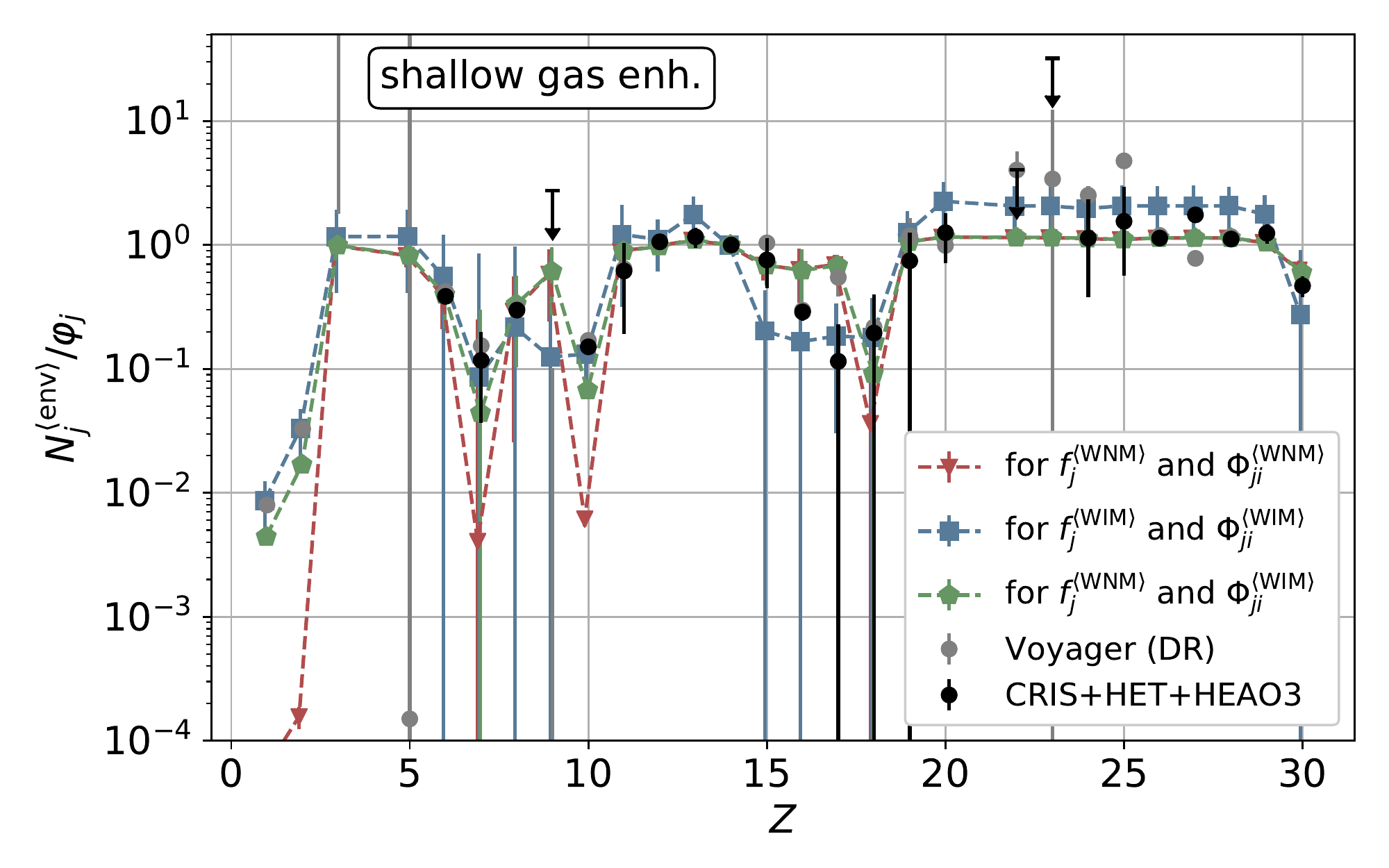}
\caption{\emph{Left:} The neutral fraction of typical gas phase elements under the influence of a strong UV radiation field. 
\emph{Right:} Using $\enhdust=486$ and $\enhmax=11$ with a shallow gas enhancement (\ref{eq:H+19enhanc}), the total enhancement of elements (similar to the lower panel in Fig.~\ref{fig:bestfit_Hanusch}) is shown for the WNM (red), the WIM (blue) as well as a mixed environment (green) that features the ionization fractions of the WIM, but the gas fractions of the WNM.}
\label{fig:fakeWIM_Hanusch}
\end{figure}
To improve the LECR fingerprint, however, would require a significant change of the ionization \emph{and} gas fractions, so that the neutral environment ahead of the shock becomes similar to the WIM. The problem is that for many dust grains energies of a few eV are not enough to liberate their surface atoms --- only the absorption of extreme UV to $\gamma$-ray photon lead to destructive process \cite{Jones2004}. Even though neon has a first ionization potential of $21.6\,\text{eV}$, significantly above that of hydrogen, a UV flux $\gtrsim 0.1\,\text{erg}\,\text{cm}^{-2}\,\text{s}^{-1}$ at the injection sight is already sufficient to ionize all of the gas elements via multiple excitation. For such a case, we show in the Fig.~\ref{fig:fakeWIM_Hanusch} the fingerprint that would be obtained for a ``mixed'' environment, i.e.\ the ionization fractions of the WIM, but the gas fractions of the WNM. We see that indeed the mismatch in hydrogen, helium and neon is removed by ionization, and we remain with the over-enhancement of sulfur and chlorine. On the other hand, the mixed case fixes the problem with the phosphorus abundance we see in the pure WIM contribution, and moreover the abundances of calcium and the iron group elements are even better reproduced. A detailed treatment of the effect of the UV field of the SNR on ionization and gas fractions contains currently too many unknowns to be solved in detail, but we want to point out that a significant contribution by the WNM to Galactic cosmic ray production cannot be excluded if ionization of the gas by the supernova is taken into account. In fact, it seems that a mix of a UV-ionized WNM with an about equally strong contribution from warm ionized ISM environments could fix the problems for all, sulfur, chlorine \textit{and} phosphorus, and produce a generally better fingerprint than one environment alone.

Our model is based on a detailed physical differentiation of processes regarding elements in the ISM gas phase, using elaborate program packages to calculate the ionization states in the ISM and published results from PIC simulations to turn these into nonthermal element abundances observable in cosmic rays. 
We even included time dependent evolution of ionization states in the post shock environment, but if the Bell instability drives the diffusion in these sources, DSA injection timescales are of the order of minutes to days, too short for a significant change in ionization states. The results we obtain for cosmic ray abundances are hereby not very dependent on different yet possible $A/Q$ scalings and maximum values for the element enhancement obtained in PIC simulations by different groups, neither they seem to strongly depend on gas temperature or density. Various types of the ionized medium at temperatures around $\sim 10^4$ Kelvin (including HII regions), with densities differing over orders of magnitude, all lead to comparable results. This strengthens our confidence that some simplifying assumptions we introduced for HII regions do not affect the validity of our results either: The restriction to a typical O6 star as a central heating source, where a mixture of different types of O and B stars would have been more reasonable, or the simplification that the SNR is centered at the center of the HII region, which may apply approximately in many, but definitely not in all cases.

So, ironically, it seems that in spite of all the effort we have put into a detailed treatment of DSA injection from the gas-phase, the process found to be most affecting the results is the injection via dust grains. This, however, leads to the main caveat of our model, because here we rely on crude simplifications: We do not differentiate between different grain masses or structures, neither we include the details of sputtering of individual atoms off grains during the acceleration. Instead, we follow the basic idea of Epstein \cite{Epstein1980}, further elaborated by Ellison, Drury and Mayer \cite{EllisonDruryMeyer1997}, that dust grains behave like ions of a high mass-to-charge ratio, and that the sputtered grain material receives the same velocity as its parental grain and remains ionized to be kept in the acceleration process. Due to the lack of knowledge on the details of the dust acceleration it seems reasonable here to condense all uncertainties into a single parameter, $\enhdust$. Certainly this entire approach cannot be considered more that a first (if not zeroth) order approximation, so we may also take our encouraging results to motivate modern, state-of-the-art simulation that could shed more light on the details dust grain shock-acceleration.

Given this caveat, it is almost surprising how well we can reproduce observed LECR element abundances. 
Though, the reduced $\chi^2$ values obtained have to be taken with a grain of salt as the actual number of degrees of freedom is highly uncertain, but somewhere in the range $1\leq \nu\leq 18$. But even for little degrees of freedom, the best fit regimes in (D)WIM and HII-regions yield $\chi^2_\nu<1$ where it should be expected from statistics to be $\sim 1$. Hence, we seem to have overestimated the uncertainties of the gas fractions in the various warm ionized ISM phases, as these are included in the errors used to determine the reduced $\chi^2$. Turning this around it leads to an expectation that, if our model can be further corroborated and all vagueness in it be replaced by hard simulation results, it might be possible to use our model to infer this gas fractions \textit{for ionized ISM phases} from cosmic ray data with a higher accuracy than possible by conventional methods. An intriguing possibility not only astroparticle physics can learn from conventional astrophysics, but also vice versa, cosmic ray observations may allow to fill remaining white spots in our knowledge about astrophysics.

From the point of view of cosmic ray physics, however, the most intriguing applications of our model will be to extend beyond the low energy regime, and to use it for predictions of cosmic ray compositions for various sources that may contribute to the knee, second knee or even the ultra-high energy regime. We want to briefly summarize our main findings which affect the fundamentals of DSA and this extend beyond the application to SNR in the ISM:
\begin{itemize}
    \item We confirm that gas element enhancements scaling with $\enhgas\propto (A/Q)$ work just as well as by $\enhgas\propto (A/Q)^{3/2}$.
    \item The $A/Q$-scaling of the gas enhancement most likely saturates around a maximum value $\enhmax \sim 30$.
    \item A much larger enhancement, $\enhdust \gtrsim 300$ must apply to elements which are injected via charged dust grains.
\end{itemize}
These basic insights can be applied to predict nonthermal element abundances resulting from shocks in \textit{any} astrophysical environment for which we know, or at least can reasonably estimate the density and temperature, as well as gas and dust fractions for all relevant elements. We want to briefly describe two examples for such future applications.

A) As argued in Ref.~\cite{Thoudam:2016syr}, cosmic rays at the second knee around an energy of 100 PeV might be explained by a Galactic contribution resulting from supernovae with Wolf-Rayet (WR) star progenitors, likely to be be identified with the supernova types Ib/c \cite{Crowther2007, Groh+2013}. In this scenario, cosmic ray acceleration does not mainly take place in the ISM, but in the supernova shock while it crosses the extended wind zone of the progenitor WR star \cite{BiermannCassinelli1993}, a regime with elemental composition very different from the regular ISM \cite[see references in][]{Thoudam:2016syr}. In previous treatments, the thermal element abundances in these wind zone have been shifted to cosmic ray abundances simply by applying the known enhancements from regular cosmic rays --- a very questionable method. Our model will allow to derive the enhancement factors in detail from the temperature and density profile of WR wind zones and thus may make more accurate predictions for the cosmic ray composition around the second knee, where currently many different experiments are providing data \cite[see][for references]{Thoudam:2016syr}. It also affects the self-consistency of the results presented here as we have to check for a possible contribution of such WR supernovae in the LECR regime -- an interesting question that we want to consider in an upcoming study.

B) Turning to the ultra-high energy cosmic ray regime, we have proposed in another work that radio galaxies may be the main sources in this regime, with a particular role being given to the nearby radio galaxy Centaurus A \cite{Eichmann+2018, Eichmann2019_proc}. Based on our current work, however, we have to reject the proposal in that paper of a gas-phase enhancement $\eta_{\rm gas}\sim 10^3$ for iron, required to explain the data at the highest energies around 100 EeV. Rather to achieve such a high enhancement, the contribution of dust grains needs to be taken into account. Only in a neutral or partly ionized environment are heavy elements in large parts locked in dust grains, so the only scenario which may allow to reach such high enhancements in radio galaxies seems to be the interaction of the radio galaxy jet with dusty molecular clouds. Luckily, evidence for jet-cloud interactions have been found for nearby radio galaxies, among them Centaurus~A \cite{Hardcastle+2003} and PKS2153-69 \cite{Young+2005}, which both count to the set of radio galaxies expected to deliver the strongest contribution to ultra-high energy cosmic rays \cite{RachenEichmannICRC2019}. So, the proposal of our above-mentioned work that Centaurus A provides an unusually heavy composition that might explain the data delivered by the Pierre Auger Observatory \cite{AugerComposition2014} may be considered reasonable, albeit on different grounds than originally thought. Accurate predictions beyond such qualitative considerations certainly require a detailed treatment of the physics of jet-cloud interactions with particular focus on the formation of strong, J-type shocks.

\acknowledgments

We like to thank D.~Caprioli for useful comments and hints concerning the injection process, and C.~Morisset for his support with respect to the pyCloudy package. In addition, we thank the anonymous referee for useful comments that improved the original manuscript. Some of the results in this paper have been derived using the software packages
Numpy \cite{vanDerWalt2011}, Matplotlib \cite{Hunter:2007} as well as \href{https://chiantipy.readthedocs.io/en/latest/index.html}{ChiantiPy}\footnote{\href{https://chiantipy.readthedocs.io/en/latest/index.html}{https://chiantipy.readthedocs.io/en/latest/index.html}}  which is based on the Chianti database \cite{Chianti9_2019, Chianti1_1997} and \href{https://github.com/Morisset/pyCloudy}{pyCloudy}\footnote{\href{https://github.com/Morisset/pyCloudy}{https://github.com/Morisset/pyCloudy}} that deals with the input and output files of Cloudy \cite{Cloudy2017}.

\bibliographystyle{JHEP}
\addcontentsline{toc}{section}{Bibliography}
\bibliography{references}

\providecommand{\href}[2]{#2}\begingroup\raggedright\begin{thebibliography}{10}

\bibitem{Matthiae2019}
G.~{Matthiae}, {\it {Knee, ankle and GZK in historical perspective}},  {\em
  Nuclear and Particle Physics Proceedings} {\bf 306-308} (2019) 98--107.

\bibitem{KulikovKhristiansen1959}
G.~{Kulikov} and G.~{Khristiansen}, {\it {On the Size Spectrum of Extensive Air
  Showers}},  {\em Soviet Physics JETP} {\bf 8} (1959), no.~3 441.

\bibitem{Bird+1994}
D.~J. {Bird}, S.~C. {Corbato}, H.~Y. {Dai}, B.~R. {Dawson}, J.~W. {Elbert},
  et~al., {\it {The Cosmic-Ray Energy Spectrum Observed by the Fly's Eye}},
  {\em \apj} {\bf 424} (1994) 491.

\bibitem{KASKADE-Grande2006}
A.~{Haungs}, W.~D. {Apel}, F.~{Badea}, K.~{Bekk}, A.~{Bercuci}, et~al., {\it
  {Investigating the 2nd knee: The KASCADE-Grande experiment}},  in {\em
  Journal of Physics Conference Series}, vol.~47 of {\em Journal of Physics
  Conference Series}, pp.~238--247, 2006.

\bibitem{Hoerandel2007}
J.~R. {H{\"o}randel}, {\it {Cosmic Rays from the Knee to the Second Knee:.
  {}10$^{14}$ to {}10$^{18}$ eV}},  {\em Modern Physics Letters A} {\bf 22}
  (2007), no.~21 1533--1551, [\href{http://arxiv.org/abs/astro-ph/0611387}{{\tt
  astro-ph/0611387}}].

\bibitem{LagageCesarsky1983}
P.~O. {Lagage} and C.~J. {Cesarsky}, {\it {Cosmic-ray shock acceleration in the
  presence of self-excited waves}},  {\em \aap} {\bf 118} (1983), no.~2
  223--228.

\bibitem{Caprioli+2010_contrSNR}
D.~{Caprioli}, E.~{Amato}, and P.~{Blasi}, {\it {The contribution of supernova
  remnants to the galactic cosmic ray spectrum}},  {\em Astroparticle Physics}
  {\bf 33} (Apr., 2010) 160--168, [\href{http://arxiv.org/abs/0912.2964}{{\tt
  arXiv:0912.2964}}].

\bibitem{Ptuskin1993}
V.~S. {Ptuskin}, S.~I. {Rogovaya}, V.~N. {Zirakashvili}, L.~G. {Chuvilgin},
  G.~B. {Khristiansen}, et~al., {\it {Diffusion and drift of very high energy
  cosmic rays in galactic magnetic fields}},  {\em \aap} {\bf 268} (1993),
  no.~2 726--735.

\bibitem{Candia+2002}
J.~{Candia}, E.~{Roulet}, and L.~N. {Epele}, {\it {Turbulent diffusion and
  drift in galactic magnetic fields and the explanation of the knee in the
  cosmic ray spectrum}},  {\em Journal of High Energy Physics} {\bf 2002}
  (2002), no.~12 033, [\href{http://arxiv.org/abs/astro-ph/0206336}{{\tt
  astro-ph/0206336}}].

\bibitem{Thoudam:2016syr}
S.~Thoudam, J.~P. Rachen, A.~van Vliet, A.~Achterberg, S.~Buitink, et~al., {\it
  {Cosmic-ray Energy Spectrum and Composition up to the Ankle: the Case for a
  Second Galactic Component}},  {\em Astron. Astrophys.} {\bf 595} (2016) A33,
  [\href{http://arxiv.org/abs/1605.03111}{{\tt arXiv:1605.03111}}].

\bibitem{Berezinsky+2005}
V.~{Berezinsky}, A.~Z. {Gazizov}, and S.~I. {Grigorieva}, {\it {Dip in UHECR
  spectrum as signature of proton interaction with CMB}},  {\em Physics Letters
  B} {\bf 612} (2005), no.~3-4 147--153,
  [\href{http://arxiv.org/abs/astro-ph/0502550}{{\tt astro-ph/0502550}}].

\bibitem{UFA2015}
M.~{Unger}, G.~R. {Farrar}, and L.~A. {Anchordoqui}, {\it {Origin of the ankle
  in the ultrahigh energy cosmic ray spectrum, and of the extragalactic protons
  below it}},  {\em \prd} {\bf 92} (2015), no.~12 123001,
  [\href{http://arxiv.org/abs/1505.02153}{{\tt arXiv:1505.02153}}].

\bibitem{Kang+1997}
H.~Kang, J.~P. Rachen, and P.~L. Biermann, {\it {Contributions to the Cosmic
  Ray Flux above the Ankle: Clusters of Galaxies}},  {\em Mon. Not. Roy.
  Astron. Soc.} {\bf 286} (1997) 257,
  [\href{http://arxiv.org/abs/astro-ph/9608071}{{\tt astro-ph/9608071}}].

\bibitem{Murase+2009}
K.~{Murase}, S.~{Inoue}, and K.~{Asano}, {\it {Cosmic Rays above the 2ND Knee
  from Clusters of Galaxies}},  {\em International Journal of Modern Physics D}
  {\bf 18} (2009), no.~10 1609--1614.

\bibitem{Eichmann_2019}
B.~Eichmann, {\it {High Energy Cosmic Rays from Fanaroff-Riley radio
  galaxies}},  {\em Journal of Cosmology and Astroparticle Physics} {\bf 2019}
  (2019), no.~05 009.

\bibitem{Kachelriess+2017}
M.~{Kachelrie{\ss}}, O.~{Kalashev}, S.~{Ostapchenko}, and D.~V. {Semikoz}, {\it
  {Minimal model for extragalactic cosmic rays and neutrinos}},  {\em \prd}
  {\bf 96} (2017), no.~8 083006, [\href{http://arxiv.org/abs/1704.06893}{{\tt
  arXiv:1704.06893}}].

\bibitem{FangMurase2018}
K.~{Fang} and K.~{Murase}, {\it {Linking high-energy cosmic particles by
  black-hole jets embedded in large-scale structures}},  {\em Nature Physics}
  {\bf 14} (2018), no.~4 396--398, [\href{http://arxiv.org/abs/1704.00015}{{\tt
  arXiv:1704.00015}}].

\bibitem{Gaisser+1993}
T.~K. {Gaisser}, T.~{Stanev}, S.~{Tilav}, S.~C. {Corbato}, H.~Y. {Dai}, et~al.,
  {\it {Cosmic-ray composition around {}10$^{18}$ eV}},  {\em \prd} {\bf 47}
  (1993), no.~5 1919--1932.

\bibitem{Rachen:1993gf}
J.~P. Rachen, T.~Stanev, and P.~L. Biermann, {\it {Extragalactic
  Ultrahigh-Energy Cosmic Rays. 2. Comparison with Experimental Data}},  {\em
  Astron. Astrophys.} {\bf 273} (1993) 377,
  [\href{http://arxiv.org/abs/astro-ph/9302005}{{\tt astro-ph/9302005}}].

\bibitem{Kascade2003}
{\bf {KASCADE}} Collaboration, T.~{Antoni} et~al., {\it {The cosmic-ray
  experiment KASCADE}},  {\em Nuclear Instruments and Methods in Physics
  Research A} {\bf 513} (2003), no.~3 490--510.

\bibitem{KascadeGrande2004}
G.~{Navarra}, T.~{Antoni}, W.~D. {Apel}, F.~{Badea}, K.~{Bekk}, et~al., {\it
  {KASCADE-Grande: a large acceptance, high-resolution cosmic-ray detector up
  to 10 $^{18}$ eV}},  {\em Nuclear Instruments and Methods in Physics Research
  A} {\bf 518} (2004), no.~1-2 207--209.

\bibitem{AugerComposition2014}
{\bf {Pierre Auger}} Collaboration, A.~Aab et~al., {\it {Depth of Maximum of
  Air-Shower Profiles at the Pierre Auger Observatory. I. Measurements at
  Energies above $10^{17.8}$ eV}},  {\em {Phys. Rev. D}} {\bf 90} (2014)
  122005, [\href{http://arxiv.org/abs/1409.4809}{{\tt arXiv:1409.4809}}].

\bibitem{TAComposition2018}
{\bf Telescope Array} Collaboration, R.~U. {Abbasi} et~al., {\it {Depth of
  Ultra High Energy Cosmic Ray Induced Air Shower Maxima Measured by the
  Telescope Array Black Rock and Long Ridge FADC Fluorescence Detectors and
  Surface Array in Hybrid Mode}},  {\em \apj} {\bf 858} (2018), no.~2 76.

\bibitem{Buitink+2014}
S.~{Buitink}, A.~{Corstanje}, J.~E. {Enriquez}, H.~{Falcke}, J.~R.
  {H{\"o}randel}, et~al., {\it {Method for high precision reconstruction of air
  shower X$_{max}$ using two-dimensional radio intensity profiles}},  {\em
  \prd} {\bf 90} (2014), no.~8 082003,
  [\href{http://arxiv.org/abs/1408.7001}{{\tt arXiv:1408.7001}}].

\bibitem{Huege2016}
T.~{Huege}, {\it {Radio detection of cosmic ray air showers in the digital
  era}},  {\em \physrep} {\bf 620} (2016) 1--52,
  [\href{http://arxiv.org/abs/1601.07426}{{\tt arXiv:1601.07426}}].

\bibitem{Hoerandel2004}
J.~R. {H{\"o}randel}, {\it {Models of the knee in the energy spectrum of cosmic
  rays}},  {\em Astroparticle Physics} {\bf 21} (2004), no.~3 241--265,
  [\href{http://arxiv.org/abs/astro-ph/0402356}{{\tt astro-ph/0402356}}].

\bibitem{GaisserStanevTilav2013}
T.~K. {Gaisser}, T.~{Stanev}, and S.~{Tilav}, {\it {Cosmic ray energy spectrum
  from measurements of air showers}},  {\em Frontiers of Physics} {\bf 8}
  (2013), no.~6 748--758, [\href{http://arxiv.org/abs/1303.3565}{{\tt
  arXiv:1303.3565}}].

\bibitem{Drury:1983zz}
L.~O. Drury, {\it {An Introduction to the Theory of Diffusive Shock
  Acceleration of Energetic Particles in Tenuous Plasmas}},  {\em Rept. Prog.
  Phys.} {\bf 46} (1983) 973--1027.

\bibitem{MeyerDruryEllison1997}
J.-P. {Meyer}, L.~O. {Drury}, and D.~C. {Ellison}, {\it {Galactic Cosmic Rays
  from Supernova Remnants. I. A Cosmic-Ray Composition Controlled by Volatility
  and Mass-to-Charge Ratio}},  {\em \apj} {\bf 487} (1997), no.~1 182--196,
  [\href{http://arxiv.org/abs/astro-ph/9704267}{{\tt astro-ph/9704267}}].

\bibitem{EllisonDruryMeyer1997}
D.~C. {Ellison}, L.~O. {Drury}, and J.-P. {Meyer}, {\it {Galactic Cosmic Rays
  from Supernova Remnants. II. Shock Acceleration of Gas and Dust}},  {\em
  \apj} {\bf 487} (1997), no.~1 197--217,
  [\href{http://arxiv.org/abs/astro-ph/9704293}{{\tt astro-ph/9704293}}].

\bibitem{Malkov1998}
M.~A. {Malkov}, {\it {Ion leakage from quasiparallel collisionless shocks:
  Implications for injection and shock dissipation}},  {\em \pre} {\bf 58}
  (1998), no.~4 4911--4928, [\href{http://arxiv.org/abs/astro-ph/9806340}{{\tt
  astro-ph/9806340}}].

\bibitem{Caprioli+2017}
D.~{Caprioli}, D.~T. {Yi}, and A.~{Spitkovsky}, {\it {Chemical Enhancements in
  Shock-Accelerated Particles: Ab initio Simulations}},  {\em \prl} {\bf 119}
  (2017), no.~17 171101, [\href{http://arxiv.org/abs/1704.08252}{{\tt
  arXiv:1704.08252}}].

\bibitem{Hanusch+2019}
A.~{Hanusch}, T.~V. {Liseykina}, and M.~{Malkov}, {\it {Acceleration of Cosmic
  Rays in Supernova Shocks: Elemental Selectivity of the Injection Mechanism}},
   {\em \apj} {\bf 872} (2019), no.~1 108,
  [\href{http://arxiv.org/abs/1803.00428}{{\tt arXiv:1803.00428}}].

\bibitem{BiermannStrittmatter1987}
P.~L. Biermann and P.~A. Strittmatter, {\it {Synchrotron Emission from Shock
  Waves in Active Galactic Nuclei}},  {\em Astrophys. J.} {\bf 322} (1987)
  643--649.

\bibitem{Rincon+2016}
F.~{Rincon}, F.~{Califano}, A.~A. {Schekochihin}, and F.~{Valentini}, {\it
  {Turbulent dynamo in a collisionless plasma}},  {\em Proceedings of the
  National Academy of Science} {\bf 113} (2016), no.~15 3950--3953,
  [\href{http://arxiv.org/abs/1512.06455}{{\tt arXiv:1512.06455}}].

\bibitem{BrandenburgSubramanian2005}
A.~{Brandenburg} and K.~{Subramanian}, {\it {Astrophysical magnetic fields and
  nonlinear dynamo theory}},  {\em \physrep} {\bf 417} (2005), no.~1-4 1--209,
  [\href{http://arxiv.org/abs/astro-ph/0405052}{{\tt astro-ph/0405052}}].

\bibitem{TobiasCattaneo2008}
S.~M. {Tobias} and F.~{Cattaneo}, {\it {Dynamo action in complex flows: the
  quick and the fast}},  {\em Journal of Fluid Mechanics} {\bf 601} (2008)
  101--122.

\bibitem{IMAGINE2018}
F.~{Boulanger}, T.~{En{\ss}lin}, A.~{Fletcher}, P.~{Girichides},
  S.~{Hackstein}, et~al., {\it {IMAGINE: a comprehensive view of the
  interstellar medium, Galactic magnetic fields and cosmic rays}},  {\em \jcap}
  {\bf 2018} (2018), no.~8 049, [\href{http://arxiv.org/abs/1805.02496}{{\tt
  arXiv:1805.02496}}].

\bibitem{BrandenburgNordlund2011}
A.~{Brandenburg} and {\r{A}}.~{Nordlund}, {\it {Astrophysical turbulence
  modeling}},  {\em Reports on Progress in Physics} {\bf 74} (2011), no.~4
  046901, [\href{http://arxiv.org/abs/0912.1340}{{\tt arXiv:0912.1340}}].

\bibitem{Seta+2020}
A.~{Seta}, P.~J. {Bushby}, A.~{Shukurov}, and T.~S. {Wood}, {\it {Saturation
  mechanism of the fluctuation dynamo at Pr$_{M}$ {\ensuremath{\geq}} 1}},
  {\em Physical Review Fluids} {\bf 5} (2020), no.~4 043702.

\bibitem{ChamandyShukurov2020}
L.~{Chamandy} and A.~{Shukurov}, {\it {Parameters of the Supernova-Driven
  Interstellar Turbulence}},  {\em Galaxies} {\bf 8} (2020), no.~3 56,
  [\href{http://arxiv.org/abs/2007.14159}{{\tt arXiv:2007.14159}}].

\bibitem{Bell2004}
A.~R. {Bell}, {\it {Turbulent amplification of magnetic field and diffusive
  shock acceleration of cosmic rays}},  {\em \mnras} {\bf 353} (2004), no.~2
  550--558.

\bibitem{Caprioli2012}
D.~{Caprioli}, {\it {Cosmic-ray acceleration in supernova remnants: non-linear
  theory revised}},  {\em \jcap} {\bf 2012} (2012), no.~7 038,
  [\href{http://arxiv.org/abs/1206.1360}{{\tt arXiv:1206.1360}}].

\bibitem{RachenManifesto2019}
J.~P. {Rachen}, {\it {Origin of Ultra-High Energy Cosmic Rays: The Cosmological
  Manifesto}},  in {\em 36th International Cosmic Ray Conference (ICRC2019)},
  vol.~36 of {\em International Cosmic Ray Conference}, p.~397, 2019.
\newblock \href{http://arxiv.org/abs/1909.00356}{{\tt arXiv:1909.00356}}.

\bibitem{Simpson1983}
J.~A. {Simpson}, {\it {Elemental and Isotopic Composition of the Galactic
  Cosmic Rays}},  {\em Annual Review of Nuclear and Particle Science} {\bf 33}
  (1983) 323--382.

\bibitem{Voyager2016}
A.~C. {Cummings}, E.~C. {Stone}, B.~C. {Heikkila}, N.~{Lal}, W.~R. {Webber},
  G.~{J{\'o}hannesson}, I.~V. {Moskalenko}, E.~{Orlando}, and T.~A. {Porter},
  {\it {Galactic Cosmic Rays in the Local Interstellar Medium: Voyager 1
  Observations and Model Results}},  {\em \apj} {\bf 831} (2016), no.~1 18.

\bibitem{HuntHirashita2009}
L.~K. {Hunt} and H.~{Hirashita}, {\it {The size-density relation of
  extragalactic H II regions}},  {\em \aap} {\bf 507} (2009), no.~3 1327--1343,
  [\href{http://arxiv.org/abs/0910.2804}{{\tt arXiv:0910.2804}}].

\bibitem{Crowther2013}
P.~A. {Crowther}, {\it {On the association between core-collapse supernovae and
  H ii regions}},  {\em \mnras} {\bf 428} (2013), no.~3 1927--1943,
  [\href{http://arxiv.org/abs/1210.1126}{{\tt arXiv:1210.1126}}].

\bibitem{GeyerWalker2018}
M.~{Geyer} and M.~A. {Walker}, {\it {The character of the warm ionized
  medium}},  {\em \mnras} {\bf 481} (2018), no.~2 1609--1623.

\bibitem{HeilesTroland_2003}
C.~{Heiles} and T.~H. {Troland}, {\it {The Millennium Arecibo 21 Centimeter
  Absorption-Line Survey. II. Properties of the Warm and Cold Neutral Media}},
  {\em \apj} {\bf 586} (2003), no.~2 1067--1093,
  [\href{http://arxiv.org/abs/astro-ph/0207105}{{\tt astro-ph/0207105}}].

\bibitem{Ferriere2001_ISMreview}
K.~M. {Ferri{\`e}re}, {\it {The interstellar environment of our galaxy}},  {\em
  Reviews of Modern Physics} {\bf 73} (2001), no.~4 1031--1066,
  [\href{http://arxiv.org/abs/astro-ph/0106359}{{\tt astro-ph/0106359}}].

\bibitem{Black1987}
J.~H. {Black}, {\it Heating and cooling of the interstellar gas},  in {\em
  Interstellar Processes} (D.~J. Hollenbach and H.~A. Thronson, eds.),
  (Dordrecht), pp.~731--744, Springer Netherlands, 1987.

\bibitem{OsterbrockFerland2006_book}
D.~E. {Osterbrock} and G.~J. {Ferland}, {\em {Astrophysics of gaseous nebulae
  and active galactic nuclei}}.
\newblock {University Science Books}, 2006.

\bibitem{2009A&A...498..915D}
K.~P. {Dere}, E.~{Landi}, P.~R. {Young}, G.~{Del Zanna}, M.~{Landini}, and
  H.~E. {Mason}, {\it {CHIANTI - an atomic database for emission lines. IX.
  Ionization rates, recombination rates, ionization equilibria for the elements
  hydrogen through zinc and updated atomic data}},  {\em \aap} {\bf 498}
  (2009), no.~3 915--929.

\bibitem{TrueloveMcKee1999}
J.~K. {Truelove} and C.~F. {McKee}, {\it {Evolution of Nonradiative Supernova
  Remnants}},  {\em \apjs} {\bf 120} (1999), no.~2 299--326.

\bibitem{Sauer+2008}
D.~N. {Sauer}, P.~A. {Mazzali}, S.~{Blondin}, M.~{Stehle}, S.~{Benetti},
  et~al., {\it {Properties of the ultraviolet flux of Type Ia supernovae: an
  analysis with synthetic spectra of SN 2001ep and SN 2001eh}},  {\em \mnras}
  {\bf 391} (2008), no.~4 1605--1618,
  [\href{http://arxiv.org/abs/0803.0871}{{\tt arXiv:0803.0871}}].

\bibitem{Lucy1999}
L.~B. {Lucy}, {\it {Improved Monte Carlo techniques for the spectral synthesis
  of supernovae}},  {\em \aap} {\bf 345} (1999) 211--220.

\bibitem{SavageMathis1979}
B.~D. {Savage} and J.~S. {Mathis}, {\it {Observed properties of interstellar
  dust.}},  {\em \araa} {\bf 17} (1979) 73--111.

\bibitem{Draine2003}
B.~T. {Draine}, {\it {Interstellar Dust Grains}},  {\em \araa} {\bf 41} (2003)
  241--289, [\href{http://arxiv.org/abs/astro-ph/0304489}{{\tt
  astro-ph/0304489}}].

\bibitem{PugetLeger1989}
J.~L. {Puget} and A.~{Leger}, {\it {A new component of the interstellar matter:
  small grains and large aromatic molecules.}},  {\em \araa} {\bf 27} (1989)
  161--198.

\bibitem{Tielens2008}
A.~G.~G.~M. {Tielens}, {\it {Interstellar polycyclic aromatic hydrocarbon
  molecules.}},  {\em \araa} {\bf 46} (2008) 289--337.

\bibitem{Tarakeshwar+2019}
P.~{Tarakeshwar}, P.~R. {Buseck}, and F.~X. {Timmes}, {\it {On the Structure,
  Magnetic Properties, and Infrared Spectra of Iron Pseudocarbynes in the
  Interstellar Medium}},  {\em \apj} {\bf 879} (2019), no.~1 2.

\bibitem{Tielens1998}
A.~G.~G.~M. {Tielens}, {\it {Interstellar Depletions and the Life Cycle of
  Interstellar Dust}},  {\em \apj} {\bf 499} (1998), no.~1 267--272.

\bibitem{WeingartnerDraine2001}
J.~C. {Weingartner} and B.~T. {Draine}, {\it {Photoelectric Emission from
  Interstellar Dust: Grain Charging and Gas Heating}},  {\em \apjs} {\bf 134}
  (2001), no.~2 263--281, [\href{http://arxiv.org/abs/astro-ph/9907251}{{\tt
  astro-ph/9907251}}].

\bibitem{Jenkins2009}
E.~B. {Jenkins}, {\it {A Unified Representation of Gas-Phase Element Depletions
  in the Interstellar Medium}},  {\em \apj} {\bf 700} (2009), no.~2 1299--1348,
  [\href{http://arxiv.org/abs/0905.3173}{{\tt arXiv:0905.3173}}].

\bibitem{SavageSembach1996}
B.~D. {Savage} and K.~R. {Sembach}, {\it {Interstellar Abundances from
  Absorption-Line Observations with the Hubble Space Telescope}},  {\em \araa}
  {\bf 34} (1996) 279--330.

\bibitem{Barker+1984_Al}
E.~S. {Barker}, P.~M. {Lugger}, E.~J. {Weiler}, and D.~G. {York}, {\it
  {Abundance of interstellar aluminum.}},  {\em \apj} {\bf 280} (1984)
  600--607.

\bibitem{Crinklaw+1994_Ca}
G.~{Crinklaw}, S.~R. {Federman}, and C.~L. {Joseph}, {\it {The Depletion of
  Calcium in the Interstellar Medium}},  {\em \apj} {\bf 424} (1994) 748.

\bibitem{SofiaJenkins1998}
U.~J. {Sofia} and E.~B. {Jenkins}, {\it {Interstellar Medium Absorption Profile
  Spectrograph Observations of Interstellar Neutral Argon and the Implications
  for Partially Ionized Gas}},  {\em \apj} {\bf 499} (1998), no.~2 951--965,
  [\href{http://arxiv.org/abs/astro-ph/9712260}{{\tt astro-ph/9712260}}].

\bibitem{Kemp+2002_NaK}
S.~N. {Kemp}, B.~{Bates}, J.~E. {Beckman}, C.~J. {Killow}, R.~{Barrena}, D.~C.
  {Kennedy}, and J.~{Rodr{\'\i}guez Alamo}, {\it {A study of the behaviour of
  the NaI/KI column density ratio in the interstellar medium using the Na
  ultraviolet doublet}},  {\em \mnras} {\bf 333} (2002), no.~3 561--574.

\bibitem{Kimura+2003_LIC}
H.~{Kimura}, I.~{Mann}, and E.~K. {Jessberger}, {\it {Elemental Abundances and
  Mass Densities of Dust and Gas in the Local Interstellar Cloud}},  {\em \apj}
  {\bf 582} (2003), no.~2 846--858.

\bibitem{Schramm+1989}
L.~S. {Schramm}, D.~E. {Brownlee}, and M.~M. {Wheelock}, {\it {Major Element
  Composition of Stratospheric Micrometeorites}},  {\em Meteoritics} {\bf 24}
  (1989), no.~2 99.

\bibitem{isotopesTableOnline}
S.~{Chu}, L.~{Ekström}, and R.~{Firestone}, ``{WWW Table of Radioactive
  Isotopes}.'' Database version 1999-02-28 from URL:
  http://nucleardata.nuclear.lu.se/toi/.

\bibitem{PalmeLoddersJones2014}
H.~{Palme}, K.~{Lodders}, and A.~{Jones}, {\it {Solar System Abundances of the
  Elements}},  in {\em {Treatise on Geochemistry}} (A.~M. {Davis}, ed.),
  vol.~2, pp.~15--36, Elsevier, 2014.

\bibitem{NievaPrzybilla2012}
M.~F. {Nieva} and N.~{Przybilla}, {\it {Present-day cosmic abundances. A
  comprehensive study of nearby early B-type stars and implications for stellar
  and Galactic evolution and interstellar dust models}},  {\em \aap} {\bf 539}
  (2012) A143, [\href{http://arxiv.org/abs/1203.5787}{{\tt arXiv:1203.5787}}].

\bibitem{DraineMcKee1993}
B.~T. {Draine} and C.~F. {McKee}, {\it {Theory of interstellar shocks.}},  {\em
  \araa} {\bf 31} (1993) 373--432.

\bibitem{ACE-CRIS2018}
M.~H. {Israel}, K.~A. {Lave}, M.~E. {Wiedenbeck}, W.~R. {Binns}, E.~R.
  {Christian}, et~al., {\it {Elemental Composition at the Cosmic-Ray Source
  Derived from the ACE-CRIS Instrument. I. $_{6}$C to $_{28}$Ni}},  {\em \apj}
  {\bf 865} (2018), no.~1 69.

\bibitem{ULYSSES1996}
M.~A. {Duvernois} and M.~R. {Thayer}, {\it {The Elemental Composition of the
  Galactic Cosmic-Ray Source: ULYSSES High-Energy Telescope Results}},  {\em
  \apj} {\bf 465} (1996) 982.

\bibitem{1981ICRC....9..118P}
C.~{Perron}, J.~J. {Engelmann}, P.~{Goret}, E.~{Juliusson}, L.~{Koch-Miramond},
  et~al., {\it {Interpretation of the elemental abundances measured by the
  French-Danish experiment on HEAO-3 - Interstellar propagation and derivation
  of source abundances}},  in {\em International Cosmic Ray Conference}, vol.~9
  of {\em International Cosmic Ray Conference}, pp.~118--121, 1981.

\bibitem{Starrfield+2020}
S.~{Starrfield}, M.~{Bose}, C.~{Iliadis}, W.~R. {Hix}, C.~E. {Woodward}, and
  R.~M. {Wagner}, {\it {Carbon-Oxygen Classical Novae Are Galactic $^{7}$Li
  Producers as well as Potential Supernova Ia Progenitors}},  {\em \apj} {\bf
  895} (2020), no.~1 70, [\href{http://arxiv.org/abs/1910.00575}{{\tt
  arXiv:1910.00575}}].

\bibitem{Duric2000proc}
N.~{Duric}, {\it {Supernova remnants and the ISM in M31 and M33}},  in {\em
  Proceedings 232. WE-Heraeus Seminar} (E.~M. {Berkhuijsen}, R.~{Beck}, and
  R.~A.~M. {Walterbos}, eds.), pp.~127--130, Dec., 2000.

\bibitem{Asvarov2014}
A.~I. {Asvarov}, {\it {Size distribution of supernova remnants and the
  interstellar medium: the case of M 33}},  {\em \aap} {\bf 561} (2014) A70,
  [\href{http://arxiv.org/abs/1311.5166}{{\tt arXiv:1311.5166}}].

\bibitem{Jones2004}
A.~P. {Jones}, {\it {Dust Destruction Processes}},  in {\em Astrophysics of
  Dust} (A.~N. {Witt}, G.~C. {Clayton}, and B.~T. {Draine}, eds.), vol.~309 of
  {\em Astronomical Society of the Pacific Conference Series}, p.~347, 2004.

\bibitem{Epstein1980}
R.~I. {Epstein}, {\it {The acceleration of interstellar grains and the
  composition of the cosmic rays}},  {\em \mnras} {\bf 193} (1980) 723--729.

\bibitem{Crowther2007}
P.~A. {Crowther}, {\it {Physical Properties of Wolf-Rayet Stars}},  {\em \araa}
  {\bf 45} (2007), no.~1 177--219,
  [\href{http://arxiv.org/abs/astro-ph/0610356}{{\tt astro-ph/0610356}}].

\bibitem{Groh+2013}
J.~H. {Groh}, C.~{Georgy}, and S.~{Ekstr{\"o}m}, {\it {Progenitors of supernova
  Ibc: a single Wolf-Rayet star as the possible progenitor of the SN Ib
  iPTF13bvn}},  {\em \aap} {\bf 558} (2013) L1,
  [\href{http://arxiv.org/abs/1307.8434}{{\tt arXiv:1307.8434}}].

\bibitem{BiermannCassinelli1993}
P.~L. {Biermann} and J.~P. {Cassinelli}, {\it {Cosmic rays. II. Evidence for a
  magnetic rotator Wolf-Rayet star origin}},  {\em \aap} {\bf 277} (1993) 691,
  [\href{http://arxiv.org/abs/astro-ph/9305003}{{\tt astro-ph/9305003}}].

\bibitem{Eichmann+2018}
B.~Eichmann, J.~Rachen, L.~Merten, A.~van Vliet, and J.~B. Tjus, {\it
  {Ultra-high-energy cosmic rays from radio galaxies}},  {\em Journal of
  Cosmology and Astroparticle Physics} {\bf 2018} (2018), no.~02 036.

\bibitem{Eichmann2019_proc}
B.~{Eichmann}, {\it {High-Energy Cosmic Rays from Radio Galaxies}},  in {\em
  European Physical Journal Web of Conferences}, vol.~210 of {\em European
  Physical Journal Web of Conferences}, p.~04001, 2019.
\newblock \href{http://arxiv.org/abs/1812.09475}{{\tt arXiv:1812.09475}}.

\bibitem{Hardcastle+2003}
M.~J. {Hardcastle}, D.~M. {Worrall}, R.~P. {Kraft}, W.~R. {Forman}, C.~{Jones},
  and S.~S. {Murray}, {\it {Radio and X-Ray Observations of the Jet in
  Centaurus A}},  {\em \apj} {\bf 593} (2003), no.~1 169--183,
  [\href{http://arxiv.org/abs/astro-ph/0304443}{{\tt astro-ph/0304443}}].

\bibitem{Young+2005}
A.~J. {Young}, A.~S. {Wilson}, S.~J. {Tingay}, and S.~{Heinz}, {\it {The Halo,
  Hot Spots, and Jet/Cloud Interaction of PKS 2153-69}},  {\em \apj} {\bf 622}
  (2005), no.~2 830--841, [\href{http://arxiv.org/abs/astro-ph/0412547}{{\tt
  astro-ph/0412547}}].

\bibitem{RachenEichmannICRC2019}
J.~P. {Rachen} and B.~{Eichmann}, {\it {A parametrized catalog of radio
  galaxies as UHECR sources}},  in {\em 36th International Cosmic Ray
  Conference (ICRC2019)}, vol.~36 of {\em International Cosmic Ray Conference},
  p.~396, 2019.
\newblock \href{http://arxiv.org/abs/1909.00261}{{\tt arXiv:1909.00261}}.

\bibitem{vanDerWalt2011}
S.~van~der Walt, S.~C. Colbert, and G.~Varoquaux, {\it {The NumPy Array: a
  Structure for Efficient Numerical Computation}},
  \href{http://arxiv.org/abs/1102.1523}{{\tt arXiv:1102.1523}}.

\bibitem{Hunter:2007}
J.~D. Hunter, {\it {Matplotlib: A 2D Graphics Environment}},  {\em Comput. Sci.
  Eng.} {\bf 9} (2007), no.~3 90--95.

\bibitem{Chianti9_2019}
K.~P. Dere, G.~D. Zanna, P.~R. Young, E.~Landi, and R.~S. Sutherland, {\it
  {CHIANTI}{\textemdash}an atomic database for emission lines. {XV}. version 9,
  improvements for the x-ray satellite lines},  {\em The Astrophysical Journal
  Supplement Series} {\bf 241} (2019), no.~2 22.

\bibitem{Chianti1_1997}
K.~P. {Dere}, E.~{Landi}, H.~E. {Mason}, B.~C. {Monsignori Fossi}, and P.~R.
  {Young}, {\it {CHIANTI - an atomic database for emission lines}},  {\em
  \aaps} {\bf 125} (1997) 149--173.

\bibitem{Cloudy2017}
G.~J. {Ferland}, M.~{Chatzikos}, F.~{Guzm{\'a}n}, M.~L. {Lykins}, P.~A.~M. {van
  Hoof}, et~al., {\it {The 2017 Release Cloudy}},  {\em \rmxaa} {\bf 53} (2017)
  385--438, [\href{http://arxiv.org/abs/1705.10877}{{\tt arXiv:1705.10877}}].

\end{thebibliography}\endgroup

\appendix

\section{Simulation setup for the upstream environment}
\label{app:cloudySim}

The ionization fractions $\ionfracUp{}$ in the upstream medium is one of the crucial parameters where the characteristics of the ISM environment enter. To account for them, we need to distinguish between the neutral and the ionized phases of the ISM, as well as HII-regions. All of the ISM phases are supposed to be in thermal equilibrium, so that we define the temperature as given in Table \ref{AParameters}. In addition to collisional ionization we include the ionization due to the cosmic microwave background, the radiation field of the ISM and the cosmic ray background which are provided by \texttt{Cloudy}. The latter is in particular important for the neutral media of the ISM. For these environments it is important to consider that they do not contain a source of ionization, but become ionized from the outside. Therefore, these environments are simulated using a typical thickness $h^{\langle \mathrm{env} \rangle}$. Here, we used their scale height, i.e. $h^{\langle \mathrm{CNM} \rangle}=140\,\text{pc}$ and $h^{\langle \mathrm{WNM} \rangle}=400\,\text{pc}$, although these are rather upper limits, but the exact value has no influence on the results as long as $h^{\langle \mathrm{env} \rangle}\gtrsim 1\,\text{pc}$. Thus, the radiation can only ionize at the very outside of the environment. Further, we suppose that the radiation is extinguished by the gas in the ISM, so that little radiation exists at around a few Ryd --- a column density of the Galactic plane of $10^{22}\,\text{cm}^{-2}$ without leakage is used as suggested by the \text{Cloudy} manual. 
The ionized ISM phases likely contain a source of ionization, so that we consider a simple (and fast) one-zone model with vanishing thickness, and a vanishing extinction of the ISM radiation. 
In contrast to these ISM phases, is the HII-region not in thermal equilibrium, so that its temperature and ionization fraction depends on the ionizing star. Here, we use that the radiation field of the star is well approximated by its blackbody radiation for a given temperature $T_*$ and photon luminosity $Q_*(H)$. As the ionization fraction strongly depends on the distance $d_*$ to the star, we suppose that the center of the SNR coincides with the location of the star, so that only those particles that correspond to the actual shock radius need to be taken into account. Hence, we consider a small (one-zone) volume element at a distance $d_*=r_{\rm sh}$ in order to determine $\ionfracUp[HII]{}$ close to the shock surface. This is in contrast to the ISM phases not constant in time due to the shock evolution. Note that this approach presumes that the interior gas at distances $< d^*$ does not absorb the radiation significantly as it is predominantly ionized.

\section{Adiabatic losses}
\label{app:AdiabaticLosses}

Supposing that DSA happens on much shorter timescales than the adiabatic losses by the expansion of the SNR, we can disentangle these two processes and apply that the kinetic equation of particles \emph{at given momentum $p_0$} in the acceleration region is given by
\begin{equation}
     \frac{\partial N_j(t)}{\partial t}=S(t)-\frac{N_j(t)}{\tau_{\rm ad}(t)} \,,
 \label{eq:DGL_adiabaticLosses}
 \end{equation}
 Here, $S(t)=\left[\frac{dN_j(p=p_0)}{dp}\right]_{\Delta t}/\Delta t$ provides the accelerated particle spectrum according to Eq.~\ref{eq:dNdp}. Further, the adiabatic loss timescale can be approximated by 
 \begin{equation}
     \tau_{\rm ad}(t) \,\simeq\, \frac{3}{\mathbf{\nabla\cdot v}_{\rm sh}} \;=\; \frac{r_{\rm sh}(t)}{\beta_{\rm sh}(t)c}
     \,\simeq\,\begin{cases}
     \frac{r_{\rm ch}}{\beta_{\rm ch}c}\left( \frac{t}{t_{\rm ch}} \right)\,\quad&\text{ for } t<t_{\rm ch}\,,\\
     2.50\,\frac{r_{\rm ch}}{\beta_{\rm ch}c} \left( \frac{t}{t_{\rm ch}} \right)\,\quad&\text{ for } t\geq t_{\rm ch}\,,
     \end{cases}
 \end{equation} 
 in the given case of a uniformly expanding sphere, whose temporal development is described by Eq.~\ref{eq:EDphaseEQ} and \ref{eq:STphaseEQ}. In order to provide an analytical solution, we use the temporal development in the limit of small/ large timescales with respect to $t_{\rm ch}$. Applying a standard solution procedure, where we account that $N_j(t=0)=0$, the solution of the first order, linear differential equation (\ref{eq:DGL_adiabaticLosses}) is given by
 \begin{equation}
     \numspec[]{j}(t) \;=\; \left( \frac{t}{t_{\rm ch}} \right)^{-\alpha_{\rm ad}(t)}\!\int_0^t\diff t'\,\,S(t')\,\left( \frac{t'}{t_{\rm ch}} \right)^{\alpha_{\rm ad}(t')}\,,
 \end{equation}
 with
 \begin{equation}
     \alpha_{\rm ad}(t)= \begin{cases}
     1\,\quad&\text{ for } t<t_{\rm ch}\,,\\
     0.4\,\quad&\text{ for } t\geq t_{\rm ch}\,.
     \end{cases}
 \end{equation} 
 Thus, the total number of accelerated particles at a time $t$, that enter the expanding sphere at $t'$ gets reduced by a factor
 \begin{equation}
     \Lambda_{\rm ad}(t,t') = \left( \frac{t}{t_{\rm ch}} \right)^{-\alpha_{\rm ad}(t)}\,\left( \frac{t'}{t_{\rm ch}} \right)^{\alpha_{\rm ad}(t')}\,.
 \end{equation}

\section{Galactic element abundances and gas fractions}
\label{app:GasFrac}

\subsection{General approach to gas fraction estimates}
\label{app:GasFrac_approach}

Gas fractions of elements are very difficult to measure, and the results of such measurements are even more difficult to generalize. Even within the rich data material presented by Jenkins \cite{Jenkins2009} there is significant spread between individual measurements and systematic uncertainties regarding the identification of typical ISM phases with the abstract depletion factor $F_*$, in particular when it comes to extrapolations in the ionized ISM phases for that no data exist. For elements not included in this study, we often rely only on depletion data gained on a few, not necessarily characteristic lines of sight, and extrapolations are mostly gained by physical plausibility and qualitative insights, as for example from micrometeorites. In short, if we want to provide a list of gas fractions for all elements up to Zn and for all ISM phases, we have to span a range from hard data constraints to physically based estimates to wild speculation. How can this be done in a mathematically sound way?

Although this is not a Bayesian paper, it is the Bayesian way of thinking which allows us to master this span. What we need to do is to provide best guess values for each gas fraction, and to assign a number to them which expresses the confidence we have in this guess. In other words, we need to think about the plausibility of our values, and plausibility is, in Bayesian thinking, the same as probability. In other words, our task is to find an appropriate prior distribution function for each gas fraction in each environment. Whether this is then used in a conventional way to estimate error bars (as we do it here), or in a Bayesian way to marginalize over the gas fractions which are in principle only nuisance parameters for our analysis, is up to the reader. 

In order to find a proper prior probability distribution function (p.d.f., for which we use the symbol \pdf), we first note an interesting similarity between our best data, i.e., the depletion data compiled and analysed by Jenkins, and the ``psychology of guesses'': The spread of the data appears roughly constant in logarithmic representation regardless of the order of magnitude of the best fit depletion values, in the same way as we usually scale the estimated error assigned to a guess with the order of magnitude of the quantity we guess. This means that in our problem we do not master mean $\mu$ and standard deviation $\sigma$ of the distribution independently, but that rather the their quotient, known as the \textit{coefficient of variation}, $\cv = \sigma/\mu$, is the parameter that determines the above-mentioned confidence or plausibility assigned to our values. And as the spread in the data seems to be symmetric around their mean in logarithmic representation, it also seems that the \textit{truncated lognormal distribution}, the properly normalized p.d.f. of which is
\begin{equation}\label{eq:trunc_lognormal}
    \pdf\big(\fraction{gas}\big) = \left.\frac{1}{\fraction{gas}}\exp\left(-\frac{\big(\ln\fraction{gas}-\tilde\mu\big)^2}{\phantom{\fraction{gas}}\tilde\sigma^2\phantom{\fraction{gas}}}\right)\right/ \tilde\sigma\sqrt{\frac{\pi}{2}}\,\erfc\left(\frac{\tilde\mu}{\phantom{\big|}\tilde\sigma\sqrt{2}\phantom{\big|}}\right)
\end{equation}
defined in and normalised over the physical range of gas fractions, $0\le\fraction{gas}\le1$, is an excellent choice for a prior pdf. The parameters of this distribution are linked to our best guess $\mu$ and coefficient of variation $\cv$ in the following simple way
\begin{eqnarray}
\tilde{\mu} &=& \ln\mu - {\textstyle\frac12} \ln\left(\cv^2+1\right)\\
\tilde{\sigma} &=& \sqrt{\ln\left(\cv^2+1\right)}\quad.
\end{eqnarray}
Note that neither $\tilde\mu$ and $\tilde\sigma$ are the mean and standard deviation of the distribution Eq.~\ref{eq:trunc_lognormal}, nor are our best guess value $\mu$ and $\sigma\equiv\mu\times\cv$ -- the latter have these properties only for the \textit{non-truncated} lognormal distribution. This complication, however, should not bother us given the nature of our considerations here, it just has to be kept in mind. 

\begin{table}
\begin{small}
\caption{\label{galactic_abundances} Cosmic abundances and bet guess gas fractions of elements used in this work. Logarithmic abundances $[X]$ are normalised to $[H]=12.0$ and taken from Ref. \cite{PalmeLoddersJones2014}. For each element we give the most relevant isotopes considered in this paper with the abundance measured on Earth. See Sect.~\ref{sec:abundances} for details and the relation to normalised abundances. For gas fractions, the best guess $\mu$ for each element and environment is given together with an integer confidence value $\icv$, see Sect.~\ref{sec:dust} and \ref{app:GasFrac_approach} for details. Note the remarks on individual elements in Sect.~\ref{app:GasFrac_individual}.}
\vspace{4.2pt}
\begin{tabular}{lrlr|@{$\qquad$}rc@{$\qquad$}rc@{$\qquad$}rc@{$\qquad$}rc}\hline
symbol &  $Z$ 	& $A$		& $[X]$  	&\mcol{2}{l}{\fraction[CNM]{}} 	& \mcol{2}{l}{\fraction[WNM]{}}  	& \mcol{2}{l}{\fraction[WIM]{}}	& \mcol{2}{l}{\fraction[HIM]{}}\\
	&		& isotopes	& solar	& $\mu$   & \icv			& $\mu$    & \icv			& $\mu$   & \icv			& $\mu$   & \icv  		\\\hline
 H 	&  $  1$ 	& $1$		& $12.00$ 	& $1.00$ & $\infty$		&  $1.00$ & $\infty$		& $1.00$ & $\infty$  		& $1.00$ & $\infty$	\\
 He 	&  $  2$ 	& $4$		& $10.93$ 	& $1.00$ & $\infty$		&  $1.00$ & $\infty$  		& $1.00$ & $\infty$  		& $1.00$ & $\infty$  	\\
 Li 	&  $  3$ 	& $6,7$		& $  3.27$ 	& $0.05$ & $1$       		&  $0.15$ & $1$       		& $0.50$ & $0$       		& $0.50$ & $0$       	\\
 Be 	&  $  4$ 	& $9$		& $  1.38$ 	& $0.05$ & $1$       		&  $0.15$ & $1$       		& $0.50$ & $0$       		& $0.50$ & $0$       	\\
 B 	&  $  5$ 	& $10,11$		& $  2.70$ 	& $0.10$ & $1$       		&  $0.30$ & $1$       		& $0.50$ & $0$       		& $0.50$ & $0$       	\\
 C 	&  $  6$ 	& $12,13$		& $  8.50$	& $0.62$ & $3$       		&  $0.68$ & $3$       		& $0.78$ & $3$       		& $0.85$ & $2$       	\\
 N 	&  $  7$ 	& $14$		& $  7.86$ 	& $1.00$ & $1$       		&  $1.00$ & $1$       		& $1.00$ & $1$       		& $1.00$ & $1$       	\\
 O 	&  $  8$ 	& $16$		& $  8.73$ 	& $0.60$ & $1$       		&  $0.75$ & $1$       		& $0.95$ & $1$     		& $1.00$ & $3$   	\\
 F 	&  $  9$ 	& $19$		& $  4.56$ 	& $0.20$ & $1$       		&  $0.50$ & $0$       		& $1.00$ & $1$     		& $1.00$ & $\infty$  	\\
 Ne 	&  $10$ 	& $20,22$		& $  8.05$ 	& $1.00$ & $\infty$		&  $1.00$ & $\infty$  		& $1.00$ & $\infty$  		& $1.00$ & $\infty$  	\\
 Na 	&  $11$ 	& $23$		& $  6.30$ 	& $0.10$ & $1$       		&  $0.25$ & $1$       		& $0.50$ & $0$       		& $0.50$ & $0$       	\\
 Mg   &  $12$ 	& $24{-}26$	& $  7.54$ 	& $0.05$ & $3$       		&  $0.17$ & $3$       		& $0.54$ & $3$       		& $1.00$ & $3$       	\\
 Al     &  $13$	& $27$		& $  6.47$	& $0.02$ & $2$       		&  $0.06$ & $2$       		& $0.25$ & $1$       		& $0.50$ & $0$       	\\
 Si    	&  $14$	& $28{-}30$	& $  7.52$	& $0.05$ & $2$       		&  $0.15$ & $2$       		& $0.60$ & $2$       		& $1.00$ & $3$       	\\
 P     	&  $15$ 	& $31$		& $  5.46$	& $0.20$ & $2$       		&  $0.45$ & $2$       		& $1.00$ & $10$     		& $1.00$ & $\infty$	\\      
 S     	&  $16$ 	& $32{-}34$	& $  7.17$ 	& $0.30$ & $1$       		&  $0.50$ & $0$       		& $1.00$ & $1$     		& $1.00$ & $10$     	\\
 Cl    	&  $17$ 	& $35,37$		& $  5.50$	& $0.20$ & $2$       		&  $0.45$ & $2$       		& $1.00$ & $10$     		& $1.00$ & $\infty$     \\
 Ar    	&  $18$ 	& $40$		& $  6.50$ 	& $0.50$ & $0$       		&  $1.00$ & $2$       		& $1.00$ & $10$       		& $1.00$ & $\infty$	\\
 K     	&  $19$	& $39,41$		& $  5.11$ 	& $0.03$ & $1$       		&  $0.10$ & $1$       		& $0.50$ & $0$       		& $0.50$ & $0$       	\\
 Ca	&  $20$ 	& $40,42,44$	& $  6.33$ 	& $10^{-4}$ & $1$        		&  $10^{-3}$ & $1$			& $0.01$ & $1$  		& $0.50$ & $0$		\\
 Sc &  $21$ 	& $45$		& $  3.10$ 	& $0.02$ & $1$       		&  $0.10$ & $1$       		& $0.50$ & $0$       		& $0.50$ & $0$       	\\
 Ti     &  $22$ 	& $46{-}50$	& $  4.90$	& $10^{-3}$ & $1$        		&  $0.01$ & $1$			& $0.10$ & $1$         		& $0.80$ & $1$		\\
 V 	&  $23$ 	& $51$		& $  4.00$	& $10^{-3}$ & $1$        		&  $0.01$ & $1$			& $0.10$ & $1$         		& $0.50$ & $0$		\\
 Cr   	&  $24$ 	& $50,52{-}54$	& $  5.64$	& $0.01$ & $3$        		&  $0.03$ & $2$			& $0.15$ & $2$         		& $0.80$ & $1$		\\	
 Mn 	&  $25$ 	& $55$		& $  5.37$	& $0.02$ & $1$        		&  $0.05$ & $1$			& $0.10$ & $1$         		& $0.50$ & $0$		\\
 Fe	&  $26$ 	& $54,56,57$	& $  7.48$ & $0.01$ & $10$        		&  $0.02$ & $3$			& $0.10$ & $3$         		& $0.50$ & $2$		\\
 Co 	&  $27$ 	& $59$		& $  4.92$ & $0.01$ & $2$        		&  $0.02$ & $1$			& $0.10$ & $1$         		& $0.50$ & $0$		\\ 
 Ni  	&  $28$ 	& $58,60{-}64$	& $  6.23$ & $0.01$ & $10$        		&  $0.02$ & $2$			& $0.10$ & $2$         		& $0.60$ & $1$		\\ 
 Cu  	&  $29$ 	& $63,65$	& $  4.21$ & $0.05$ & $3$        		&  $0.11$ & $3$			& $0.25$ & $3$         		& $0.55$ & $2$		\\ 
 Zn  	&  $30$ 	& $64,66{-}68$	& $  4.62$ & $0.30$ & $2$        		&  $0.55$ & $2$			& $1.00$ & $1$         		& $1.00$ & $10$		\\ 
 \end{tabular}
\end{small}
 \end{table}

In this paper, we find it convenient to use the inverse coefficient of variation, $\icv=\mu/\sigma$, to classify the confidence of our guesses by assigning integer numbers to it. The case $\icv=1$ may be considered an ``educated guess'', e.g., $\mu=0.05$ with $\icv=1$ means ``we know that this element behaves largely refractory in this ISM phase''. The values $2$, $3$ or $10$ for \icv\ usually mean that our choice for $\mu$ is based on measurements, referring to the multiple of a classical 1-sigma error. We use $\icv=10$ also in connection to $\mu=1$ to express a high (but not complete) confidence that the element is almost entirely in the gas phase. In this case ($\mu=1$) we also allow for $\icv=\infty$ to express complete confidence for a fully gaseous phase, like for light noble gases or hydrogen, for which it is empirically known that it appears in molecules deposited on dust grains only to a negligible fraction. Also the assignment $\icv=0$ is meaningful and used to express complete ignorance in cases where no meaningful guess can be made; we then always set $\mu=0.5$, and note that the distribution Eq.~\ref{eq:trunc_lognormal} indeed approaches a flat distribution over the interval $[0,1]$ for $\icv\to 0$.

\subsection{Remarks on individual elements}
\label{app:GasFrac_individual}

\begin{description}

\item[Hydrogen (H)] is known to be bound in dust, but only to a fraction ${\sim}10^{-4}$ of its total abundance. As this is much below our accuracy goal of $1\%$, we set $\fraction{H}=1$ for all ISM phases with no uncertainty. 

\item[Helium (He) and neon (Ne)] are light noble gases, chemically inert and extremely volatile. It is generally assumed that they reside to 100\% in the gas phase, so we also adopt $\fraction{He,Ne}=1$ in all ISM phases. The primordial element He is second in cosmic abundance and in LECR as well, Ne is \#5 in solar abundance and falls back to \#8 in LECR source abundance behind Mg, Si and Fe. Reproducing the LECR abundance of pure gas phase elements compared is an important test for our acceleration model and a gauge for our free parameters. 

\item[Lithium (Li), beryllium (Be) and boron (B)] are extremely rare elements, which are\break strongly enhanced in LECR because they are produced as spallation products of heavier cosmic rays during propagation. It is thus not meaningful to consider these elements in this paper which is mainly about source abundances, they are added here just for completeness. Our priors rely on just a few gas-phase column data points on Li and B, for Be we just assume the same properties as for Li.

\item[Carbon (C)] is \#4 in both solar and LECR abundance. While it was long thought to be highly volatile and thus present mostly in the ISM gas phase, we know today that it is depleted in the neutral ISM by about $30\%$, roughly half of which is not in classical dust grains, but in so-called polycyclic aromatic hydrocarbon (PAH) macro-molecules that are extremely stable against collisions and thus likely survive even hot ISM phases \cite[and references therein]{PugetLeger1989, Tielens2008}. In fact, carbon depletion shows only weak dependence on the Jenkins depletion factor $F_*$, hence it is comparable in all ISM phases.  

\item[Nitrogen (N)] is usually assumed to be a gas phase element, but absorption line data rather suggest $\fraction{N} \!\approx 0.77$ with no trend in ISM phase temperatures. As this is physically not plausible, and it is worth to mention that while the zero-slope is well supported by the data, the offset from $\fraction{gas}=1$ is only supported on a $1\sigma$ level with the main uncertainty given by the unknown true abundance of N in the ISM. Obviously, there would be no problem to understand a constant gas fraction of 1, so we use a broad prior around this physically most plausible value that still assigns a significant probability to $\fraction{N}\!=0.77$. N is \#6 in solar abundances, but only \#9 in LECR behind Mg, Si, and Fe.

\item[Oxygen (O) and sulfur (S)] are the two elements of main group VI in the periodic table and known in the ISM community as ``troublemakers'', because they tend to provide contradictory information about their abundance and depletion properties, repeatedly feeding speculations that the ISM composition may not be well represented by the solar values. The problem is that both elements are highly volatile on the one hand, but on the other hand have a strong tendency to be chemically bound to refractory elements in form of oxides or sulfides. They are both found in micrometeorites with an atomic fraction that suggests $\fraction{dust}\sim0.5$, and only recently also absorption line measurements seem to confirm also $\fraction[CNM]{O,S}\sim 0.5$ for very cold and dense phases of the ISM where dust is formed. In warmer phases these elements seem to evaporate largely back into the gas phase, and often data suggest gas densities larger than what the solar abundance values would allow (i.e., $\fraction{gas}>1$). For both elements, we leave the possibility that some of it remains bound in the core grains of dust even in ionized phases, but assign to $\fraction{O,S}\simeq 1$ the highest probability. In spite of all scepticism, we do not exchange the solar abundance values for potential higher cosmic abundances, as there is no agreement on this in the ISM community. Regarding total abundance, O is ranking \#3 and S ranking \#10 in both solar and LECR abundances.

\item[Fluorine (F), phosphorus (P) and chlorine (Cl)] show a similar behaviour as O and S, from significantly depleted in cold ISM phases to fully in the gas phase for the HIM. Also here, absorption line data often lead to unphysical gas fractions, in particular $\fraction{Cl} > 1$ for warm environments. Although these elements are of little or no relevance for the total cosmic ray flux, it turns out in our investigation that they are quite relevant for the ``LECR fingerprint'', which we use to determine the sited of Galactic cosmic ray production. So we may consider the possibility that as for S and O, the depletion of Cl and also P might be underestimated, thus gas fractions in the WNM and WIM might be slightly, but systematically larger than given in Table \ref{galactic_abundances}.

\item[Sodium (Na), magnesium (Mg), aluminum (Al) and silicon (Si)] form together with Ne the ``intermediate mass group'' of cosmic rays, and contain with Mg and Si elements \#5 and \#6 in the LECR abundance ranking---in solar abundances they are  \#7 and \#8. These two elements behave similar regarding gas phase depletion, with $\fraction[CNM]{gas} \le 0.1$ in cold phases and $\fraction[WNM]{gas}\gtrsim 0.1$ in the WNM. In the ionized phases they both seem to return largely 
into the gas phase, which fits together with the fact that they are mostly found in loose grain structures in micrometeorites. A similar behaviour can be assumed for Na, but while for Mg and Si very good data on gas-phase depletion in the cold and warm interstellar medium exist, little is known about Na beyond the CNM and only crude guesses are possible for its gas fraction in other phases. In contrast to these elements, Al is known to be highly refractory and there is evidence that it is strongly depleted in the local interstellar cloud \cite{Kimura+2003_LIC}, which belongs to the WNM. The fact that it is found in the dense cores of micrometeorites lets us suppose that may be depleted even in ionized phases, but due to the poor data constraints we have to keep our priors broad there. 

\item[Argon (Ar)] is worth mentioning on its own, as it is one of the more abundant elements in between the intermediate mass group and iron. Although it is a noble gas we cannot exclude that it is to some degree depleted from the gas phase and condenses on grain surfaces by some mechanism. Absorption line measurements even suggest $\fraction[CNM]{Ar}\approx 0.3$, but this may be explained by the possibility that Ar can ``hide'' in an ionized state rather than by deposition on dust \cite{SofiaJenkins1998}. Our scepticism is justified by a possible depletion trend seen for the heavier noble gas Krypton, where a condensation on dust is rendered a possible explanation. In this paper we keep $\fraction[CNM]{Ar}$ open ($\icv{=}\,0$), and assume for other environments that $\fraction{Ar}=1$ has the highest probability to be correct with depletion still considered possible the more, the lower the temperature is. 

\item[Potassium (K) and scandium (Sc)] behave essentially like the elements of the Na-Si group. For K gas-phase abundances it has been shown that they are proportional to those of Na, but generally a factor ${\sim}3$ more suppressed \cite{Kemp+2002_NaK}. For Sc we simply assume similarity to Al based on similar chemical properties of both elements. Neither element plays a role for cosmic rays due to their low abundances, they are included here only for completeness. 

\item[Calcium (Ca)] is the other relevant in-between element besides Ar. It is again worth mentioning on its own because it is the strongest depleted element that exists. In all phases of the ISM except the HIM, $\fraction{Ca}\ll 0.1$, and as it is not unimportant for cosmic rays ranking \#11 in LECR source abundances, it may serve as a gauge for the efficiency of dust grain acceleration. Our assumption that Ca may return significantly into the gas phase in the HIM emerges from micrometeorite analysis where Ca is mostly found in coarse structures. If this can be taken as a hint that it resides mostly in the mantles of dust grains, it may largely evaporate in hot ISM phases, but again we keep $\fraction{Ca}$ open ($\icv{=}\,0$). 

\item[Iron (Fe) and the iron group (Ti, V, Cr, Mn, Co, Ni, Cu)] form the heaviest part of the cosmic ray spectrum, with Fe, ranking \#9 in solar and \#7 in LECR abundances, vastly dominating this group. All these elements have in common that they show low gas fractions $\fraction{gas}\lesssim 0.1$ except in the HIM; for the latter, many elements largely return into the gas phase, but for Fe, Ni and Cu our priors essentially exclude $\fraction{gas}=1$. This result, obtained from absorption line measurements, is consistent with the expectation that these are the elements forming the resilient grain cores of dust that may exist in the HIM (with Cu usually not mentioned because of its low total abundance). A particularity of iron is that from micrometeorite composition analysis one would expect $\fraction{dust}\sim 0.5$, which is in clear contradiction to the low gas-phase fractions measured in dense neutral ISM phases. A possible solution for this may be that a significant part of the cosmic iron could reside in large molecules called iron-pseudocarbynes \cite{Tarakeshwar+2019}. Also these molecules would have such large masses and low ionization states that we can treat them like small dust grains in our acceleration approach, but it is not obvious whether these molecules would remain stable in the HIM like PAHs. 

\item[Zinc (Zn)] is the heaviest element we still consider, and although its abundance is too low to make a significant contribution to the total cosmic ray flux, it is abundant enough that there are still good LECR data for it. It is an interesting gauge-point for our model as it is again a volatile element that resides largely in the gas phase for ionized environments. 

\end{description}

\section{Additional results on the low energy abundances}
\label{app:AddResults}
\begin{figure}[htb]
\centering 
\includegraphics[width=.99\linewidth]{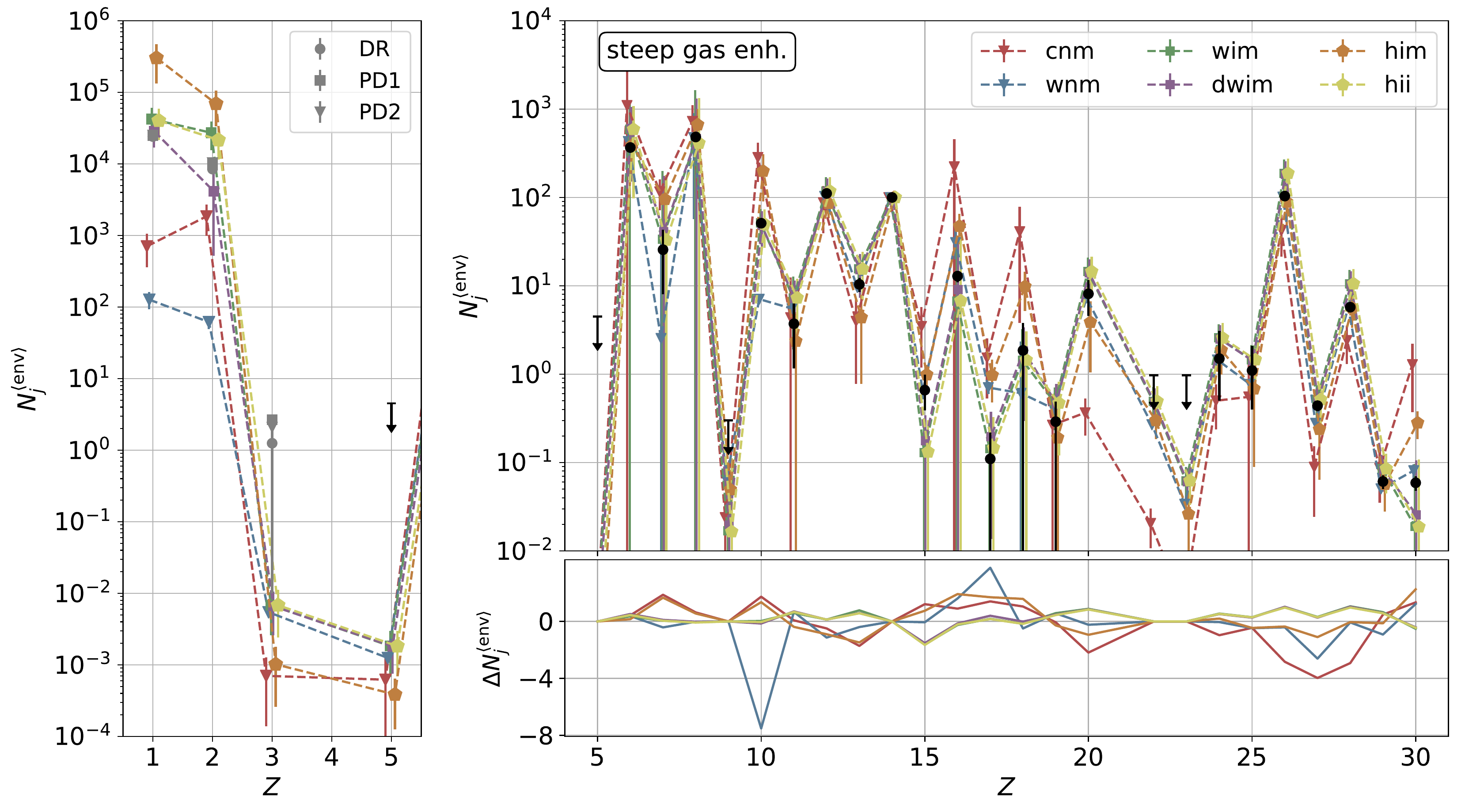}
\includegraphics[width=.99\linewidth]{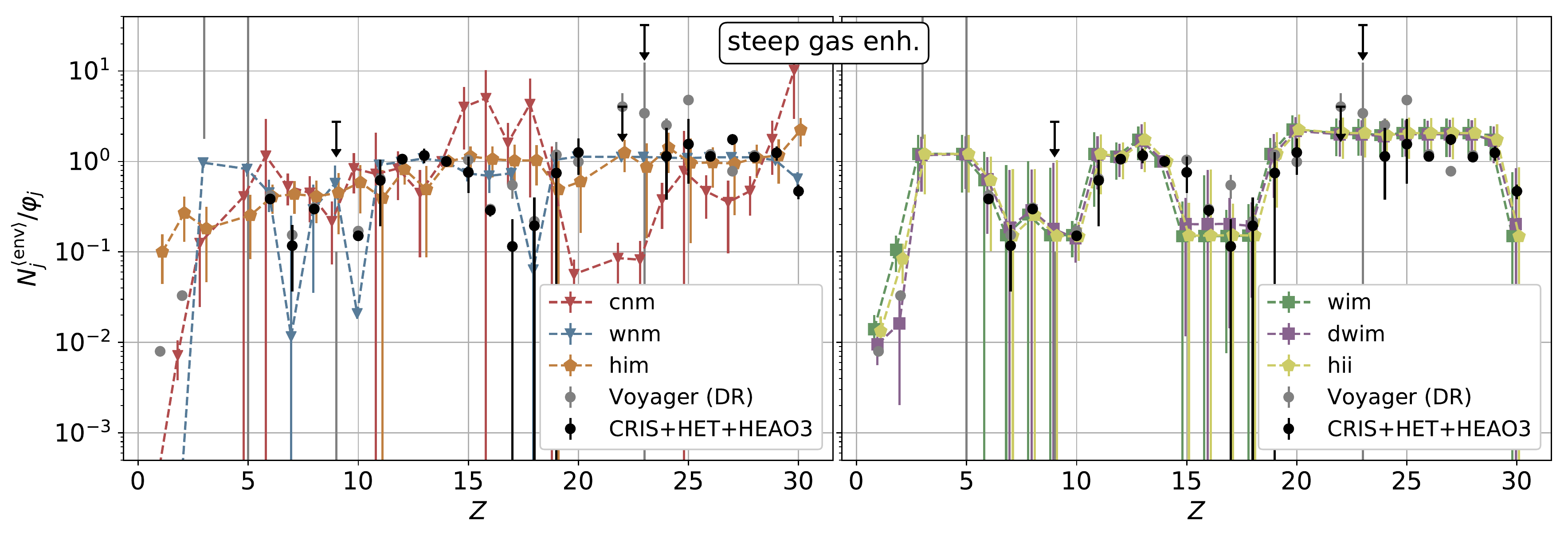} 
\caption{The best-fit model using the steep gas enhancement (\ref{eq:C+17enhanc}); see the caption of Fig.~\ref{fig:bestfit_Hanusch} for more information.\label{fig:bestfit_Caprioli}}
\end{figure}

\begin{figure}[h]
\includegraphics[width=.99\linewidth]{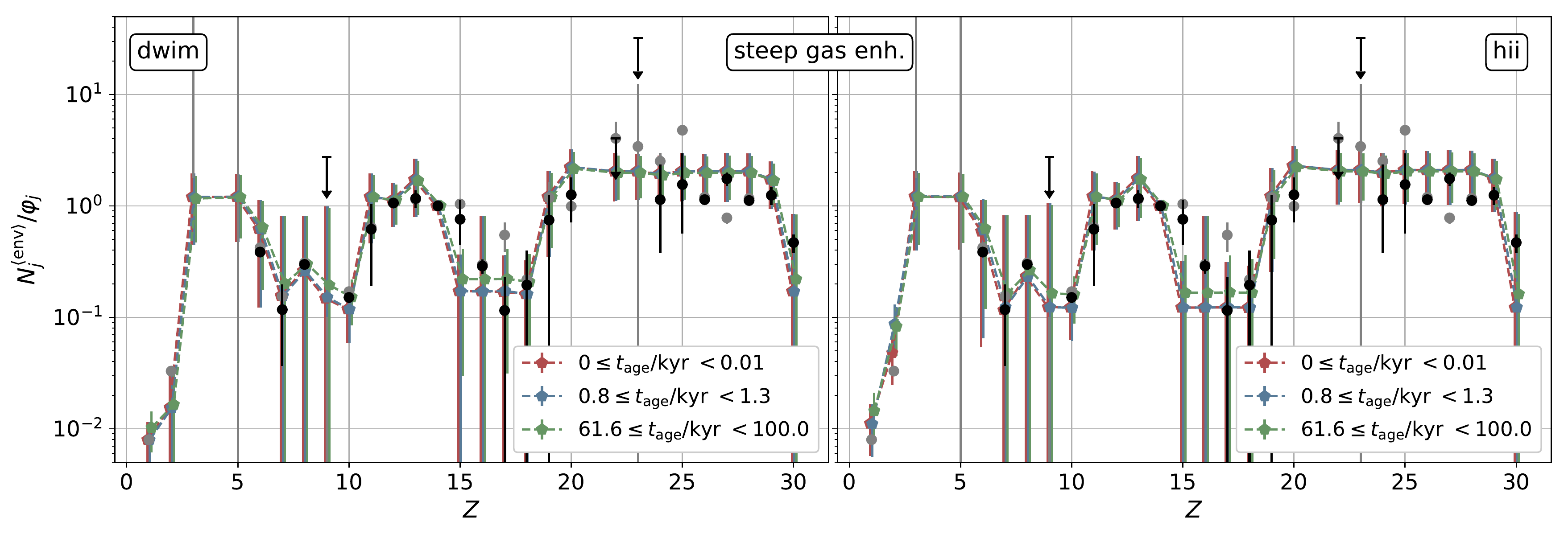} 
\caption{The enhancement of elements for the DWIM (\emph{left}) and HII-regions (\emph{right}) at three different time intervals using the best-fit model for the steep gas enhancement (\ref{eq:C+17enhanc}). See the caption of Fig.~\ref{fig:bestfit_Evol_Hanusch} for more information.
\label{fig:bestfit_hiiEvol_Hanusch}}
\end{figure}

For completeness, we present here the results on the LECR abundances that we omitted from the main text.
Figure~\ref{fig:bestfit_Caprioli} provides the predictions on the LECR abundances in the case of the best-fit model using the steep gas enhancement, and Fig.~\ref{fig:bestfit_hiiEvol_Hanusch} shows the results enhancement factors obtained for different evolution times of the SNR shock, like in Fig.~\ref{fig:bestfit_Evol_Hanusch}, but here for the steep gas enhancement model. These results do not add anything beyond what have been discussed in Sect.~\ref{sec:results} with respect to Fig.~\ref{fig:bestfit_Hanusch} already.  

\end{document}